\documentclass[a4paper, 11 pt , english]{article}
\usepackage{ulem}
\usepackage[english]{babel}
\usepackage{amsthm}
\usepackage{pstricks,pst-node,pst-tree}
\usepackage{amssymb}
\usepackage[utf8]{inputenc}
\usepackage[T1]{fontenc} 
\usepackage{fancybox} 
\usepackage{alltt}
\usepackage{graphicx}
\usepackage{bm}
\usepackage{lmodern}                      
\usepackage{babel}
\usepackage[top = 3cm, bottom = 3 cm, right = 3cm, left = 3 cm]{geometry}
\usepackage{appendix}
\usepackage{float}
\usepackage{xcolor}
\usepackage{pgf,tikz,pgfplots}
\pgfplotsset{compat=1.15}
\usepackage{mathrsfs}
\usepackage{dsfont}
\usetikzlibrary{arrows}
\usepackage{graphicx}
\usepackage{amsmath}
\usepackage{amssymb}
\usepackage{mathtools}              
\usepackage[T1]{fontenc}      
\usepackage{lmodern}          
\usepackage[utf8]{inputenc}      
\usepackage{babel}
\usepackage[linesnumbered, french]{algorithm2e}
\usepackage{physics}
\usepackage{caption}
\usepackage{subcaption} 
\usepackage{caption}
\usepackage{enumitem}
\usepackage{mathtools}
\usepackage{empheq}
\usepackage[most]{tcolorbox}
\usepackage{stmaryrd}
\usepackage{graphicx}
\newtheorem{Theorem}{Theorem}[section]
\newtheorem{Definition}[Theorem]{Definition}
\newtheorem{Proposition}[Theorem]{Proposition}
\newtheorem{Assumption}[Theorem]{Assumption}
\newtheorem{Lemma}[Theorem]{Lemma}

\newtheorem{Remark}[Theorem]{Remark}

\newtheorem{Condition}[Theorem]{Condition}

\def \be{\begin{eqnarray}}
\def \ee{\end{eqnarray}}
\def \b*{\begin{align*}}
\def \e*{\end{align*}}

\def \E{\mathbb{E}}

\def \N{\mathbb{N}}
\def \P{\mathbb{P}}

\def \R{\mathbb{R}}

\def \W{\mathbb{W}}
\def \X{\mathbb{X}}

\def \[{[\,\!\![}
\def \]{]\,\!\!]}

\def \upGMad{\overline{G}^{\rm M}_{\rm ad}}
\def \lowGMad{\underline{G}^{\rm M}_{\rm ad}}
\def \upGMmiad{\overline{G}^{{\rm M}, {\rm m}_1}_{\rm ad}}
\def \lowGMmiad{\underline{G}^{{\rm M}, {\rm m}_1}_{\rm ad}}
\def \upGmiad{\overline{G}^{{\rm m}_1}_{\rm ad}}
\def \lowGmiad{\underline{G}^{{\rm m}_1}_{\rm ad}}

\def \upGMc{\overline{G}^{\rm M}}
\def \lowGMc{\underline{G}^{\rm M}}

\def \apGMcl{\hat{G}^{\rm M}}
\def \apGMad{\hat{G}_{\rm ad}^{\rm M}}
\def \apGMmiad{\hat{G}_{\rm ad}^{ {\rm M}, {\rm m}_1} }

\def \Bad{B_{\W_p^{\rm ad}}}
\def \BadM{B^{\rm M}_{\W_p^{\rm ad}}}
\def \BadMmi{B^{{\rm M}, {\rm m}_1}_{\W_p^{\rm ad}}}
\def \Badmi{B^{{\rm m}_1}_{\W_p^{\rm ad}}}

\def \Bcl{B_{\W_p}}
\def \BclM{B^{\rm M}_{\W_p}}

\def \Nad{\mathbf{N}_{\rm ad}}

\def \1{{\bf 1}}

\def \proof{{\noindent \bf Proof. }}

\def \ep{\hbox{ }\hfill$\Box$}

\def\Ac{{\cal A}}
\def\Bc{{\cal B}}

\def\Lc{{\cal L}}

\def\Nc{{\cal N}}
\def\Kc{{\cal K}}

\def\Pc{{\cal P}}

\def\namedlabel#1#2{\begingroup
    #2%
    \def\@currentlabel{#2}%
    \phantomsection\label{#1}\endgroup
}

\def\eps{\varepsilon}

\usepackage{epsfig}
\usepackage{fullpage}
\usepackage{fancyhdr}
\usepackage{moreverb}
\usepackage{xspace}
\usepackage[colorlinks,hyperindex,bookmarks,linkcolor=blue,citecolor=blue,urlcolor=blue]{hyperref}

\usepackage{wrapfig}
\usepackage{epsf}
\usepackage[maxnames = 50]{biblatex}
\DeclareFieldFormat{title}{#1}
\DeclareFieldFormat[article]{title}{#1}
\DeclareFieldFormat[inproceedings]{title}{#1}
\DeclareFieldFormat[incollection]{title}{#1}
\DeclareFieldFormat[article]{journaltitle}{\textit{#1}}
\DeclareFieldFormat[inproceedings]{booktitle}{\textit{#1}}
\DeclareFieldFormat[incollection]{booktitle}{\textit{#1}}
\DeclareFieldFormat[book]{title}{\textit{#1}}
\AtEveryBibitem{%
  \ifentrytype{article}{\clearfield{publisher}}{}%
}
\AtEveryBibitem{%
  \iffieldundef{doi}{}{\clearfield{url}}%
}
\AtBeginBibliography{%
  \sloppy
  \emergencystretch=3em
}

\title{\bf First-order Martingale model risk 
             \\ and semi-static hedging}
\author{Nathan Sauldubois\thanks{New York University, 
                                            Tandon School of Engineering, 
                                            ns6982@nyu.edu.}\and 
             Nizar Touzi\thanks{New York University, 
                                            Tandon School of Engineering, 
                                            nizar.touzi@nyu.edu.}
           }
\date{\today}

\begin{document}

\maketitle
\vspace{-6mm}

\begin{abstract}	
We investigate model risk distributionally robust sensitivities for functionals on the Wasserstein space when the underlying model is either constrained to the martingale class and/or is subject to constraints on the first marginal law. Our results extend the findings of Bartl, Drapeau, Obloj \& Wiesel \cite{bartl2021sensitivity} and Bartl \& Wiesel \cite{bartlsensitivityadapted} by introducing the minimization of the distributionally robust problem with respect to semi-static hedging strategies. We provide explicit characterizations of the model risk (first-order) optimal semi-static hedging strategies. The distributional robustness is analyzed both in terms of the adapted Wasserstein metric and the more relevant standard Wasserstein metric. 
\end{abstract}

\noindent{\bf MSC2020.} 49K45, 49Q22.

\vspace{3mm}
\noindent{\bf Keywords.} Distributionally robust optimization, adapted Wasserstein distance, optimal transport.
\vspace{-2mm}

\section{Introduction}

Modelling plays a crucial role in providing a simple representation of phenomena. 
Random modelling consists of choosing a probability measure $\mu$ on an underlying state space $\X$ hosting the object of interest $X$. 
It also involves specifying a criterion $g$, which is often derived from a decision-making mechanism, typically involving optimization or equilibrium. 
Throughout this article, $g$ is a scalar map acting on the space of probability measures, so that $g(\mu)$ represents the resulting criterion from the chosen model $\mu$.
\vspace{-2mm}

\paragraph*{Model risk through sensitivity analysis.} 
A fundamental question is how to quantify the risk associated with choosing a given model $\mu$ relative to the criterion $g$. 
This question is traditionally addressed through sensitivity analysis within a finite-dimensional set of deviations. 
More precisely, the model $\mu$ is often chosen from a parametrized family $\{\mu(\theta), \theta\in\Theta\}$ where $\Theta\subset\R^k$ for some finite $k\ge 1$. 
This reduces the problem to selecting a parameter $\theta$ consistent with current information and historical data.
Consistency is achieved by combining calibration and statistical estimation. 
Model risk is then measured by evaluating the sensitivities $\partial_\theta g(\mu(\theta))$; see \citeauthor{greeks}\cite{greeks}. 
Naturally, a good model should be such that this vector of sensitivities is small in an appropriate sense. 
However, this approach can become cumbersome due to the high dimensionality of $\theta$.
As a possible remedy, a large part of the literature has focused on the robust criterion evaluation, which explores the worst-case scenario
$$
\sup_{|\theta'-\theta|\le r} g(\mu(\theta')),
$$
for some choice of finite-dimensional norm $|\cdot|$. 
This problem is appealing as it reduces model risk evaluation to a scalar function of a single variable. 
In particular, we may analyze its sensitivity at the origin, which provides the first-order correction for the worst deviation from the parameter $\theta$. 
However, from a conceptual viewpoint, the superfluous nonlinear map $\theta\longmapsto\mu(\theta)$ should not play a crucial role in this problem, as our main interest lies in the chosen model $\mu(\theta)$. 
Nonetheless, it may add nontrivial technical difficulties and restrict exploration to a prescribed finite-dimensional family of models $\{\mu(\theta),\theta\in\Theta\}$.
\vspace{-2mm}

\paragraph*{Model risk through distributionally robust optimization.} 
An original perspective, well studied in the literature, considers the robust evaluation over a neighborhood of models. 
This worst-case formulation, called distributionally robust optimization (DRO), requires introducing a substitute for our finite-dimensional distance. 
Many choices have been considered in the literature: the Kullback divergence in \citeauthor{lam_robust_2016} \cite{lam_robust_2016}, the total variation distance in \citeauthor{farokhi_distributionally-robust_2023} \cite{farokhi_distributionally-robust_2023}, or a criterion based on cumulative distribution functions in \citeauthor{bayraktar_nonparametric_2023} \cite{bayraktar_nonparametric_2023}. 
Following \citeauthor{mohajerin_esfahani_data-driven_2018} \cite{mohajerin_esfahani_data-driven_2018} and \citeauthor{blanchet_quantifying_2016} \cite{blanchet_quantifying_2016}, we consider distributionally robust optimization based on the $p$-Wasserstein distance\footnote{See also \citeauthor{neufeld_robust_2024} \cite{neufeld_robust_2024}, \citeauthor{fuhrmann_wasserstein_2023} \cite{fuhrmann_wasserstein_2023}, \citeauthor{blanchet_statistical_2023} \cite{blanchet_statistical_2023}, \citeauthor{nendel_parametric_2022} \cite{nendel_parametric_2022}, \citeauthor{blanchet_distributionally_2018} \cite{blanchet_distributionally_2018}, \citeauthor{blanchet_robust_2019} \cite{blanchet_robust_2019}, \citeauthor{lam_robust_2016} \cite{lam_robust_2016}, \citeauthor{blanchet_distributionally_2024} \cite{blanchet_distributionally_2024}, \citeauthor{lanzetti_first-order_2022} \cite{lanzetti_first-order_2022}.}
\vspace{-2mm}
$$
G(r):=\sup_{ \W_p ( \mu, \mu' ) \le r} g(\mu').
\vspace{-2mm}
$$
The one-variable scalar function $G$ has many desirable properties. 
For instance, if $g$ is concave then $\rho\longmapsto G(\rho^{1/p})$ is also concave and has left and right derivatives at every point, a property inherited by $G$ at every $r>0$, though the behaviour at zero remains undetermined. 
The existence of the sensitivity at the origin, representing the worst model risk, was established in the remarkable paper by \citeauthor{bartl2021sensitivity} \cite{bartl2021sensitivity} with an appealing expression in the context of a specific example of concave maps $g$ as
\vspace{-2mm}
$$
G'(0) = \|\partial_x\delta_m g \|_{\mathbb{L}^{p'}(\mu)} := \E^{\mu} [ \vert \partial_x\delta_m g\vert^{p'} ]^{1/p'} ,
~~\mbox{with}~~
\frac1p+\frac{1}{p'}=1.
\vspace{-1.5mm}
$$ 
Here $\delta_m$ denotes the linear functional derivative in the set of probability measures, and $\partial_x\delta_m$ is the Wasserstein gradient, which coincides with the Lions derivative under mild regularity and growth conditions, see \citeauthor{CarmonaDelarue} \cite{CarmonaDelarue}. 
We also refer to the subsequent work by \citeauthor{bartlsensitivityadapted} \cite{bartlsensitivityadapted}, who instead consider the adapted Wasserstein distance $\W^{\rm ad}_p$ to extend to dynamic problems such as optimal stopping, motivated by American options in finance, see \citeauthor{backhoff_adapted_2020} \cite{backhoff_adapted_2020}, \citeauthor{pflug_multistage_2014} \cite{pflug_multistage_2014}, \citeauthor{backhoff_all_2020} \cite{backhoff_all_2020}. 
\vspace{-1mm}

\paragraph*{Our contribution: Model risk hedging.} 
The above discussion applies broadly to all engineering models. 
We now specialize the discussion to financial modeling, focusing on model risk management by introducing hedging instruments to decrease overall model sensitivities. 
The main objective of this article is to analyze model risk reduction via a subset of zero-cost hedging instruments $\mathfrak{h}\in\mathfrak{H}$. 
Here $\mathfrak{h}$ is a map on $\X$ satisfying $\E^{\mu'}[\mathfrak{h}]=0$ for all $\mu'$, reflecting the zero-cost condition. 
In this article, we focus on the context $\X=\R^d\times\R^d$, with $X_1,X_2$ the projection coordinates representing the prices of an underlying security at two points in time. 
Typical examples of such zero-cost hedging strategies are the following:
\begin{itemize}
\item all returns from buy-and-hold strategies $\mathfrak{H}_{\rm M}:= \{h(X_1)\cdot(X_2-X_1):h\in C^0_b(\R^d, \R^d) \}$;
\item all returns from a Vanilla payoff with maturity $T=1,2$, inducing the set of zero-cost strategies $\mathfrak{H}_{\rm m_T}:= \{f(X_T)-\mu_T(f):f\in C^0_b(\R^d) \}$, where $\mu_T=\mu\circ X_T^{-1}$ denotes the $T$-th marginal of $\mu$. 
\end{itemize}

Given a subset $\mathfrak{H}$ of hedging instruments, we define the upper and lower model distributionally robust hedging problems by
\vspace{-2mm}
$$
\overline{G}^{\mathfrak{H}}(r)
:=
\inf_{\mathfrak{h}\in\mathfrak{H}}\sup_{ \W_p ( \mu, \mu' ) \le r} \{g(\mu')+\mu'(\mathfrak{h}) \},
\qquad
\underline{G}^{\mathfrak{H}}(r)
:=
\sup_{  \W_p ( \mu, \mu' ) \le r} 
\inf_{\mathfrak{h}\in\mathfrak{H}} \{g(\mu')+\mu'(\mathfrak{h}) \}.
\vspace{-2mm}
$$ 
We similarly define the corresponding problems $\overline{G}^{\mathfrak{H}}_{\rm ad}$ and $\underline{G}^{\mathfrak{H}}_{\rm ad}$ by replacing the Wasserstein distance with its adapted version.
 
Our first main result is that, throughout all the examples of hedging instrument sets $\mathfrak{H}$ considered in this article, the lower and upper hedging distributionally robust values are differentiable at the origin with
\vspace{-2mm}
$$
\left. \overline{G}^{\mathfrak{H}} \right.'(0)
=
\left. \underline{G}^{\mathfrak{H}} \right. '(0)
=
\|\partial_x(\delta_m g+\hat{\mathfrak{h}}) \|_{\mathbb{L}^{p'}(\mu)},
\vspace{-2mm}
$$ 
where $\hat{\mathfrak{h}}$ is the unique minimizer of $\|\partial_x(\delta_m g+\mathfrak{h}) \|_{\mathbb{L}^{p'}(\mu)}$ among all $\mathfrak{h}$ ranging in an appropriate relaxation of $\mathfrak{H}$. The zero-cost strategy $\hat{\mathfrak{h}}$ represents the model risk-optimal hedging and is characterized explicitly or quasi-explicitly in our different hedging situations. 

A similar result is proved for the corresponding $\overline{G}^{\mathfrak{H}}_{\rm ad}$ and $\underline{G}^{\mathfrak{H}}_{\rm ad}$, with corresponding optimal hedging strategies $\hat{\mathfrak{h}}_{\rm ad}$. We emphasize that a similar result for $\underline{G}^{\mathfrak{H}_{\rm M}}_{\rm ad}$ appeared in the parallel work by \citeauthor{jiang2024sensitivity} \cite{jiang2024sensitivity}.

\paragraph*{Practical aspects.} 
The hedging strategies introduced in this article aim to minimize the model risk of a given financial position. 
The hedging strategies $\mathfrak{H}_{\rm M}$ restrict deviations in the corresponding DRO problem to models satisfying the martingale property, \textit{i.e.}, models that are arbitrage-free. 
Similarly, hedging strategies $\mathfrak{H}_{{\rm m}_T}$ restrict the deviations in the corresponding DRO problem to models calibrated to a prescribed $T$-marginal. 
Considering the hedging set $\mathfrak{H}_{{\rm M,m}_T}:=\mathfrak{H}_{\rm M}\cup\mathfrak{H}_{{\rm m}_T}$ induces deviations to arbitrage-free models with a calibrated $T$-marginal. 

\noindent We also analyze the adapted Wasserstein distance, which, by construction, yields lower model risk sensitivity and thus serves as a less conservative measure of risk uncertainty. 
However, restricting neighbouring models to those that are bi-causal (or causal in the case of causal optimal transport) with the reference model $\mu$ has no convincing foundation from the modeling viewpoint: why would one restrict deviations to satisfy the causality property with respect to the very model whose relevance is being questioned? For this reason, this article studies both sensitivities. 
From the technical perspective, the mathematical analysis of the martingale restriction is much easier under the adapted Wasserstein metric. 
Our numerical findings in Section \ref{sec:numerics} show that the martingale Wasserstein sensitivity is significantly higher than the corresponding sensitivity under the adapted Wasserstein metric.

\section{Notations and Definitions}
Throughout this article, we denote $S := \R^d$ and $\X := S \times S$ for some integer $d\ge 1$, both endowed with the corresponding canonical Euclidean structure and the associated norm $x\longmapsto\vert x \vert :=\sqrt{x\!\cdot\!x}$. 
An element of $\X$ is written as $ x := (x_1, x_2) \in \X$, for which we define 
\begin{equation}\label{eqdef:n and n_ad}
 \mathbf{N} (x)  := \frac{\nabla(\vert \cdot \vert^{p'}) (x) }{p'} 
 = \frac{x}{ \vert x \vert^{2 - p'}} \,\, , \quad 
 \mathbf{N}_{\rm ad} (x) := 
 \begin{bmatrix} 
 \mathbf{N} (x_1)
\\[3pt]
 \mathbf{N} (x_2)
\end{bmatrix}
,
\end{equation}
where  for $i = 1, 2$, $\mathbf{N} ( x_i ) := \frac{1}{p'} \nabla(\vert \cdot \vert^{p'}) (x_i) = \frac{1}{ \vert x_i \vert^{2 - p'}} x_i $. 
We abuse the same notation $\mathbf{N} (x_i) $ and $ \mathbf{N} (x) $ because the space that $ \mathbf{N} $ is defined on is implied by the variable $x$ or $x_i$. 
Let $C_b^1(S,S)$ denote the set of bounded functions $f:S\to S$ with bounded gradients. 
We define $C_b^1(S)$ similarly for real-valued functions. 
When no ambiguity is possible, the domain and codomain will be omitted from the notation.
For an open subset $U\subset S$ we denote 
\begin{equation}\label{eqdefLloc_sobo}
\text{ $W^{1, 1}_{{\rm loc}}(U)$ the set of weakly differentiable functions $f$ with $f, \partial_x f$ locally integrable.}
\end{equation}

\noindent Let $w_0,w_1:S\longrightarrow\R_+$ be two weights, let $w = (w_0, w_1)$ and let  $\infty > p>1$, we introduce the corresponding subset with finite weighted norm $\|\cdot\|_{p, w}$:
$$
\mathbf{W}^{p}(U, w) 
:= \!
\big\{ f \!\in\!  W^{1, 1}_{{\rm loc}}(U):  \|f \|_{p, w} < \infty \big\},
\, \text{where}~ 
\| f \|_{p, w} 
:= 
\! \Big( \int_U \!\!\big(|\partial_x f|^p w_1 \!+\! |f|^p w_0\big) \mathrm{d}x \Big)^{\frac1p},
$$
together with the following weighted Sobolev spaces:
\begin{equation}\label{eqdef:sobolev}
\begin{array}{l} 
W^p (U,w):~\mbox{the $\| \cdot \|_{p, w}-$completion of $\mathbf{W}^{p}( U, w)$,}
\\
H^p(U,w):~\mbox{the $\| \cdot \|_{p, w}-$completion of $\mathbf{W}^{p}( U, w)\cap C^{\infty}(U)$.}
\end{array}
\end{equation}
A scalar function is said to have $p$-polynomial growth if it is uniformly bounded by $C(1+\vert x\vert^p)$ for some constant $C$. 
As usual, we denote by $p'$ the conjugate exponent, \textit{i.e.} $ \frac{1}{p} + \frac{1}{p'} = 1$.

Let $\Pc(\X)$ be the collection of all probability measures on $\X$, with the subset of measures with finite $p$-th moment:
$$
\Pc_p ( \X ):=\big\{\mu\in\Pc (  \X ) :~\E^\mu[|X|^p]<\infty\big\}.
$$
\noindent
We introduce the projection coordinates $(X,X')$ on $\X\times \X$ defined by $X(x,x')=x$ and $X'(x,x')=x'$ for all $x,x'\in \X$. 
For $\mu,\mu'\in\Pc ( \X  ) \times \Pc ( \X  )$, we define the set of all couplings
$$
\Pi(\mu,\mu')
:=
\{ \pi\in\Pc ( \X \times \X ) :\pi\circ X^{-1}=\mu~\mbox{and}~\pi\circ {X'}^{-1}=\mu' \},
$$ 
The $p-$Wasserstein distance between $\mu$ and $\mu'$ is defined as:
\begin{eqnarray*}
\W_p(\mu,\mu')
:=
\inf_{\pi\in\Pi(\mu,\mu')}\E^\pi\Big[\big|X-X'\big|^p\Big]^{\frac1p},
&\mbox{for all}&
\mu,\mu'\in\Pc_p(\X)
.
\end{eqnarray*}
Throughout this article, the continuity of a function \(g:\Pc_p(\X)\to\R\) is understood with respect to the \(p\)-Wasserstein distance \(\W_p\).
For the $p-$Wasserstein distance, we denote the corresponding ball of radius $r$ by
\begin{equation*} 
\Bcl (\mu,r)
:=
\big\{\mu'\in\Pc_p(\X):\W_p(\mu,\mu')\le r \big\}
\end{equation*}
Finally, for a normed space $F$, we denote by $L^p_F(\mu)$ the set of all $F-$valued measurable maps with $\int |f(x)|_F^p \mu(\mathrm{d}x)<\infty$,  and we denote by $\mathbb{L}^p_F(\mu)$ the corresponding quotient space. 
Note that $ \mathbb{L}^p_F(\mu)$ is a Banach space while $L^p_F(\mu)$ is not. 
For simplicity, whenever the underlying domain and codomain are clear from the context, we write $\mathbb{L}^p(\mu):=\mathbb{L}^p_F(\mu)$.
\begin{Definition}\label{ch01def:coercivity}
Let $(E, \vert \cdot \vert)$ be a normed vector space and let $ f : E \rightarrow \R $. 
The function $f$ is coercive if it satisfies 
\begin{equation*}
    f(x) \xrightarrow[\vert x \vert \rightarrow \infty]{} \infty.
\end{equation*}
\end{Definition}

In the dynamic setting considered in this article, we define the time steps on the state space $\X$ through the projection coordinate maps on $\X \times \X$: 
$$ 
X_i (x, x') = x_i
~\mbox{and}~ 
X'_i (x,x') = x'_i,
~i=1, 2,
~\mbox{for all}~
(x, x')\in \X\times\X.
$$ 
For $\mu \in \Pc_p ( \X )$, the corresponding marginals are denoted $ \mu_i = \mu \circ X_i^{-1} $ for $ i=1, 2$ and we define $\Pi_i (\mu_i)$ as the set of all probability measures on $\X$ with $i-$th marginal being equal to $\mu_i$ for $ i = 1, 2$ 
\begin{equation*}
\Pi_i (\mu_i) := \{ \mu' \in \Pc(\X) \, : \, \mu'\circ X_i^{-1} = \mu_i \, \}.
\end{equation*}
Define ${\rm M} $ as the set of all probability measures on $\X$ that are martingales, 
\begin{equation*}
\rm M
:=
 \{\mu\in\Pc (\X) :~\E^\mu[X_2|X_1]=X_1,~\mu-\mbox{almost surely (\textit{a.s.})} \},
\end{equation*}
and note that for $\mu\in \Pc_p (\X)$, we have
\begin{equation}\label{eqdef:otimes}
\mu\in \rm M
\mbox{ if and only if} 
\E^\mu [h^\otimes ]=0,
~\mbox{for all}~
h\in \mathbb{L}^{p'}(\mu_1),
\end{equation}
where
$
h^\otimes(x)
:=h(x_1)\cdot(x_2-x_1)
$.
\begin{Definition}\label{def:compatible}
A probability measure $ \P \in \mathcal{P}( \X \times \X ) $ is causal if $\mathbb{F}^{X} := \sigma( X ) $ is compatible with $\mathbb{F}^{X'} :=  \sigma( X' ) $, in the sense that for all bounded Borel-measurable $ f : \X \rightarrow \R $ and $ g : S \rightarrow \R $,
$$
\E^{\P} \big[ 
f ( X_1, X_2 ) g ( X'_1 )  \vert  X_1 \big] 
=
\E^{\P} \big[ 
f ( X_1, X_2 )  \vert  X_1 \big] 
\,
\E^{\P} \big[ 
g ( X'_1 )  \vert  X_1 \big] 
.
$$

\end{Definition}
\noindent We introduce the set of {\it bi-causal} couplings 
$$
\Pi^{\text{bc}}(\mu,\mu')
:=
\{\pi\in \Pi ( \mu, \mu' ) \,\, \text{such that} \,\,  \pi \,\, \text{and} \,\,\pi \circ ( X', X )^{-1} \,\, \text{are causal}  \},
$$
together with the corresponding {\it adapted Wasserstein} distance 
\begin{eqnarray*}
\W^{{\rm ad}}_p(\mu,\mu')
:=
\inf_{\pi\in\Pi^{\text{bc}}(\mu,\mu')}\E^\pi\Big[\big|X-X'\big|^p\Big]^{\frac1p},
&\mbox{for all}&
\mu,\mu'\in\Pc_p(\X).
\end{eqnarray*}
We denote the corresponding ball of radius $r$ by
$$
\Bad (\mu,r)
:=
\big\{\mu'\in\Pc_p(\X):\W^{{\rm ad}}_p(\mu,\mu')\le r\big\}.$$ 
For the sake of convenience, in the following, for the distance  $\mathbf{d} \in \{ \W_p, \W_p^{\rm ad} \}$, we denote by 
\begin{equation}\label{eqdef:balls_deviations}
B^{\rm M}_{\mathbf{d}}( \mu, r ) = B_{\mathbf{d}}( \mu, r ) \cap {\rm M} 
\,\, 
\text{and}
\,\,
B^{ {\rm M}, {\rm m}_1}_{\mathbf{d}}( \mu, r ) = B_{\mathbf{d}}( \mu, r ) \cap {\rm M} \cap \Pi_1 (\mu_1) 
\end{equation}
where $ B_{\mathbf{d}}( \mu, r )$ is the ball centered at $\mu$, of radius $r$ with respect to the distance $\mathbf{d}$.
Throughout this article, we use the notation $ \E_1^\mu [ \cdot ] :=  \E^\mu [  \cdot \vert X_1 ]$ and for a $C^2$ function  $\phi : \X \rightarrow \R $, we have 
\begin{equation}\label{eqdef:caus_grad,J}
\partial_x^{\rm ad} \phi 
:= 
\begin{bmatrix} 
\E ^\mu_1 [ \partial_{x_1} \phi  ]
\\
\partial_{x_2} \phi 
\end{bmatrix},
~~
J_1 := \begin{bmatrix}
\text{Id}_S \\
0 
\end{bmatrix}
,~~J_2 := \begin{bmatrix}
0 \\
\text{Id}_S
\end{bmatrix}
~~\mbox{and}~~J := J_2 - J_1.
\end{equation}
Throughout this article, we consider a function $g:\Pc_p(\X)\longrightarrow\R$ with appropriate smoothness in the following sense. 
\begin{Definition}\label{def:lin derivative}
We say that $g$ has a linear functional derivative if there exists a continuous function $\delta_mg:\Pc_p(\X) \times \X \longrightarrow\R$, with $p-$polynomial growth in $x$, locally uniformly in $\mu$, such that for all $\mu,\mu'\in\Pc_p(\X)$ and denoting $\bar\mu^\lambda:=\mu+\lambda(\mu'-\mu)$, we have
$$
\frac{g (\bar\mu^\lambda )-g(\mu)}{\lambda}
\longrightarrow
\langle\delta_mg(\mu, \cdot),\mu'-\mu\rangle
:=
\int_\X \delta_mg(\mu,x)(\mu'-\mu)(\mathrm{d}x),
~\mbox{as}~\lambda\searrow 0.
$$
\end{Definition}

Clearly, the linear functional derivative is defined up to an additive constant which will be irrelevant throughout this article. 
In the linear case $ g ( \mu ) = \int_{\X} f \mathrm{d} \mu$, for some continuous map $f$ with $p-$polynomial growth, the linear functional derivative is the constant (in $\mu$) map $ \delta_m g ( \mu, x ) = f(x) $ for all $\mu \in \Pc_p ( \X ) $ , $ x \in \X $.   We note that, under technical conditions, the Lions' derivative coincides with the Wasserstein gradient and is given by $\partial_x\delta_m$, see \citeauthor{CarmonaDelarue} \cite{CarmonaDelarue}. 
This definition is equivalent to the existence of such a continuous function $\delta_mg$ such that the following holds:
\begin{equation*}
g(\mu')-g(\mu)
=
\int_0^1 \!\!\!\int_\X \delta_mg(\bar\mu^\lambda,x)(\mu'-\mu)(\mathrm{d}x) \mathrm{d}\lambda
\,\,\, \mbox{for all} \,\,\,
\mu,\mu'\in\Pc_p ( \X ).
\end{equation*}

\section{Main Results}

Our objective is to obtain the first-order correction at the origin $r=0$ of the lower and upper distributionally robust evaluations of some criterion $g(\mu)$ under various subsets of hedging instruments. 

\begin{Assumption} \label{ass:on g}
The mapping $g$ has a linear functional derivative such that $\delta_mg$ is $C^1$ in $x$, and $\partial_x\delta_mg$ is jointly continuous, with $(p-1)-$polynomial growth in the $x-$variable locally in the $\mu-$variable.
\end{Assumption}
\noindent We start with a set of dynamic buy-and-hold strategies and we define 
\vspace{-1mm}
\begin{equation}\label{eqdef:upbar G}
\upGMc ( r )  := \inf_{ h \in C^1_b } \sup_{ \mu' \in \Bcl ( \mu, r )  }  g( \mu' )  \!+\! \E^{\mu'} [ h^{\otimes} ]
\,\, \text{and} \,\, 
\upGMad ( r )  := \inf_{ h \in C^1_b} \sup_{ \mu' \in \Bad ( \mu, r )  }  g( \mu' )  \!+\!  \E^{\mu'} [ h^{\otimes} ],
\vspace{-3mm}
\end{equation}
where $\Bcl(\mu, r) $ and $ \Bad ( \mu, r )$ are the closed balls for the $p-$Wasserstein distance and the adapted $p-$Wasserstein distance, and $h^{\otimes}$ is defined by \eqref{eqdef:otimes}.
Inverting the infimum and the supremum in Equation \eqref{eqdef:upbar G} leads to the lower distributionally robust problems:
\vspace{-1mm}
\begin{equation}\label{eqdef:lowbar G}
\lowGMc ( r ) := \sup_{ \mu' \in \BclM ( \mu, r ) } g( \mu' ) 
~~\text{and}~~
\lowGMad ( r ):= \sup_{ \mu' \in \BadM ( \mu, r)  } g( \mu'), 
\vspace{-2mm}
\end{equation}
where $\BadM, \BclM$ are defined by \eqref{eqdef:balls_deviations}. 
Hence, the infimum over all dynamic hedging strategies restricts the deviations to those martingale models in the ball. 

\begin{Remark}\label{rem:adaptedornot} {\rm
We address here the main question of whether the distributionally robust value function should be defined through the deviation from the reference model in the sense of the Wasserstein distance \textbf{or} in the sense of the adapted Wasserstein distance.

\noindent $\bullet$ In the remarkable work by \citeauthor{backhoff_estimating_2022} \cite{backhoff_estimating_2022}, it is shown that the adapted Wasserstein distance is more suitable for dynamic optimization problems such as optimal control and optimal stopping: the adapted Wasserstein distance is the coarsest topology that guarantees the continuity of optimal stopping problems with bounded continuous rewards, and it restores the continuity of (bounded) portfolio optimization problems. 
For this reason, it is tempting to consider the deviations from the starting model $\mu$ within the adapted Wasserstein ball, as in \citeauthor{bartlsensitivityadapted} \cite{bartlsensitivityadapted}. 
The corresponding robust evaluation problems are denoted $\upGMad$ and $\lowGMad$ in Equations \eqref{eqdef:upbar G}–\eqref{eqdef:lowbar G}.

\noindent $\bullet$ However, {\it we believe that there is no reason to restrict to such deviations}, as the adaptedness condition is imposed with reference to the initial model whose accuracy is under question; this is precisely the primary reason why the distributionally robust problem is considered in this work. 

\noindent $\bullet$ Consequently, this article also considers the distributionally robust evaluations $\upGMc$ and $\lowGMc$ introduced in Equations \eqref{eqdef:upbar G}–\eqref{eqdef:lowbar G} based on the classical Wasserstein distance. Our numerical findings show that this more conservative viewpoint induces significantly higher model risk sensitivity.
}
\vspace{-2mm}
\end{Remark}
Notice that both $\lowGMc$ and $\lowGMad$ are non-decreasing and $\lowGMc\ge \lowGMad$. 
As these functions are equal at the origin, it follows that
\vspace{-2mm}
\begin{eqnarray*}
0 \;\le\; \left.\lowGMad\right.^\prime(0)
   \;\le\; \left. \lowGMc \right.^\prime(0),
   &\mbox{provided that these derivatives exist.}
\vspace{-2mm}
\end{eqnarray*}

\subsection{Adapted Wasserstein Model Deviation}

We start with the maps $\upGMad$ and $\lowGMad $ whose study turns out to be more tractable. 
This is consistent with the previous literature justifying the suitability of the adapted Wasserstein topology for dynamic problems, as shown in \citeauthor{backhoff_adapted_2020} \cite{backhoff_adapted_2020}, \citeauthor{backhoff_all_2020} \cite{backhoff_all_2020} and \citeauthor{margheriti_sur_2020} \cite{margheriti_sur_2020}.
Throughout this section, as well as in the corresponding proofs, we assume that $\mu \in {\rm M}\cap \Pc_p(\X)$.

\begin{Proposition}\label{prop:order 1 expansion martingale adapted case}
Under Assumption \ref{ass:on g}, $\upGMad $ and $\lowGMad$ are differentiable at the origin and 
\vspace{-3mm}
\begin{equation}\label{eq:deriv mart adapted}
\left. \upGMad \right. ' \!  ( 0 ) 
\! = \! \left. \lowGMad \right. ' \! ( 0 ) 
\! =\! \! \! \!
\inf_{h \in \mathbb{L}^{p'}( \mu_1)} 
U_{{\rm ad}}^{\rm M} ( h ),
\,\mbox{ where }\,
U_{{\rm ad}}^{\rm M}(h) 
:=
\| \partial_x^{\rm ad} \delta_m g + J h ( X_1 ) \|_{\mathbb{L}^{p'}(\mu)},
\vspace{-3mm}
\end{equation}
and $J, \partial_x^{\rm ad}$ defined by Equation \eqref{eqdef:caus_grad,J}. 
Moreover, $U^{\rm M}_{{\rm ad}}$ is convex and coercive (in the sense of Definition \ref{ch01def:coercivity}); hence, the optimization problem \eqref{eq:deriv mart adapted} admits a solution $h_{\rm ad, M}$ characterized by the first-order condition 
\vspace{-2mm}
\begin{equation}\label{eq:first order condition martingale}
\mathbf{N}\big( 
\E ^\mu_1 [ \partial_{x_1} \delta_m g ]
-
h_{\rm ad, M}(X_1) 
\big)
=
\E^\mu_1 \big[  
\mathbf{N} 
(
\partial_{x_2} \delta_m g 
+
h_{\rm ad, M}(X_1) 
)
\big]
\vspace{-2mm}
,\end{equation}
where $ \mathbf{N} $ is defined by \eqref{eqdef:n and n_ad}.
In particular, for $ p = 2 $, $
h_{\rm ad, M}
=
\frac{1}{2}
\E^\mu_1 \big[ ( \partial_{x_1}-\partial_{x_2})  \delta_m g  \big]
$
and 
\vspace{-2mm}
$$
\left. \upGMad \right. '  ( 0 ) 
= \left.\lowGMad \right. ' ( 0 ) 
=
\E^\mu
\Big[ 
\frac{1}{2}
\big| \E^\mu_1[ (\partial_{x_1}+\partial_{x_2})\delta_m g]\big|^2
+
\big| \partial_{x_2} \delta_m g  - \E^\mu_1[ \partial_{x_2} \delta_m g ] \big|^2
\Big]^{1/2}
\vspace{-2mm}
.$$
\end{Proposition}

\begin{Remark}
{\rm 
\noindent
$\bullet$ A similar result was obtained independently by \citeauthor{jiang2024sensitivity} \cite{jiang2024sensitivity} for a linear map $g$. 

\noindent
$\bullet$ The optimal map $h_{\rm ad, M}$ is the {\it first-order hedge against model risk} in the following sense: the zero-cost buy-and-hold strategy consisting of holding $h_{\rm ad, M}$ shares of the underlying asset at time $1$ induces the smallest first-order correction for the distributionally robust criterion.

\noindent
$\bullet$ It is also possible to consider deviations with respect to causal couplings, as done in \citeauthor{jiang2024sensitivity} \cite{jiang2024sensitivity},  \citeauthor{jiang2024duality} \cite{jiang2024duality} and
\citeauthor{han2022distributionally} \cite{han2022distributionally}. In this case, the result is exactly the same. Indeed, defining 
\vspace{-2mm}
\begin{equation*}
\overline{G}^{\rm M}_{{\mathbf{c}}} ( r ) := \inf_{ h \in C^1_b (S, S) } \sup_{ \mu' \in B_{\mathbf{d}_{\mathbf{c}}} ( \mu, r ) } g( \mu' ) + \E^{\mu'} [ h^{\otimes} ],
\vspace{-2mm}
\end{equation*}
where $\mathbf{d}_{\mathbf{c}} $ is the $p-$causal Wasserstein distance, defined for $ \mu, \mu' \in \Pc_p(\X)$ by  
\vspace{-1mm}
$$
\mathbf{d}_{\mathbf{c}} ( \mu, \mu' )
=
\inf_{\substack{ \pi \in \Pi_{\mathbf{c}}(\mu, \mu') }} \E^\pi [ \vert X - X'\vert^p]^{1/p},
\vspace{-2mm}
$$
where $ \Pi_{\mathbf{c}}(\mu, \mu') $ is the set of causal couplings (see Definition \ref{def:compatible}). Again, inverting the infimum and supremum yields
\vspace{-2mm}
\begin{equation*}
\underline{G}^{\rm M}_{{\mathbf{c}}} ( r ):= \sup_{ \mu' \in {\rm M} \cap B_{\mathbf{d}_{\mathbf{c}}} ( \mu, r )  } g( \mu').
\vspace{-2mm}
\end{equation*}
}
\end{Remark}

\begin{Proposition}\label{prop:order 1 expansion martingale causal case}
Under Assumption \ref{ass:on g}, $\overline{G}^{\rm M}_{ {\mathbf{c}} } $ and $\underline{G}^{\rm M}_{ {\mathbf{c}} }$ are differentiable at the origin and 
\vspace{-1mm}
\begin{equation*}
\left. \overline{G}^{\rm M}_{ {\mathbf{c}} } \right. '  ( 0 ) 
= \left. \underline{G}^{\rm M}_{ {\mathbf{c}} } \right. ' ( 0 ) 
=
\left. \upGMad \right. '  ( 0 ) 
= \left. \lowGMad \right. ' ( 0 ) 
.\end{equation*}
\vspace{-2mm}
\end{Proposition}
\vspace{-2mm}

\subsection{Wasserstein Model Deviation}

We next turn to the more involved distributionally robust evaluation under the classical Wasserstein metric. 
See Remark \ref{rem:adaptedornot} for the relevance of this formulation. 
Further technical conditions are required in this setting.
Throughout this section, as well as in the corresponding proofs, we assume that $\mu \in {\rm M}\cap \Pc_p(\X)$.

\begin{Assumption} \label{ass: mu converse inequality}

\begin{enumerate}[label=\textnormal{(\roman*)}]
\item \label{cond:density}  The probability measure $\mu$ is absolutely continuous with respect to (w.r.t) the Lebesgue measure; we denote by $q$ and $q_1$ the densities of $\mu$ and $\mu_1$, respectively. 
\item \label{cond:support}  The boundaries $\partial\Omega$ and $\partial\Omega_1$ of the supports $\Omega$ and $\Omega_1$ of $\mu$ and $\mu_1$ are Lipschitz-continuous. 
\item \label{cond:reg density} The maps $q$ and $v_{p'}:= q_1\E^\mu_1\big[ \vert X_2-X_1 \vert^{p'}\big]$ have a continuous version; moreover, $q$ is $C^1$ in $x_1$ with both $q$ and $\partial_{x_1}q$ bounded.  
\end{enumerate}
\end{Assumption}
Our last main result involves the following classical Wasserstein version of the optimal first-order model risk hedging problem. Letting $h^\otimes$ be defined by \eqref{eqdef:otimes}, we have
\begin{equation} \label{eq:deriv martingale class}
\inf_{h \in H^{p'}_{\mu_1}  }
\!\!U^{\rm M} (h),
\, \mbox{where} \, 
U^{\rm M}(h) \!:=\!
\big\| \partial_x(\delta_m g(\mu, \cdot ) \! + \! h^\otimes ) \big\|_{\mathbb{L}^{p'}\! (\mu)},\,
\text{$H^{p'}_{\mu_1} := H^{p'}(\Omega_1, w)$,}
\end{equation}
with $H^{p'}(\Omega_1, w)$ defined by \eqref{eqdef:sobolev}, with $w=(w_0,w_1)=(q_1, v_{p'})$. We start by proving the existence of a minimizer.

\begin{Proposition} \label{prop:optim problem wass}
Let Assumptions \ref{ass:on g} and \ref{ass: mu converse inequality} hold. 
Then $H^{p'}_{\mu_1}$ is a reflexive Banach space and $U^{\rm M}$ is convex and coercive. 
Hence, the optimization problem \eqref{eq:deriv martingale class} has a unique solution $h_{\rm M} \in H^{p'}_{\mu_1}$.
\end{Proposition}

\begin{Remark} \label{rem:sobolev}{\rm
In Equation \eqref{eq:deriv martingale class}, the infimum is taken over $H^{p'}( \Omega_1, w) $ and not $W^{p'}(\Omega_1,w)$. 
These weighted Sobolev spaces are different in general. 
Additional conditions are needed for equality; see \citeauthor{duoandikoetxea_forty_2013} \cite{duoandikoetxea_forty_2013}, \citeauthor{tolle_uniqueness_2012} \cite{tolle_uniqueness_2012}, and \citeauthor{1998} \cite{1998} for further information on this matter. 
However, since both $w_i$ and $\frac1{w_i}$ are strictly positive and continuous, for $i=0,1$, the maps are all in $\mathbb{L}^1_{\rm loc}(\Omega_1)$. 
Using notations \ref{eqdef:sobolev}, by the remark following Equation (1.2) of \cite{1998}, the space $W^{p'}(\Omega_1, w)$ is complete and therefore $H^{p'}( \Omega_1, w)$ is the closure of $ C^{\infty} (\Omega_1) $ with respect to the norm $ \Vert \cdot \Vert_{p', w}$.
}
\end{Remark}

The existence of a solution $h_{\rm M}$ for the minimization problem \eqref{eq:deriv martingale class} is a first crucial step in the proof of our subsequent sensitivity result. 
To prove that both $\left. \lowGMc \right.'(0)$ and $\left. \upGMc \right.'(0)$ coincide with the minimum value introduced in \eqref{eq:deriv martingale class}, additional technical conditions are needed. 
Using $ \mathbf{N}$ defined by notation \eqref{eqdef:n and n_ad}, let 
$$
\Psi  :=
\partial_x ( \delta_m g \!+\! h_{\rm M}^{\otimes} ),
~T
\!=\!
\begin{bmatrix} 
       T_1\\ T_2
      \end{bmatrix} 
:= 
\frac{1}{c} \mathbf{N} ( \Psi )
~\mbox{and}~
\alpha_j := \E^\mu_1\big[ (X_{2, j} \!-\! X_{1, j} )  T_{1, j}\big],
~j=1,\ldots,d 
\,\, ,
$$
where $c$ is chosen such that $\Vert T \Vert_{\mathbb{L}^p(\mu)} = 1$.

\begin{Assumption}\label{ass:optimum wass}
For all $j=1,\ldots,d$:
\begin{enumerate}[label=\textnormal{(\roman*)}]
\item \label{cond:regularity of inputs} The map $\alpha_j$ has a version in $W^{1, 1}_{{\rm loc}} ( \Omega_1)$ (see Definition \eqref{eqdefLloc_sobo}), with $ \nabla \alpha_j$ and $\frac{\nabla q_1 }{q_1 } \!\cdot\!\alpha_j$ in $\mathbb{L}^{p} ( \mu_1)$;
\item\label{cond:domain and optimum} there exists a sequence of smooth functions $T^n_1 : \Omega \rightarrow S$ with compact support $ \Omega_n \subsetneq\Omega $, such that the map $
  \alpha^n_j := \E^\mu_1 \big[(X_{2, j} \!-\! X_{1, j} )  T^n_{1}\big]$ has a version in $W^{1, 1}_{ {\rm loc}}$ and  
$$
T_1^n \longrightarrow T_1
\text{ in } \mathbb{L}^p(\mu),
\quad
\big(\nabla\alpha_j^n,\nabla\ln q_1\cdot\alpha_j^n\big)
\longrightarrow
\big(\nabla\alpha_j,\nabla\ln q_1\cdot\alpha_j\big)
 \text{ in } \mathbb{L}^p(\mu_1),
\quad j=1,\ldots,d.
$$
\end{enumerate}
\end{Assumption}

\begin{Remark}{\rm
The last assumption is a refinement of the usual trace theorem. 
Indeed, using the first-order condition of the problem \eqref{eq:deriv martingale class}, $  \alpha_j \vert_{ \partial \Omega_1}= 0 $ for $1 \leq j \leq d$. 
Hence, by the usual trace theorem, there exists a sequence of smooth compactly supported functions $( \alpha^n )$ whose support is strictly included in $\Omega_1$ and such that $ \alpha^n \rightarrow \alpha  $ in $W^{1,p} (\Omega_1)$. 
Assumption \ref{ass:optimum wass} states that the usual trace theorem is refined and that one can construct a version $\alpha^n_j$ of $\E^\mu\big[ (X_{2, j} - X_{1, j} )  T^n_{1} \vert X_1 = x_1 \big]$ with convergence holding in some weighted Sobolev space. 
}
\end{Remark}

\begin{Proposition}\label{prop:order 1 expansion class martingale}
Let Assumptions \ref{ass:on g}, \ref{ass: mu converse inequality} and \ref{ass:optimum wass} hold. Then $\upGMc $ and $\lowGMc$ are differentiable at the origin and
$$
\left. \upGMc \right. '  ( 0 ) 
= 
\left. \lowGMc \right. ' ( 0 ) 
=
\| \partial_x (\delta_m g(\mu, \cdot)+h_{\rm M}^\otimes ) \|_{\mathbb{L}^{p'}(\mu)},
$$
where $h_{\rm M}$ is the solution of the optimal first-order model risk hedging problem \eqref{eq:deriv martingale class}.
\end{Proposition}

\begin{Remark}{\rm
We will see that the proof of Proposition \ref{prop:order 1 expansion class martingale} is more difficult than the proof of Proposition \ref{prop:order 1 expansion martingale adapted case}. 
With regard to \citeauthor{bartl2021sensitivity} \cite{bartl2021sensitivity}, the natural candidate for $ \left. G^{\rm M} \right. '(0)$ is the optimization problem \eqref{eq:deriv martingale class}, and the one for $ \left. \underline{G}^{\rm M}_{\rm ad} \right. '(0)$ is the optimization problem \eqref{eq:deriv mart adapted}. 
While the optimization problem \eqref{eq:deriv mart adapted} is a convex, coercive optimization problem over $\mathbb{L}^p (\mu)$; the one defined by Equation \eqref{eq:deriv martingale class} can be more involved as it is a problem of calculus of variation.
}
\end{Remark}
In the rest of this section, we specialize the discussion to the case $p = 2 \,\, , \,\, S = \R$, and we provide a sufficient condition to characterize the first-order model risk optimal hedge through a Fredholm integral equation.
Recall the maps $q_1$ and $v_2$ introduced in Assumption \ref{ass: mu converse inequality}.

\begin{Assumption}\label{ass:fredholm}
In addition to Assumption \ref{ass: mu converse inequality}, let $p = 2$, $S = \R$, and
\begin{enumerate}[label=\textnormal{(\roman*)}]
\item \label{cond:support 1d} $ \Omega_1 = I := [ \ell , b ]$ for some finite $ \ell < b$;
\item\label{cond:density 1d} $c:=\inf_{I} (q_1\wedge v_2) > 0$.
\end{enumerate}
\end{Assumption}
The last conditions may be weakened further at the expense of more technical effort. 
For instance, Condition ({\rm ii}) may be weakened to an appropriate integrability condition of the inverse. 
To state our last result, we introduce the kernel 
$$
K(x_1, z_1) := \big( k ( x_1 ) - k ( z_1 ) \big)^{+},
~\mbox{with}~
k (x_1):=\int_{\ell}^{x_1}\frac{\mathrm{d}\xi}{v_2(\xi)},
~~
x_1, z_1\in I,
$$ 
together with the corresponding Hilbert-Schmidt operator $ \Kc : \mathbb{L}^2 (\mu_1) \longrightarrow  \mathbb{L}^2 (\mu_1)$ defined by:
\begin{equation}\label{eqdef:kernel}
\Kc f(x_1)
:=
2\int_I K(x_1, z_1)f(z_1)\mu_1(\mathrm{d}z_1),
~~
f\in \mathbb{L}^2 (\mu_1),
~~x_1\in I.
\end{equation}
We also introduce the map: 
\begin{align}\label{eqdef:def u}
u(x_1) 
\!:=\!\!
\int_\ell^{x_1} \!\! \Big(\!(\gamma_1q_1)(z_1)
                                                              \!-\!\!\int_\ell^{z_1} \!\!\!(\gamma_2q_1) ( \xi_1 ) \mathrm{d} \xi_1 
                                                     \Big)\frac{\mathrm{d}z_1}{v_2 (z_1) },
\end{align}
where $\gamma_1
:=
\E^{\mu}_{1} \big[ ( X_2 \!-\! X_1 ) \partial_{x_1}\! \delta_m g (\mu, X) \big] $ and $
\gamma_2
:=
\E^{\mu}_{1} \big[( \partial_{x_2} \!-\! \partial_{x_1} ) \delta_m g (\mu, X) \big]$.

\begin{Proposition}\label{prop:fredholm equation}
Let Assumptions \ref{ass:on g}, \ref{ass: mu converse inequality} and \ref{ass:optimum wass} hold, and let $d=1$, $p=2$. 
Then 
\\
\noindent {\rm (i)} the unique solution $h_{\rm M}$ of the optimization problem \eqref{eq:deriv martingale class} satisfies the following integro-differential equation  
\begin{equation}\label{eqdef:fred hedge}
(h_{\rm M}'v_2) (x_1)
=
c_1 
-\gamma_1(x_1)
+
\int_{\ell}^{x_1} (\gamma_2+ 2h_{\rm M})(\xi_1)q_1 ( \xi_1 ) \mathrm{d} \xi_1,
~~\mu_1-\mbox{\textit{a.s.}}
\end{equation}

\noindent {\rm (ii)} Under the additional Assumption \ref{ass:fredholm}, the operator $I - \Kc $ is invertible.
\\

\noindent {\rm (iii)} Assume further that the maps $\phi_0 := ( I - \Kc )^{-1} (1)$ and $\phi_1 := ( I - \Kc )^{-1} (k)$ satisfy 
$$
a_0 b_1 - a_1 b_0 \neq 0 
\,\,\,\, 
\text{where}
\,\,\,\,
a_i 
:=
2\E^{\mu_1}[ \phi_i ]
,
\,\,\,
b_i 
:=
\E^{\mu}\big[ ( X_2 - X_1 )^2 \phi_i' + 2 X_1 \phi_i \big]
\,\,\, 
i=0, 1.
$$
Then $
h_{\rm M} = c_0 \phi_0+c_1\phi_1  + (  I - \Kc )^{-1} (u),
$
with constants $c_0,c_1$ determined by the linear system
\begin{eqnarray} \label{eqdef:system_cst}\!\!\!\!\!\!
        &&2\,\E^{\mu_1} [h_{\rm M} ] = \E^\mu [ (\partial_{x_1}-\partial_{x_2})\delta_mg ] 
        \label{f=1}
        \\
        &&\E^\mu [ (X_2 - X_1)^2 h'_{\rm M} + 2 X_1 h_{\rm M}  ] 
        = \E^\mu [ X_1 (\partial_{x_1}-\partial_{x_2})\delta_mg 
                        - ( X_2 - X_1)\partial_{x_1}\delta_mg ]. 
        \label{f=Id}
\end{eqnarray}
\end{Proposition}

\subsection{One-dimensional First-Marginal Constraint}

We consider the one-dimensional setting $d = 1$ and restrict attention to the case in which Vanilla payoffs at the first maturity are available. 
The minimum distributionally robust evaluation problem is then defined by:
$$
\upGmiad
(r) 
:=
\inf_{f\in C^1_b}\sup_{\mu' \in \Bad (\mu, r)} 
 g ( \mu' )+\E^{\mu'}[f(\!X_1\!)]-\mu_1(f),
~~ 
\lowGmiad(r) 
:=
\sup_{ \mu' \in \Badmi ( \mu, r ) } 
 g ( \mu' ),
 $$
where the second problem, obtained by inverting the infimum and the supremum, naturally restricts deviations to models with a fixed first marginal (where $\Badmi(\mu,r):=\Bad(\mu,r)\cap\Pi_1(\mu_1)$, in line with the notation introduced in \eqref{eqdef:balls_deviations}). 
We leave the general case with fixed first and second marginals for future work.

\noindent We also introduce the corresponding problems with additional optimal dynamic hedging:
 $$
\upGMmiad (r)
:= \!\!\!\!
\inf_{\substack{ f \in\mathbb{L}^1 (\mu_1)  \\
h \in \mathbb{L}^{p'} \left( \mu_1 \right) }}
\sup_{ \mu' \in \Bad (\mu,r)}
 \!\!\!g ( \mu' )+\E^{\mu'}[f(\!X_1\!)\!+\!h^\otimes]-\mu_1(f),
\,\,
 \lowGMmiad  (r)
:= \! \! \! \!\!\!
\sup_{ \substack{\mu' \in B_{\W_p^{\rm ad}}^{{\rm M}, {\rm m}_1} } ( \mu, r )  \\ 
} 
 g ( \mu' ),
 $$
where $ B_{\W_p^{\rm ad}}^{{\rm M}, {\rm m}_1} $ is defined by \eqref{eqdef:balls_deviations}. 
The last problems restrict model deviations to the class of arbitrage-free models calibrated to the first marginal, and are thus related to the martingale optimal transport literature.  

\begin{Proposition} \label{prop:order 1 expansion marginalandmartingale adapted case}
Under Assumptions \ref{ass:on g}, $\lowGmiad $, $\upGmiad$, $\lowGMmiad $ and $\upGMmiad$ are differentiable at the origin, and
\begin{eqnarray} \label{eqref:deriv marginal}
\left. \lowGmiad  \right.' (0)
&=&
\left. \upGmiad  \right.' (0) 
=
\inf_{ \substack{ f \in \mathbb{L}^{p'}(\mu_1) } }
U_{{\rm ad}}^{{{\rm m}_1}}(f),
\\ 
\left. \lowGMmiad  \right.' \!\!(0) 
&=&
\left. \upGMmiad  \right.' \!\!(0) 
=
\inf_{f,h \in \mathbb{L}^{p'}(\mu_1)}
U_{{\rm ad}}^{{\rm M, m}_1} \!(h, f),
\end{eqnarray}
with $J = -e_1 + e_2$, $J_1 = e_1$ (corresponding to definition \eqref{eqdef:caus_grad,J} for $d =1$), where $(e_1, e_2)$ is the canonical basis of $\R^2$, we have 
$$U_{{\rm ad}}^{{\rm M, m}_1} (h,f) 
:=
\|
\partial_x^{\rm ad} \delta_m g + h(X_1) J+ f(X_1)J_1 \|_{\mathbb{L}^{p'}(\mu)}
, $$
and $ U_{{\rm ad}}^{{{\rm m}_1}}(f)
:= U_{{\rm ad}}^{{\rm M, m}_1} (0, f) $.
Moreover, both maps $U_{{\rm ad}}^{{{\rm m}_1}}$ and $U_{{\rm ad}}^{{\rm M, m}_1}$ are convex and coercive (in the sense of Definition \ref{ch01def:coercivity}), and thus admit minimizers $f_{{\rm m}_1}$ and $( h_{{\rm M, m}_1}, f_{{\rm M, m}_1})$, respectively, characterized by the first-order conditions
$$
f_{{\rm m}_1}=- \E ^\mu_1 [ \partial_{x_1} \delta_m g ],
~
\E^\mu_1 [\mathbf{N}(\partial_{x_2} \delta_m g +h_{{\rm M, m}_1} )
                ]
=0,
~\text{and}~
f_{{\rm M, m}_1}
=
-\E ^\mu_1[ \partial_{x_1} \delta_m g ] + h_{{\rm M, m}_1},
$$
where $\mathbf{N}$ is defined by \eqref{eqdef:n and n_ad}.
\end{Proposition}

\begin{Remark}
{\rm The $d-$dimensional setting can also be addressed using the calculus of variation. 
Denoting $\gamma_1(x_1):=\E[\partial_{x_1}\delta_mg|X_1=x_1]$, we readily see that the solution $f_{{\rm m}_1}$ of the variational problem $\min_f U_{\rm ad}^{{\rm m}_1}$ is characterized by the degenerate elliptic equation ${\rm div}[q_1(\nabla f+\gamma_1)]=0$ with boundary condition $(\nabla f+\gamma_1)\cdot\vec{n}=0$, where $\vec{n}$ denotes the outward unit normal vector to the domain. 
In the general $d-$dimensional setting, it is more challenging to prove that this solution is indeed the optimal first-order static hedge. 
This extension is left for future work.
}
\end{Remark}

\subsection{Optimal Stopping Problem}\label{sec:optstop}

We now investigate the sensitivity analysis of the optimal stopping problem: 
\vspace{-1mm}
\begin{equation}\label{eqdef:optimal stopping}
g ( \mu' ) := \inf_{ \tau \in \text{ST} } \E^{\mu'} [ \ell_\tau(X) ],
\vspace{-1mm}
\end{equation}
where $\text{ST}$ is the set of all stopping times with respect to the canonical filtration.

\begin{Assumption}\label{ass:optimal stopping} 
\begin{enumerate}[label=\textnormal{(\roman*)}]
\item \label{cond:objective function} The map $\ell: \X  \times \left\lbrace	1, 2 \right\rbrace \rightarrow \R $ is adapted, with maps $\ell_1,\ell_2$ that are continuously differentiable and $(p\!-\!1)-$polynomially growing.
\item\label{cond:uniqueness optimal stopping } The optimal stopping problem \eqref{eqdef:optimal stopping} admits a unique solution $\hat\tau$. 
\end{enumerate}
\end{Assumption}

The following results extend \citeauthor{bartlsensitivityadapted} \cite{bartlsensitivityadapted}, by considering the martingale and/or marginal adapted case in the context of model deviations induced by the adapted Wasserstein metric. 
The case of deviations induced by the standard Wasserstein metric is left for future research.

\begin{Proposition} \label{prop:deriv optimal stopping mart adapted}
Let $g$ be defined by Equation \eqref{eqdef:optimal stopping} with $\ell$ satisfying Assumption \ref{ass:optimal stopping}. Then the corresponding distributionally robust optimization maps $\lowGmiad$, $ \lowGMad$, $\lowGMmiad $,  $\upGmiad$, $ \upGMad$, $\upGMmiad $ are differentiable at the origin, the following differentiability results  hold.

\noindent {\rm (i)} When $d = 1$, the model risk sensitivity under the first-marginal constraint is given by:
\vspace{-1mm}
\begin{equation} \label{eqref:deriv american option marginal}
\left. \lowGmiad \right.' (0)
=
\left. \upGmiad \right.' (0) 
=
\inf_{ f \in \mathbb{L}^{p'}( \mu_1 ) } 
\big\| \partial_{x}^{\rm ad} \ell_{\hat\tau} (X) + J_1 f(X_1) \big\|_{\mathbb{L}^{p'}(\mu)} 
=
\big\| \partial_{x_2} \ell_{\hat\tau} (X)\big\|_{\mathbb{L}^{p'}(\mu)},
\vspace{-1.5mm}
\end{equation}
with minimizer $\hat f_{ \rm  m_1}:=-\E^\mu_1 [ \partial_{x_1} \ell_{\hat\tau} ]$.
\\
{\rm (ii)} For general dimension $d$, the model risk sensitivity under the martingale constraint is:
\vspace{-1.5mm}
\begin{equation} \label{eqref:deriv american option martingale}
\left. \lowGMad \right.' (0)
=
\left. \upGMad \right.' (0) 
=
\inf_{ h \in \mathbb{L}^{p'}( \mu_1 )} 
\big\| \partial_{x}^{\rm ad} \ell_{\hat\tau} (X) + J h (X_1) \big\|_{\mathbb{L}^{p'}(\mu)}.
\vspace{-1.5mm}
\end{equation}
Letting $\mathbf{N}$ be defined by \eqref{eqdef:n and n_ad}, the optimization problem \eqref{eqref:deriv american option martingale} admits a unique minimizer $\hat h_{ \rm  M} $ satisfying $ \E^\mu_1 [ \mathbf{N}
( \partial_{x_2} \ell_{\hat\tau } 
+
\hat h_{  \rm M} ( X_1 ) 
)
]
=
\mathbf{N} ( \E^\mu_1 [  \partial_{x_1} \ell_{ \hat\tau } ] - \hat h_{ \rm M} ( X_1 )  )
$.
\\
{\rm (iii)} Assuming that $d = 1$, the model risk sensitivity in the martingale and first-marginal adapted case is given by:
\vspace{-1mm}
\begin{equation} \label{eqref:deriv american option martingale and marginal}
\begin{split}
\hspace{-1.5mm}
\left. \lowGMmiad \right.' (0) 
=
\left. \upGMmiad \right.' (0) 
&=
\inf_{f, h \in\mathbb{L}^{p'} (\mu_1)} 
\| \partial_{x}^{\rm ad} \ell_{\hat\tau} +(J h + J_1 f)( X_1 )\|_{\mathbb{L}^{p'} (\mu)} 
\\
&=
\inf_{ h \in\mathbb{L}^{p'} (\mu_1)}
\| \partial_{x_2} \ell_{\hat\tau} + h(X_1) \|_{\mathbb{L}^{p'}(\mu)} ,
\end{split}
\end{equation}
\vspace{-3mm}

\noindent with minimizer $(\hat h_{ {\rm M, \rm m}_1},\hat f_{ {\rm M, \rm m}_1} ) $ satisfying  
$$
\hat f_{ {\rm M, \rm m}_1}(X_1) 
=
\hat h_{ {\rm M, \rm m}_1}(X_1) 
-\E^\mu_1 [ 
\partial_{x_1} \ell_{ \hat\tau }
]
~\mbox{and}~
\E^\mu_1 \big[ \mathbf{N} (\partial_{x_2} \ell_{ \hat\tau } 
+ \hat h_{ {\rm M, \rm m}_1}(X_1) 
)
\big]
=
0.
\vspace{-2mm}
$$
\end{Proposition}

\section{Examples for Specific $g$}

For a function $ g : \Pc_p(\X) \rightarrow \R $ satisfying Assumption \ref{ass:on g}, the standard DRO problems under the Wasserstein and adapted Wasserstein distances, without additional constraints, are given by 
\vspace{-2mm}
\begin{equation*}
\underline{G}(r) := \sup_{ \mu' \in \Bcl (\mu, r )}  g(\mu') \,\,\, \text{and}\,\,\, \underline{G}_{\rm ad} (r) := \sup_{  \mu' \in \Bad (\mu, r )} g(\mu') 
.
\vspace{-2mm}
\end{equation*}
Adapting proofs of \citeauthor{bartl2021sensitivity} \cite{bartl2021sensitivity} and \citeauthor{bartlsensitivityadapted} \cite{bartlsensitivityadapted} yields 
\vspace{-1mm}
$$
\underline{G}'(0) = \Vert \partial_x \delta_m g \Vert_{\mathbb{L}^{p'}(\mu)} \,\, \text{and} \,\,    \underline{G}'_{\rm ad}(0) = \Vert \partial^{\rm ad}_x \delta_m g \Vert_{\mathbb{L}^{p'}(\mu)} 
.
\vspace{-1mm}
$$

\noindent In this section, we illustrate our results for a stochastic optimization problem $$g(\mu) := \inf_{a \in \Ac} \int f(x, a) \mu(\mathrm{d} x),$$ and a stochastic game $g(\mu) := \inf_{a \in \Ac} \sup_{b \in \Bc} \int f(x, a, b) \mu(\mathrm{d} x)$. 
For both of those cases, we will give sufficient conditions under which $g$ satisfies Assumption \ref{ass:on g}.

\subsection{Stochastic Optimization Problem}

Let $ k \geq 1$ be an integer, and let  $ \Ac \subset \R^k$ be a convex compact subset. 
Let $ f : \X \times \Ac \rightarrow \R$, and define
\vspace{-2mm}
\begin{equation}\label{eqdef:sto opt prob}
g(\mu) := \inf_{a \in \Ac} \int f(x, a) \mu(\mathrm{d} x) .
\vspace{-2mm}
\end{equation}

\begin{Condition}\label{ass:stochastic optimization problem}
The function $f  : \X \times \Ac \rightarrow \R$ satisfies 
\begin{enumerate}[label=\textnormal{(\roman*)}]
\item \label{cond:reg on f sto opt} $(x, a) \mapsto f(x, a)$ is differentiable on $\X \times \Ac$. 
Moreover, $(x, a) \mapsto \partial_x f(x, a) $ is continuous, and there exists $ C > 0 $ such that $\vert \partial_x f(x, a)  \vert \leq C ( 1 + \vert x \vert^{p-1} )$ for all $x \in \X$ and $a \in \Ac$.
\item\label{cond:conv on f sto opt} $a \mapsto f(x, a)$ is $\kappa-$strictly convex, uniformly in $\X$ in the sense that 
\vspace{-2mm}
$$
( \partial_a f(x, a_1) - \partial_a f(x, a_2) ) \cdot ( a_1 - a_2 ) \geq \kappa \vert a_1 - a_2 \vert^2
\,\, \text{for all } a_1, a_2 \in \Ac \,\,, x \in \X.
\vspace{-2mm}
$$
\end{enumerate}
\end{Condition}

\begin{Proposition}\label{prop:stochastic optimization linear func derivative}
Let $f$ satisfy Conditions \ref{ass:stochastic optimization problem} and $g$ be defined by Equation \eqref{eqdef:sto opt prob}. 
Then $g$ satisfies Assumption \ref{ass:on g} and 
$
\delta_m g (\mu, x) = f (x, a^{*}(\mu))$, where $a^{*}(\mu) = \text{argmin}_{a \in \Ac} \int f(x, a) \mu(\mathrm{d}x)$. 
In particular, letting  $U_{{\rm ad}}^{\rm M} (h) := \Vert \partial_x^{\rm ad} f(x, a^* (\mu) ) + J h(X_1) \Vert_{\mathbb{L}^{p'}(\mu)} $, we have 
$$
\underline{G}'(0) \!=\! \Vert \partial_x f (x, a^{*}(\mu)) \Vert_{\mathbb{L}^{p'} (\mu)} \, , \,
\underline{G}'_{\rm ad}(0) \!=\! \Vert \partial^{\rm ad}_x f (x, a^{*}(\mu) ) \Vert_{\mathbb{L}^{p'} (\mu)}
\,\text{and}\,
\left.\underline{G}^{\rm M}_{\rm ad} \right.' (0) = \! \inf_{h \in \mathbb{L}^{p'}( \mu_1)} 
U_{{\rm ad}}^{\rm M} ( h ),
$$

\end{Proposition}

\begin{Remark}
{\rm
\noindent $\bullet$ If $f$ satisfies Condition \ref{ass:stochastic optimization problem} then $f$ satisfies Assumption $1$ given in \citeauthor{bartl2021sensitivity} \cite{bartl2021sensitivity}.

\noindent $\bullet$ In \citeauthor{bartl2021sensitivity} \cite{bartl2021sensitivity} and \citeauthor{bartlsensitivityadapted} \cite{bartlsensitivityadapted}, instead of $G(r)$ and $G_{\rm ad} (r)$, authors considered the "true" DRO problems 
$$
V(r) = \inf_{a \in \Ac} \sup_{ \mu' \in \Bcl (\mu, r )}  \int f(x, a) \mu'(\mathrm{d}x) \,\, \text{and} \,\, 
V_{\rm ad} (r) = \inf_{a \in \Ac} \sup_{ \mu' \in \Bad(\mu,r) (\mu, r )}  \int f(x, a) \mu'(\mathrm{d}x)
,$$
which, except for min-max equality cases, is different from what we considered.
They proved that 
$$
V'(0) = \Vert \partial_x f(x, a^* (\mu) ) \Vert_{\mathbb{L}^{p'} (\mu) }
\,\, \text{and} \,\, 
V_{\rm ad}' (0) =\Vert \partial^{\rm ad}_x f(x, a^* (\mu) ) \Vert_{\mathbb{L}^{p'} (\mu) }
.$$
Hence, if $f$ satisfies Condition \ref{ass:stochastic optimization problem}, \ref{prop:stochastic optimization linear func derivative} states that $V$ and $\underline{G}$ are equals up to order one and so are their adapted counterparts. 
}
\end{Remark}

\noindent \textbf{Proof.} We need to check that $g$ satisfies Assumption \ref{ass:on g}. 

\noindent{\bf Step $1$}. We prove that $\mu \in \Pc_p(\X) \mapsto \text{argmin}_{a \in \Ac} \int_\X f(x, a) \mu(\mathrm{d}x) \in \Ac$ is continuous. 
Point \ref{cond:conv on f sto opt} of Condition \ref{ass:stochastic optimization problem} implies that $ a \in \Ac \mapsto \int_\X f(x, a) \mu(\mathrm{d}x) $ is strictly convex, and, since $\Ac$ is compact, for each $\mu \in \Pc(\X)$, there exists a unique minimizer $a^* (\mu) \in \Ac$. 
Furthermore, since $ (\mu, a) \in \Pc_p (\X) \times \Ac \mapsto \int_\X f(x, a) \mu(\mathrm{d}x) $ is continuous, the mapping $ \mu \mapsto a^*(\mu)$ is also continuous.

\noindent {\bf Step $2$.} We now prove the differentiability in the sense of Definition \ref{def:lin derivative}. 
Let $\mu, \mu'$ be in $ \Pc_p(\X)$. 
For $\lambda \geq 0$, define $\Gamma(\lambda) := g( \bar\mu^\lambda ) $ where $\bar\mu^\lambda :=(1 -\lambda) \mu + \lambda \mu'$. 
Let $a_\lambda := a^*(\bar\mu^\lambda)$. 
By the previous step, we have $a_\lambda \rightarrow a_0$. Moreover,  
\begin{align*}
\int_\X f(x, a_0) (\mu'-\mu)(\mathrm{d}x) \geq \frac{\Gamma(\lambda) - \Gamma(0)}\lambda
&=
\frac{1}\lambda \int_\X \big( f(x, a_\lambda) \bar\mu^\lambda (\mathrm{d}x) - \int_\X f(x, a_0) \big) 
\mu(\mathrm{d}x) 
\\
&= \frac{1}\lambda \! \int_\X \! \!f(x, a_\lambda) \!-\! f(x, a_0) \mu (\mathrm{d}x) \! 
\\
&\,\,\, \,\,\, +\int_\X \! \! f(x, a_\lambda ) ( \mu' \!- \mu) (\mathrm{d}x)
.\end{align*}
Furthermore, by optimality, $ \int_\X f(x, a_\lambda)\mu (\mathrm{d}x) \geq \int_\X  f(x, a_0) \mu (\mathrm{d}x) $. Hence, we obtain the inequality 
\begin{align*}
\int_\X f(x, a_0) (\mu'-\mu)(\mathrm{d}x) \geq \frac{\Gamma(\lambda) - \Gamma(0)}\lambda
\geq
 \int_\X f(x, a_\lambda) ( \mu' - \mu) (\mathrm{d}x)
.\end{align*}
Letting $\lambda$ go to $0$, we have proved that $ \delta_m g (\mu, x) = f (x, a^*(\mu) )$, the rest follows directly. \ep 

\subsection{Stochastic Games}
 
Let $ k, \ell \geq 1$ be two integers and let $ \Ac \subset \R^k$, $\Bc \subset \R^\ell$ be two convex compact subsets. 
Let $ f : \X \times \Ac \times \Bc \rightarrow \R$, and define
\begin{equation}\label{eqdef:sto game prob}
g(\mu) := \inf_{a \in \Ac} \sup_{b \in \Bc} \int f(x, a, b) \mu(\mathrm{d} x) .
\end{equation}

\noindent We focus on a case where min-max equality holds. 
\begin{Assumption}\label{ass:stochastic game problem}
The function $f  : \X \times \Ac \times \Bc \rightarrow \R$ satisfies 
\begin{enumerate}[label=\textnormal{(\roman*)}]
\vspace{-1.5mm}
\item \label{cond:reg on f sto gam} $(x, a, b) \mapsto f(x, a, b)$ is differentiable on $\X \times \Ac \times \Bc$. 
Moreover, $(x, a, b) \mapsto \partial_x f(x, a, b) $ is continuous, and there exists $ C > 0 $ such that $\vert \partial_x f(x, a, b)  \vert \leq C ( 1 + \vert x \vert^{p-1} )$ for all $x \in \X$, $a \in \Ac$ and $b \in \Bc$.
\vspace{-1.5mm}
\item\label{cond:conv on f sto gam} $a \mapsto f(x, a, b)$ is $\kappa-$strictly convex, uniformly in $\X \times \Bc$ in the sense that 
\vspace{-2mm}
$$
( \partial_a f(x, a_1, b) - \partial_a f(x, a_2, b) ) \cdot ( a_1 - a_2 ) \geq \kappa \vert a_1 - a_2 \vert^2
\,\, a_1, a_2 \in \Ac \, , \, b \in \Bc \, , \,   x \in \X.
\vspace{-3mm}
$$
\item\label{cond:conc on f sto gam} $b \mapsto f(x, a, b)$ is $\kappa-$strictly concave, uniformly in $\X \times \Ac$ in the sense that 
\vspace{-2mm}
$$
( \partial_b f(x, a, b_1) - \partial_b f(x, a, b_2) ) \cdot ( b_1 - b_2 ) \leq -\kappa \vert b_1 - b_2 \vert^2
\,\, b_1, b_2 \in \Bc \, , \, a \in \Ac \, , \,   x \in \X.
\vspace{-3mm}$$
\end{enumerate}
\end{Assumption}

\begin{Proposition}\label{prop:stochastic game linear func derivative}
Let $f$ satisfy Assumption \ref{ass:stochastic game problem} and let $g$ be defined by Equation \eqref{eqdef:sto game prob}. 
Then $g$ satisfies Assumption \ref{ass:on g} and 
\vspace{-4mm}
$$
\delta_m g (\mu, x) = f (x, \bar{a} (\mu) , \bar{b} (\mu)  ),
\vspace{-3mm}
$$ 
where $ ( \bar{a} (\mu) , \bar{b} (\mu) )$ is the unique saddle point of the min-max problem $g(\mu)$.
In particular, we have
\par\vspace{-8mm}
\begin{align*}
&\underline{G}'(0) = \Vert \partial_x f (x, \bar{a} (\mu) , \bar{b} (\mu)) \Vert_{\mathbb{L}^{p'} (\mu)} \,\, , \,\,
\underline{G}'_{\rm ad}(0) = \Vert \partial^{\rm ad}_x f (x, \bar{a} (\mu) , \bar{b} (\mu) ) \Vert_{\mathbb{L}^{p'} (\mu)}
\\[-1mm]
\,\, \text{and} \,\,
&\left. \underline{G}^{\rm M}_{\rm ad} \right.'(0) = 
 \inf_{h \in \mathbb{L}^{p'}( \mu_1)} 
U_{{\rm ad}}^{\rm M} ( h ),
\text{
where $U_{{\rm ad}}^{\rm M} (h) := \Vert \partial_x^{\rm ad} f (x, \bar{a} (\mu) , \bar{b} (\mu) )  + J h(X_1) \Vert_{\mathbb{L}^{p'}(\mu)} $}.
\end{align*}
\vspace{-8mm}
\end{Proposition}

\noindent \textbf{Proof.} 
By Assumption \ref{ass:stochastic game problem} and the Von Neumann Minimax Theorem \citeauthor{von1928theorie} \cite{von1928theorie}, min-max equality holds and
\vspace{-3mm}
$$g(\mu) := \inf_{a \in \Ac} \sup_{b \in \Bc} U_\mu(a,b) = \sup_{b \in \Bc} \inf_{a \in \Ac} U_\mu(a,b),
\text{ where }
U_\mu(a,b):=\int f(x,a,b)\mu(\mathrm{d}x).
\vspace{-2mm}$$

\noindent{\bf Step $1$.} We prove that given $\mu \in \Pc_p(\X)$, there exists a unique saddle point $ (\bar{a}(\mu),\bar{b}(\mu) )$. 
Following the proof of Proposition \ref{prop:stochastic optimization linear func derivative}, since $f$ satisfies Assumption \ref{ass:stochastic game problem}, the following mappings $(a, \mu) \mapsto b^* (a, \mu) := \text{argmax}_{b \in \Bc} U_\mu(a,b)$ and $(b, \mu) \mapsto a^*(b, \mu) := \text{argmin}_{a \in \Ac} U_\mu(a,b),$ are well defined and continuous (both {\rm argmax} and {\rm argmin} are singletons by strict convexity and compactness). 
Furthermore, since the supremum of a family $\kappa-$strictly convex function (as defined in Assumption \ref{ass:stochastic game problem}), is also $\kappa-$strictly convex, there exists a unique 
\vspace{-2mm}
$$ \bar{a}(\mu)  :=\text{argmin}_{a \in \Ac}  U_\mu(a,b^*(a, \mu) )
\,
\text{and} 
\,
\bar{b}(\mu) :=\text{argmax}_{b \in \Bc}  U_\mu(a^*(b, \mu), b).
\vspace{-1.5mm}$$ 
We now prove that $ ( \bar{a} (\mu ), \bar{b}(\mu ) )$ is the unique saddle point. 
The fact that $(\bar a(\mu),\bar b(\mu))$ is a saddle point follows from its definition and from strong duality. 
We now prove uniqueness. Let $(\tilde a,\tilde b)$ be another saddle point. Then $\tilde a$ is a minimizer of the convex function
$
a \longmapsto \sup_{b\in\Bc}U_\mu(a,b).
$
Indeed, every saddle point attains the value of the min-max problem. By Condition \ref{cond:conv on f sto gam}, this map is strictly convex, as a supremum of $\kappa$-strictly convex functions. Hence its minimizer is unique, and therefore $\tilde a=\bar a(\mu)$.
Similarly, $\tilde b$ is a maximizer of the concave function
$
b \longmapsto \inf_{a\in\Ac}U_\mu(a,b).
$
By Condition \ref{cond:conc on f sto gam}, this map is strictly concave, and therefore its maximizer is unique. Hence $\tilde b=\bar b(\mu)$.

\noindent This proves that the saddle point is unique. Finally, the continuity of the map
$
\mu\longmapsto(\bar a(\mu),\bar b(\mu))
$
follows from the continuity of the value functions and the uniqueness of the corresponding minimizer and maximizer.

\noindent Let $\mu' \in \Pc_p(\X)$, define $\Gamma(\lambda ) := g \big( (1- \lambda) \mu + \lambda \mu' \big) $, $ \Gamma_{\bar{a}} (\lambda) := \sup_{b \in \Bc} U_{\bar\mu^\lambda}(\bar{a}, b)$ and $ \Gamma_{\bar{b}} (\lambda) := \inf_{a \in \Ac} U_{\bar\mu^\lambda}(a, \bar{b})$, where $\bar\mu^\lambda := (1-\lambda)\mu+\lambda\mu'$.
Since $(  \bar{a}(\mu) , \bar{b} (\mu) )$ is a saddle point, we have  $\Gamma_{\bar{a}} (0) = \Gamma_{\bar{b}} (0) = g(\mu)$, and
$
\Gamma_{\bar{b}} (\lambda) \leq \Gamma(\lambda ) \leq \Gamma_{\bar{a}} (\lambda).
$
Hence 
\vspace{-2mm}
$$
\frac{\Gamma_{\bar{b}} (\lambda) - \Gamma_{\bar{b}} (0) }{\lambda}
\leq 
\frac{\Gamma (\lambda) - \Gamma (0) }{\lambda}
\leq 
\frac{\Gamma_{\bar{a}} (\lambda) - \Gamma_{\bar{a}} (0) }{\lambda}
.
\vspace{-2mm}
$$
The left-hand and right-hand sides are stochastic optimization problems satisfying Assumption \ref{ass:stochastic optimization problem}. 
Applying Proposition \ref{prop:stochastic optimization linear func derivative}, and by the convex/concavity assumption on $f$, we get, 
\vspace{-2mm}
$$
\frac{\Gamma_{\bar{b}} (\lambda) - \Gamma_{\bar{b}} (0) }{\lambda}
\rightarrow \!\! \int_\X \! f ( x, a^*( \bar{b}, \mu) , \bar{b} ) (\mu' - \mu)(\mathrm{d}x) 
\,
\text{and}
\,
\frac{\Gamma_{\bar{a}} (\lambda) - \Gamma_{\bar{a}} (0) }{\lambda}
\rightarrow \!\!\int_\X \! f (x, \bar{a}, b^*( \bar{a}, \mu) )(\mu' - \mu)(\mathrm{d}x) 
.
\vspace{-2mm}$$
Finally, by the first point, $ a^*( \bar{b}, \mu) = \bar{a}$ and $ b^*( \bar{a}, \mu) = \bar{b}$, we obtain the desired result.
\ep 
\vspace{-2mm}

\section{Numerical Illustrations} \label{sec:numerics}

In this section, we compare sensitivities for two families of models defined through a pair $(Z_1,Z_2)\sim \Nc ( 0, 1 )\otimes\Nc ( 0, 1 )$:
\vspace{-2mm}
\begin{itemize}
\item the Black-Scholes model $ \mu_{\text{BS}}^{\sigma} := \mathcal{L} \big( e^{-\frac{\sigma^2}{2} + \sigma Z_1 }, e^{ -\sigma^2+ \sigma ( Z_1 + Z_2 )} \big)  $,
\vspace{-2mm}
\item  and the Bachelier model $ \mu^{\sigma} _{\text{Bach}} := \mathcal{L} \big( \sigma Z_1, \sigma ( Z_1 + Z_2 ) \big) $.
\vspace{-3mm}
\end{itemize}

\subsubsection{Forward-start European option model risk sensitivities}

We first explore the impact of the martingale constraint on sensitivities through the example $g(\mu) := \E^\mu \big[ ( X_2 - X_1 )^{+} \big]$. 
We assume that all our results can be extended to such $g$ even though $ \partial_x \delta_m g $ is not continuous (see remark $11$ of \citeauthor{bartl2021sensitivity} \cite{bartl2021sensitivity} for further discussions and a proof). 

\noindent For this particular function $g$, it turns out that the Bachelier model induces sensitivities that are constant in the volatility parameter: 
\par\vspace{-7mm}
\begin{align*}
G' ( 0 ) &=  
\sqrt{2}\; \E^\mu[ 
					\mathds{1}_{ \lbrace \sigma ( Z_2 + Z_1 ) > \sigma Z_1 \rbrace } 
				]^{1/2} 
=
 \sqrt{2}\; \E^\mu[ 
 					\mathds{1}_{ \lbrace Z_2 + Z_1 > Z_1 \rbrace } 
 				]^{1/2} 
\\[-2mm]
G_{{\rm ad}}' ( 0) &= \Vert \partial_x^{\rm ad} \delta_m g \Vert_{\mathbb{L}^2(\mu)}
= \E^\mu \Big[  \big| \E_1^{\mu} [ \mathds{1}_{ \lbrace  Z_2 + Z_1 >  Z_1 \rbrace } ] \big|^2 + \mathds{1}_{ \lbrace  Z_2 + Z_1 >  Z_1 \rbrace } \Big]^{1/2} ,
\end{align*}
\vspace{-7mm}

\noindent and similarly under the martingale Wasserstein ball:
\par
\begin{figure}[H]
\centering

\begin{minipage}{0.48\textwidth}
\centering
\includegraphics[width=\textwidth]{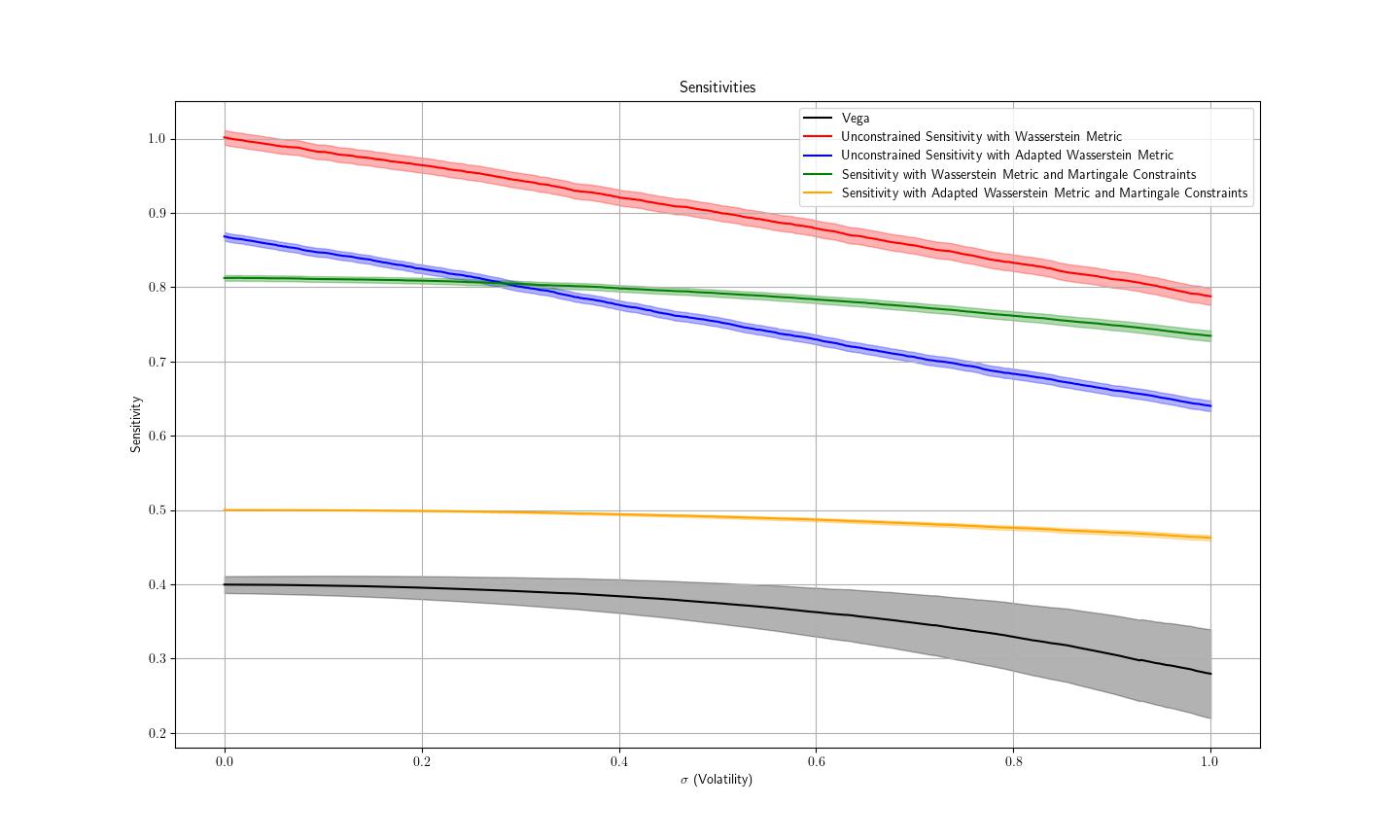}
\caption{\it \footnotesize Sensitivities in the Black-Scholes model, 
$ g ( \mu ) =  \E^{\mu} [ ( X_2 - X_1 )^+ ]$.}
\label{ch01fig:fwstart}
\end{minipage}
\hfill
\begin{minipage}{0.48\textwidth}
\centering
\includegraphics[width=\textwidth]{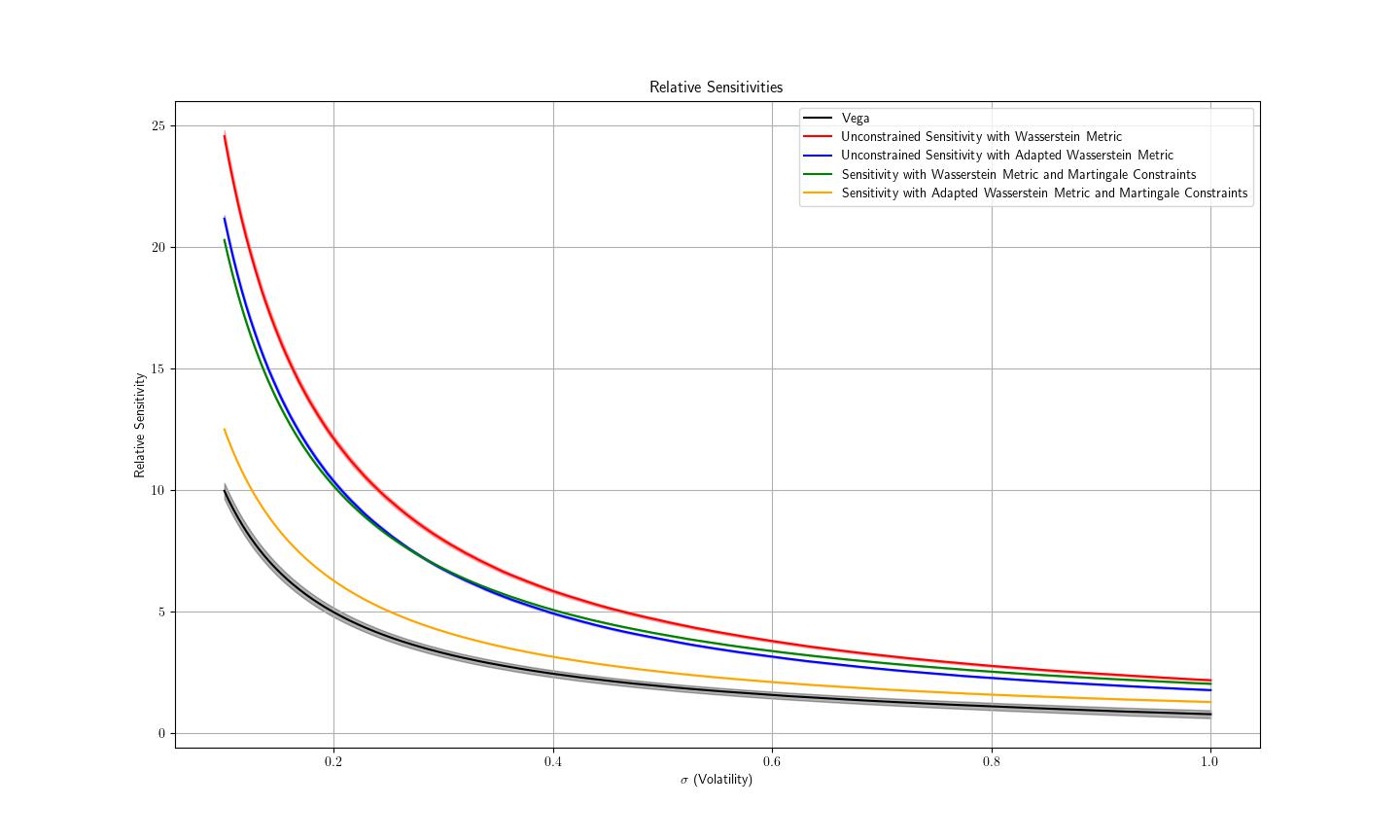}
\caption{\it \footnotesize Relative sensitivities in the Black-Scholes model for $ g \left( \mu \right) =  \E^{\mu} [ ( X_2 - X_1 )^+ ]$.}
\label{fig:relativefwstart}
\end{minipage}
\end{figure}

\vspace{-7mm}
\begin{align*}
G_{\rm M } ' ( 0 ) &= \inf_{h \in C^1_b} \E^\mu\Big[ \vert  \mathds{1}_{ \lbrace Z_2 + Z_1 > Z_1\rbrace } + \sigma h' ( \sigma Z_1 ) Z_2  - h ( \sigma Z_1 )\vert^2 + \vert  \mathds{1}_{ \lbrace Z_2 + Z_1 > Z_1 \rbrace } + h ( \sigma Z_1 )\vert^2 \Big]^{1/2} 
\\[-2mm]
&=
\inf_{h \in C^1_b} \E^\mu\Big[ \vert  \mathds{1}_{ \lbrace Z_2 + Z_1 > Z_1 \rbrace } +  h' ( Z_1 ) Z_2  - h ( Z_1 )\vert^2 + \vert  \mathds{1}_{ \lbrace Z_2 + Z_1 > Z_1 \rbrace } + h (  Z_1 )\vert^2 \Big]^{1/2} ,
\end{align*}
\vspace{-6mm}

\noindent and the adapted martingale Wasserstein version:
$$
\left.\underline{G}^{\rm M}_{\rm ad}\right.'(0)
=
\inf_{h \in \mathbb{L}^{2}(\mu_1)}
\E^\mu\Big[
\big|
\mathds{1}_{\{Z_2+Z_1>Z_1\}} - h(Z_1)
\big|^2
+
\big|
\mathds{1}_{\{Z_2+Z_1>Z_1\}} + h(Z_1)
\big|^2
\Big]^{1/2}.
$$
In the context of the Black-Scholes model, Figure \ref{ch01fig:fwstart} plots the various sensitivities together with the standard {\it Vega} $:= \partial_\sigma g(\mu^\sigma_{\rm{BS}})$ as functions of the volatility parameter, as well as the confidence interval in grey.
We observe that the sensitivity is decreasing with respect to the volatility. 
Furthermore, we observe that the martingale adapted Wasserstein sensitivity is much smaller than the other sensitivities, meaning that restricting the Wasserstein ball to adapted models leads to a less conservative robustness viewpoint. 
Finally, we observe that all sensitivities are significantly smaller than the {\it Vega}. 
Although this may seem natural, it is not guaranteed, as the {\it Vega} is the derivative with respect to the Euclidean norm, which may differ by a constant from the Wasserstein sensitivity when the set of probability measures is restricted to the log-normal distributions. 
Our subsequent example of the American put illustrates this fact.
\vspace{-5mm}
\vspace{-5mm}

To analyse the different sensitivities further, Figure \ref{fig:relativefwstart} plots the relative sensitivity, \textit{i.e.} the ratio of the sensitivities to the reference value $\E^{\mu} [ ( X_2 - X_1 )^+ ]$. 
In the current example, all sensitivities are almost the same for volatility larger than $45$\%. For smaller volatility parameter, we again see a significant difference between the Wasserstein martingale sensitivity and its adapted version. 
This illustrates the relevance of the Wasserstein martingale sensitivity and the corresponding first-order model hedge $h_{\rm M}$.

\noindent We finally explore the worst-case scenario inducing our sensitivity results. 
Such worst-case scenarios correspond to an almost optimal $\mu'$ in the ball $\Bcl(\mu, r)$ or $\Bad (\mu, r)$ with or without martingale restriction. 
By careful examination of the proofs of Propositions \ref{prop:order 1 expansion martingale adapted case}, \ref{prop:order 1 expansion class martingale} and \ref{prop:order 1 expansion marginalandmartingale adapted case}, we see that such (almost) worst-case models are obtained by displaced transport towards a direction singled out by the optimality condition. 

\begin{figure}[H]
\begin{centering}
\includegraphics[scale = 0.25]{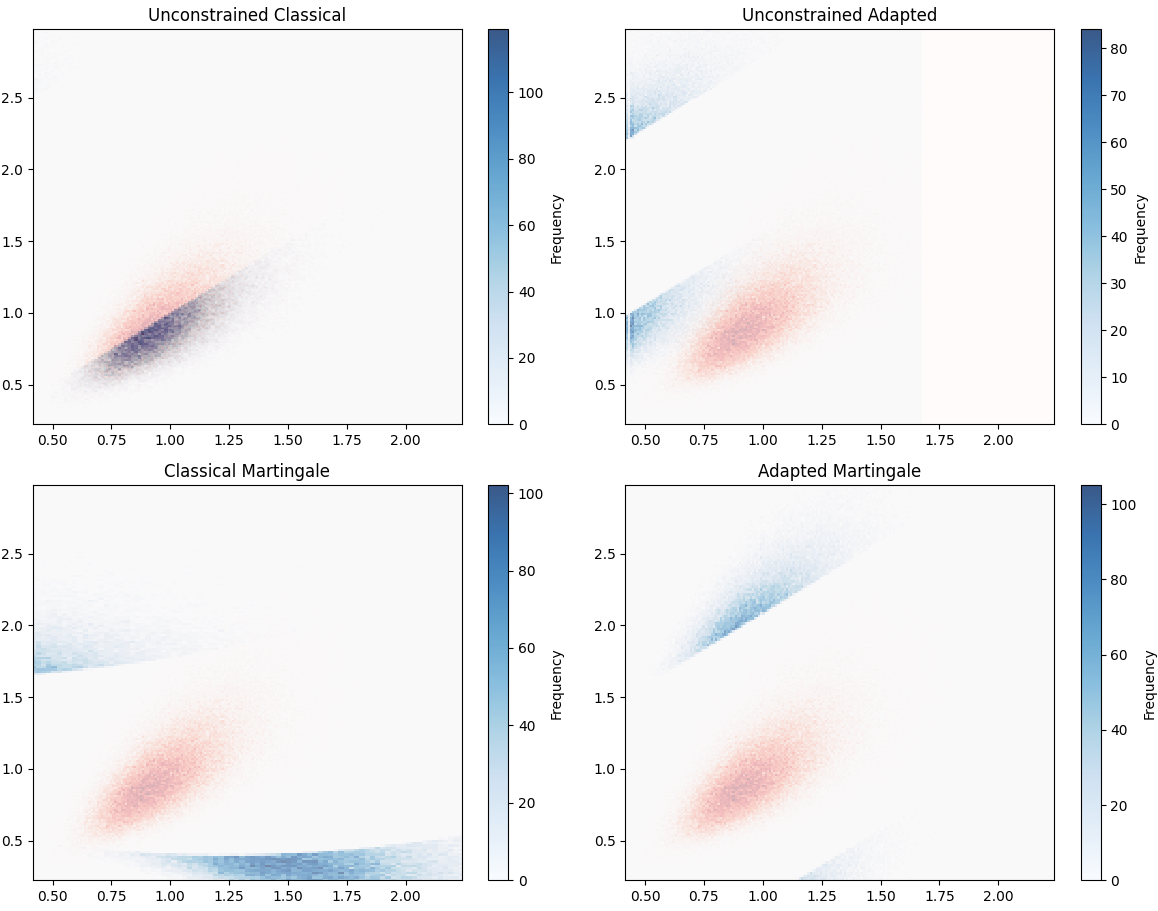}
\caption{\it \footnotesize Worst-case scenarios in the Black-Scholes model 
                   for $ g(\mu) =  \E^{\mu} [ ( X_2 - X_1 )^+ ]$, $\sigma = 0.4$ and $ r = 0.5$ 
                   (Black-Scholes distribution in red).}
                  \label{fig:worstcasefwstart}
\end{centering}
\end{figure}

\noindent Figure \ref{fig:worstcasefwstart} plots the target distribution along which the sensitivity is obtained. 
We observe that, unlike the unconstrained robust Wasserstein sensitivity, the adapted Wasserstein and the martingale-constrained cases exhibit a target distribution that concentrates mass in the tails and leaves the central part near the diagonal line $x_1=x_2$ relatively sparse.

\vspace{-3mm}

\noindent We finally compare the optimal hedging strategies $h$. In order to find $h_{\rm M}$, the optimal hedge in the classical Wasserstein setting, we need to solve the ODE implied by the first order equation (see the  {\bf Proof of Proposition \ref{prop:fredholm equation}} to see how to derive it).
For this particular case, we can choose $h_{\rm M} (x_1) = \alpha \log (x_1) + \beta$ that will be a general solution, and choose the constants such that the system \eqref{eqdef:system_cst} is satisfied, namely we obtain for $log(X) \sim \N(-\sigma^2/2, \sigma ) $
$
\mu_R = \mathbb{E}\!\left[(X -1)^+\right]$, $
p_R = \P(X>1)$, 
$
\alpha = -\frac{\mu_R/\sigma^2}{\,2 + (e^{\sigma^2}-1)/\sigma^2\,} $, 
$
\beta = -\tfrac{1}{2} (2p_R + \tfrac{1}{2}\mu_R 
   - \tfrac{1}{2} \frac{\mu_R/\sigma^2}{ 2 + (e^{\sigma^2}-1)/\sigma^2 }(e^{\sigma^2}-1))
$
In order to find $h_{\rm ad , M}$ and $h_{\rm ad , M, m_1}$ we used propositions \ref{prop:order 1 expansion martingale adapted case} and \ref{prop:order 1 expansion marginalandmartingale adapted case}. 
\begin{figure}[H]
\begin{centering}
\includegraphics[scale = 0.50]{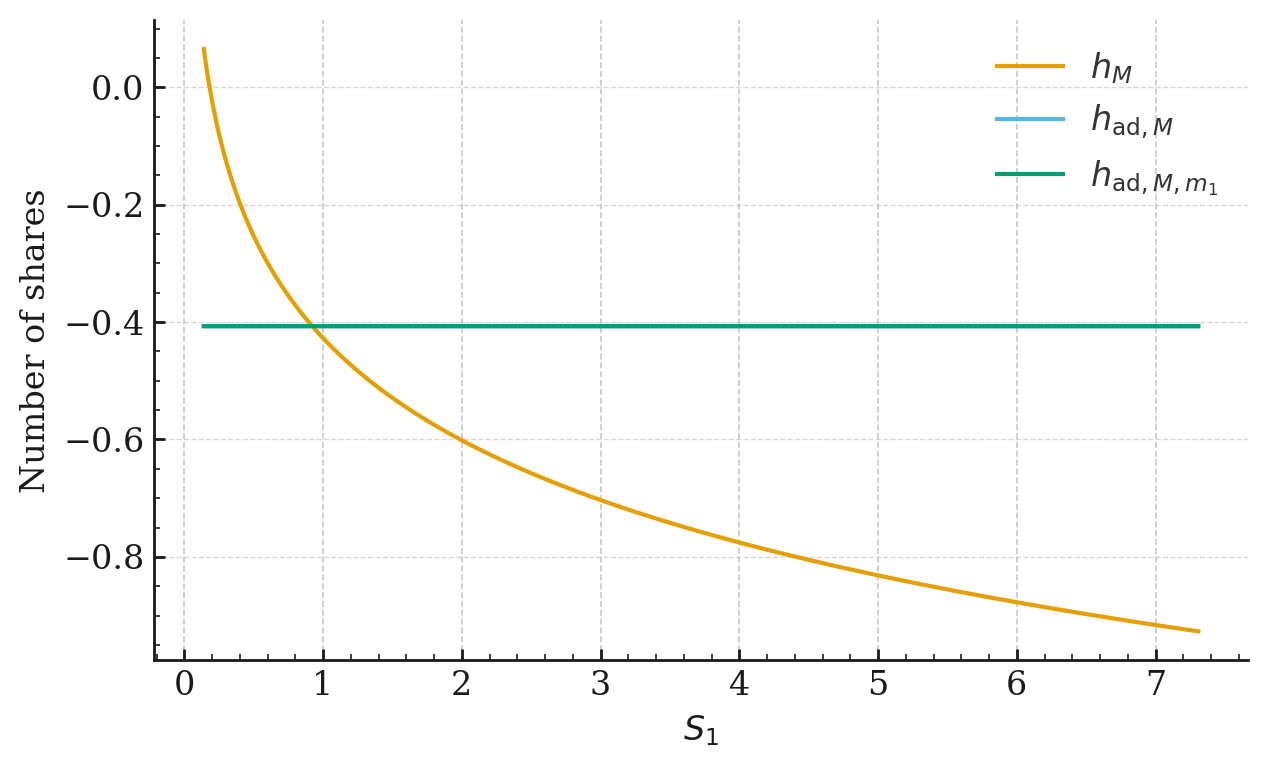}
\caption{\it \footnotesize Optimal Hedges for $ g(\mu) =  \E^{\mu} [ ( X_2 - X_1 )^+ ]$, $\sigma = 0.4$}
\end{centering}
\end{figure}
\noindent A quick computation yields $h_{\rm ad , M} = h_{\rm ad , M, m_1}= - \P( X >1)$, which is coherent with the graph.

\subsection{American put model risk sensitivities}

As a second illustration, we consider the example of the seller price for an American put option with intrinsic values $ \ell_t ( X ) = ( e^{- \rho t} K - X_t  )^+ $, $t=1,2$, with $K = 1.3$, $\rho = 0.05$ and $S_0 = 1$. 
In the present context, all sensitivities are induced by model deviations in the sense of the adapted Wasserstein ball. 
This is consistent with the results of Section \ref{sec:optstop} which are limited to this context. 
The case of deviations induced by a standard Wasserstein ball is left for future research. 

We again observe a clear discrepancy among the various sensitivities. 
The martingale constraint induces a significant change in the sensitivity. 
Combined with the first-marginal constraint, the sensitivity appears to be very flat. 
It is interesting to note that the {\it Vega} is very similar to the sensitivity with marginal, and with martingale and marginal, constraints.

\begin{figure}[H]
\begin{centering}
\includegraphics[scale = 0.29]{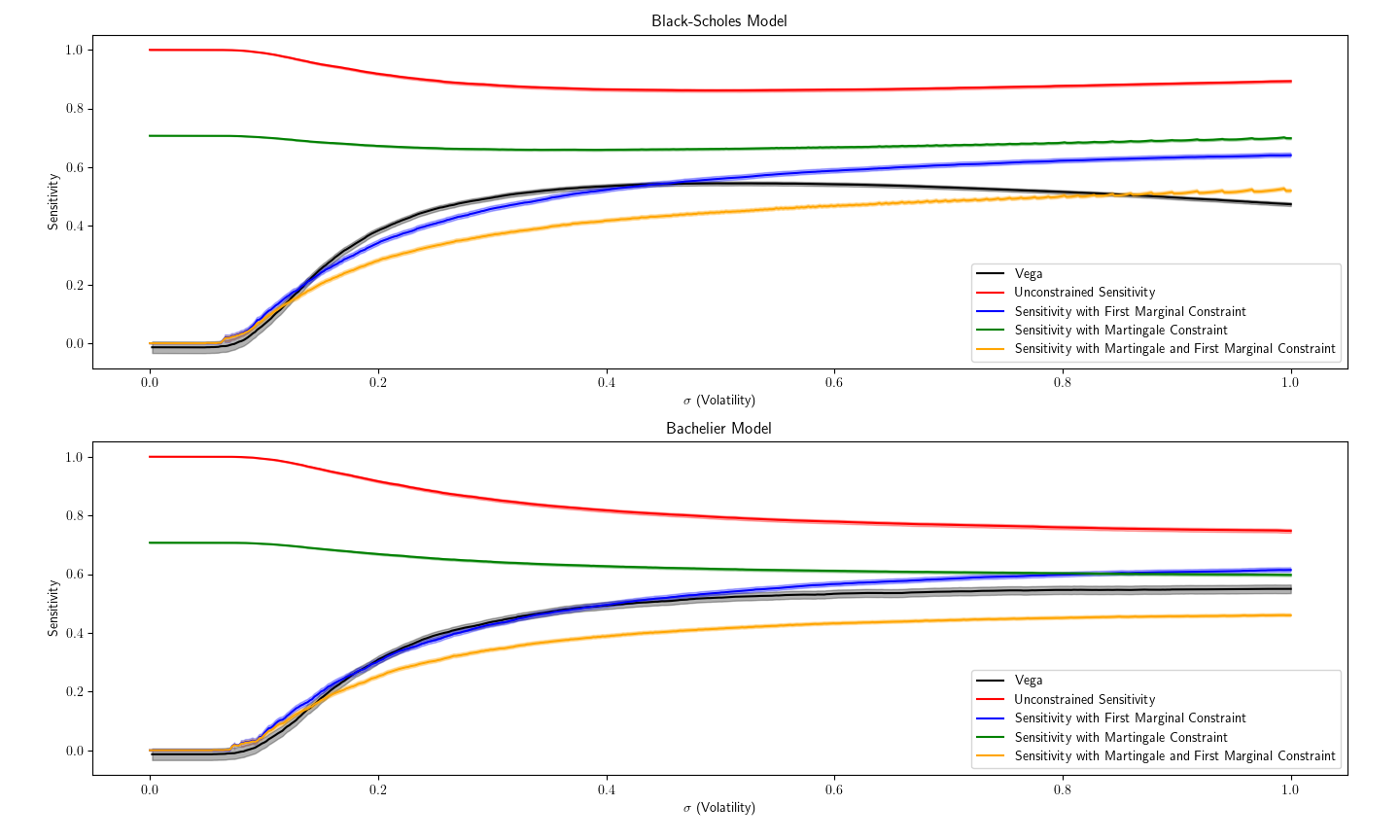}
\caption{\it  \footnotesize Sensitivities in the Black-Scholes model for the American put option 
                  \\$ g(\mu)=\inf_{\tau \in \rm ST} \E^{\mu} \big[(e^{- \rho \tau} K - X_\tau)^+\big]$.}
\end{centering}
\end{figure}

\begin{figure}[H]
\begin{centering}
\includegraphics[scale = 0.3]{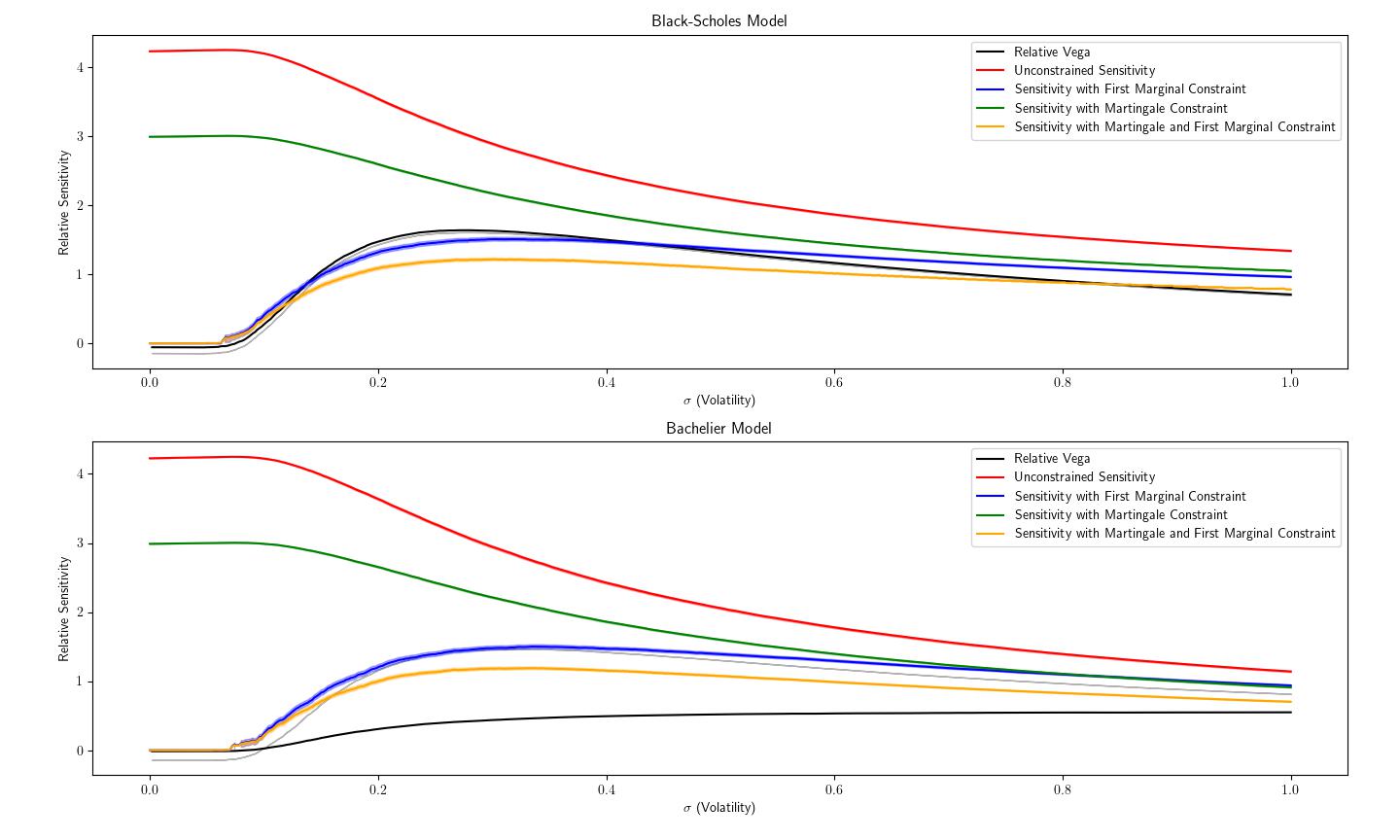}
\caption{\it  \footnotesize Relative sensitivities in the Black-Scholes model for the American put option 
                  \\$ g(\mu)=\inf_{\tau \in \rm ST} \E^{\mu} \big[(e^{-\rho\tau}K-X_\tau)^+\big]$.}
\label{fig:relative-us}
\end{centering}
\end{figure}

As in the previous section, we further explore the differences between the various sensitivities by normalizing with the reference value function. 
The resulting relative sensitivities are plotted in Figure \ref{fig:relative-us}. 
In both models, we observe that the discrepancy between the various sensitivities decreases for large values of the volatility. 

We finally plot the worst-case scenarios. 
We recall that such an (almost) worst-case model is the target distribution which is used in the proof with a displacement transport argument. 
The present example exhibits behavior completely different from the forward start European option of the previous section. 
The worst-case model does not show the mass concentration at the extremes. 
Instead, the distribution appears to be translated with similar concentration characteristics. 

\begin{figure}[H]
\begin{centering}
\includegraphics[scale = 0.3]{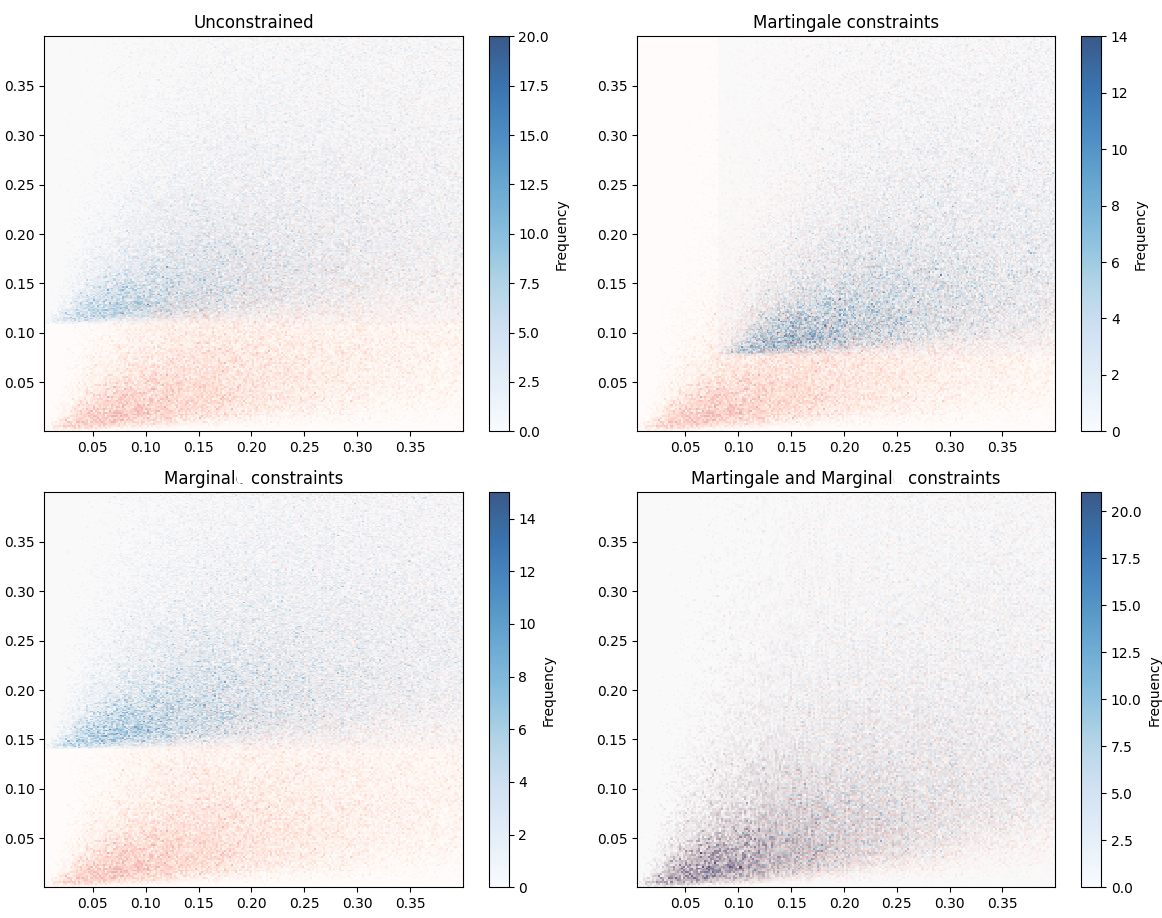}
\caption{\it  \footnotesize Worst-case scenarios in the Black-Scholes setting 
             \\ for the American put option 
              \\ $g (\mu)=\inf_{ \tau \in \rm ST} \E^{\mu} \big[(e^{-\rho\tau}K-X_\tau)^+ \big]$,  $\sigma = 0.5$ and $ r = 0.1$ 
              \\ (Black-Scholes distribution in red).}
\end{centering}
\end{figure}

\begin{figure}[H]
\begin{centering}
\includegraphics[scale = 0.3]{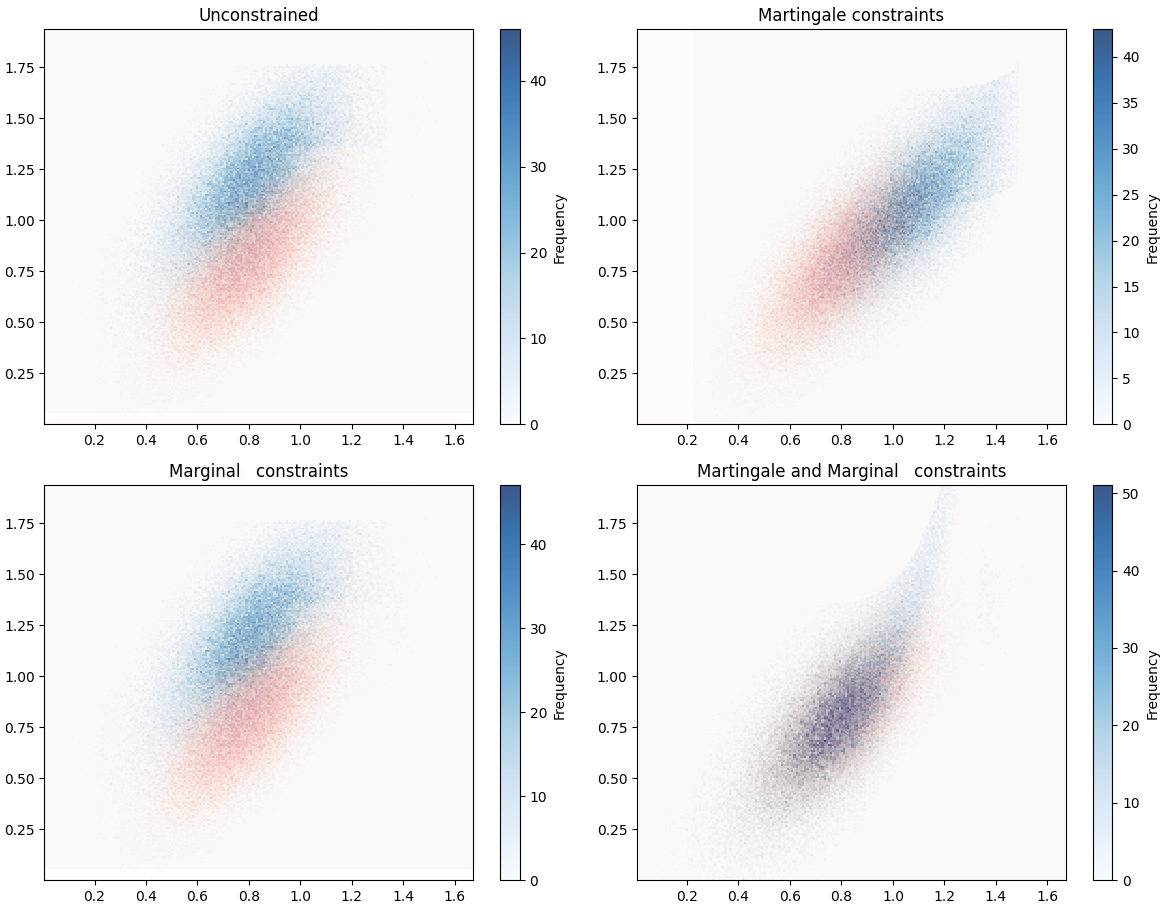}
\caption{\it  \footnotesize Worst-case scenarios in the Bachelier model for the American put option 
              \\ $g (\mu)=\inf_{ \tau \in \rm ST} \E^{\mu} \big[(e^{-\rho\tau}K-X_\tau)^+ \big]$,  $\sigma = 0.5$ and $ r = 0.1$ 
              \\ (Bachelier distribution in red).}
\end{centering}

\end{figure} 

Again, we can compare the optimal hedges for both models

\begin{figure}[H]
\centering

\begin{minipage}{0.48\textwidth}
\centering
\includegraphics[width=\textwidth]{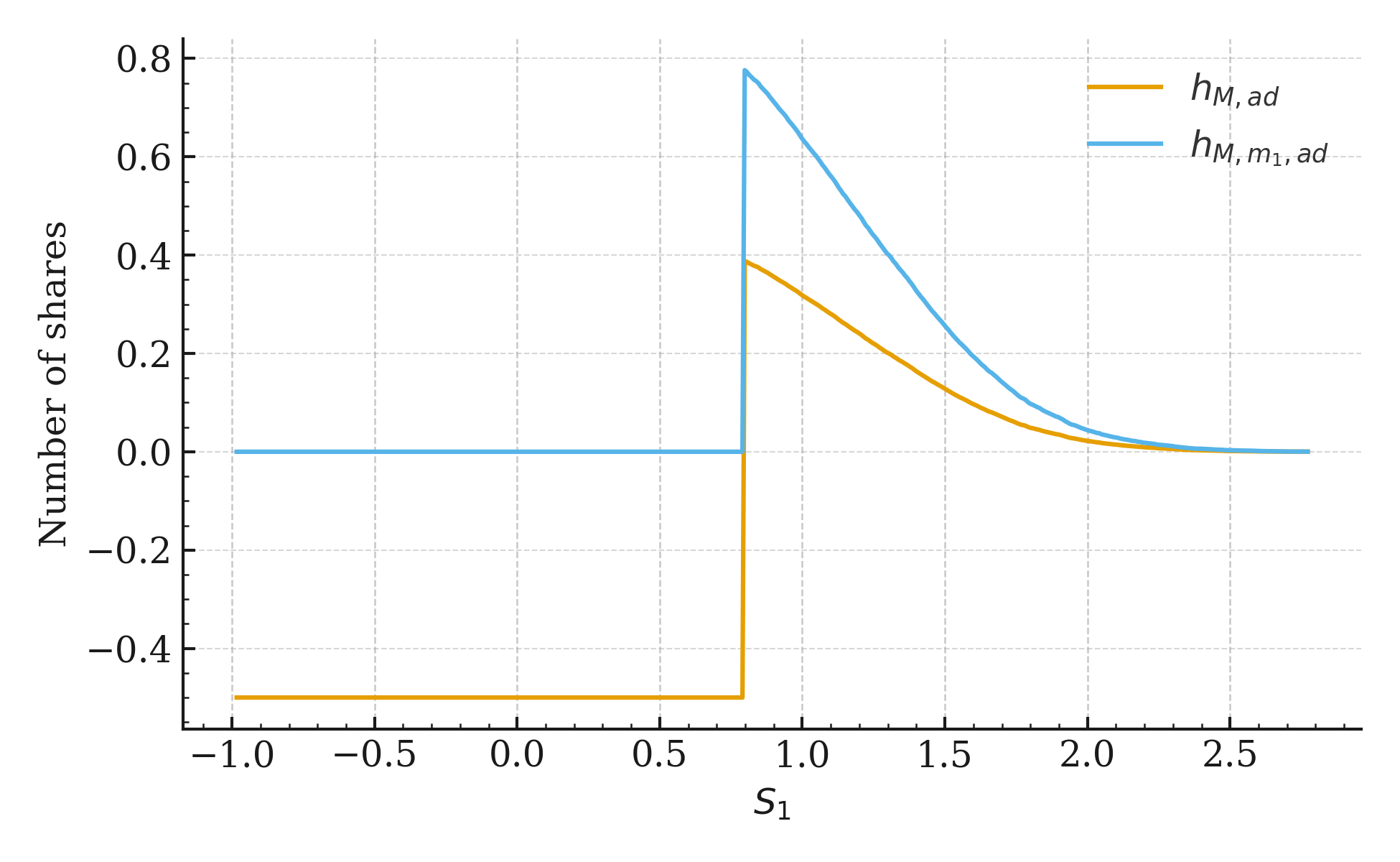}
\vspace{-10mm}
\caption*{\it \footnotesize Bachelier model, $\sigma = 0.5$.}
\end{minipage}
\hfill
\begin{minipage}{0.48\textwidth}
\centering
\includegraphics[width=\textwidth]{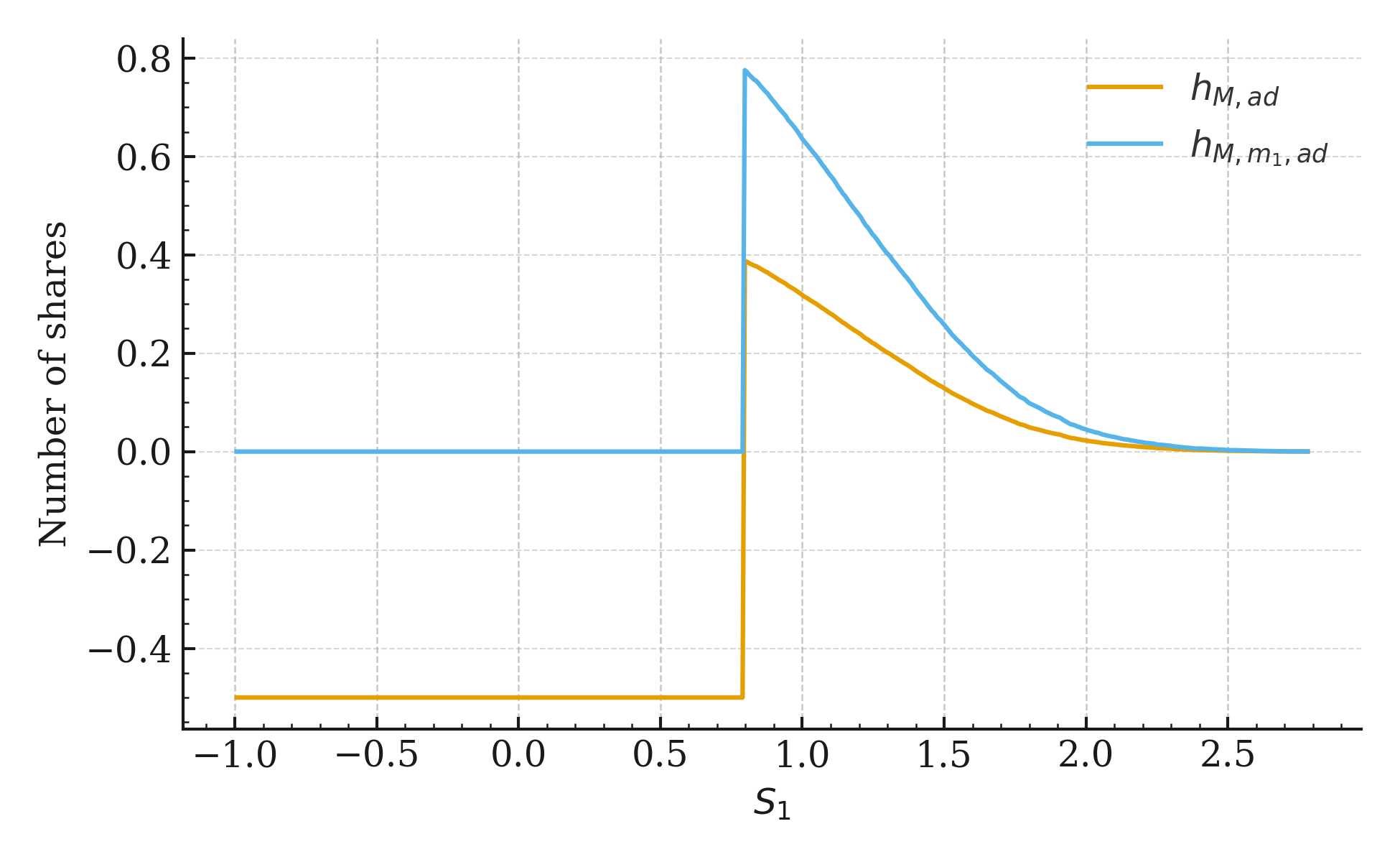}
\vspace{-10mm}
\caption*{\it \footnotesize Black-Scholes model, $\sigma = 0.5$.}
\end{minipage}

\caption{\it \footnotesize Optimal Hedges in the Bachelier and Black-Scholes models
\\ $g (\mu)=\inf_{\tau \in \rm ST} \E^{\mu} \big[(e^{-\rho\tau}K-X_\tau)^+ \big]$, $\sigma = 0.5$.}
\end{figure}

\section{Proofs}

\subsection{Reduction to the linearized problem}

For fixed $\mu\in\Pc_p(\X)$, we introduce the (linear) first-order approximation of $g$ near $\mu$, which will be used in all subsequent proofs:
\vspace{-2mm}
\begin{equation*}
\hat{g} ( \mu' ) := g ( \mu ) + \int_{\X } \delta_m g ( \mu, x ) ( \mu' - \mu ) ( \mathrm{ d } x ),
~~\mu'\in\Pc_p(\X).
\vspace{-2mm}
\end{equation*} 
We also define the corresponding distributionally robust evaluations:
\vspace{-2mm}
$$
\apGMcl (r) 
:=
\sup_{\mu'\in \BclM (\mu,r)} 
\hat{g} ( \mu' ) 
~~\mbox{and}~~
\apGMad ( r ) 
:=
\sup_{\mu'\in \BadM (\mu,r)} 
\hat{g} ( \mu' ), 
\vspace{-2mm}
$$
where $\BclM, \BadM$ are defined by \eqref{eqdef:balls_deviations}.
We first show that the maps $\lowGMad$ and $\lowGMc$ are differentiable at the origin with sensitivities related to those of $\apGMad$ and $\apGMcl$.

\begin{Lemma}\label{lemma:reducto al linear}
Under Assumption \ref{ass:on g}, the maps $\apGMcl$ and $\apGMad$ satisfy
\vspace{-2mm}
$$ \big(\lowGMc- \apGMcl\big)(r)
=
\circ(r) 
~~\text{and}~~ 
\big(\lowGMad - \apGMad\big)(r)
=
\circ(r) 
.\vspace{-2mm}
$$
Consequently, the following families have the same $\liminf$ and $\limsup$ at the origin $r\searrow 0$:

$\bullet$ $\frac{\lowGMc ( r ) - \lowGMc ( 0 )}{r}$ and $\frac{\apGMcl ( r ) - \apGMcl ( 0 )}{r}$,

$\bullet$ $\frac{\lowGMad ( r ) - \lowGMad ( 0 ) } {r}$
and $\frac{ \apGMad ( r ) - \apGMad ( 0 ) }{r}$.
\end{Lemma}

\proof We only prove it for the standard Wasserstein distance, since the adapted Wasserstein distance setting will be done following the same line of arguments.
For $\mu' \in \BclM (\mu,r)$, define $\bar{\mu}_\lambda:=( 1 - \lambda ) \mu + \lambda \mu'$ and set $R( \bar{\mu}_\lambda, x) 
:=
\delta_m g \big(\bar{\mu}_\lambda, x )
-
\delta_m g ( \mu , x )$. 
From Assumption \ref{ass:on g}, we have that
$
g( \mu' ) 
=
\hat{g}( \mu' ) 
+
\int_0^1 
\int_{\X} 
R ( \bar{\mu}_\lambda, x ) 
 ( \mu' - \mu ) ( \mathrm{ d } x )
\mathrm{d} \lambda .
$
Applying the triangle inequality and Hölder's inequalities, we obtain
\vspace*{-4mm}
\begin{equation*}
\begin{split}
\vert 
 \underline{G}^{\rm M} ( r ) - \apGMcl ( r )
\vert 
&\leq
\sup_{ \mu' \in \Bcl (\mu,r)} 
\int_0^1 \left\vert \int_{\X} 
R ( \bar{\mu}_\lambda, x )
 ( \mu' - \mu ) ( \mathrm{ d } x )
 \right\vert 
\mathrm{d} \lambda 
\\[-2mm]
&\leq 
r
\sup_{\pi\in \mathfrak{D}^{ \X \times \X }_p(\mu,r)} 
\int_0^1 
\int_0^1 
\E^\pi \big[ \vert 
 \partial_x R ( \bar{\mu}_\lambda, \bar{X}_{\lambda'} ) 
\vert^{p'} 
\big]^{\frac1{p'}}
\mathrm{d} \lambda' 
\mathrm{d} \lambda,
\vspace{-3mm}
\end{split}
\vspace{-3mm}
\end{equation*}
\vspace{-3mm}

\noindent 
where $\bar{X}_{\lambda'} = ( 1 - \lambda' )X' + \lambda' X $ for $ 0 \leq \lambda' \leq 1 $ and 
\vspace{-2mm}
$$
\mathfrak{D}^{\X \times \X}_p(\mu,r)
:=
\big\{\pi\in\Pc_p( \X \times \X):~\pi\circ X^{-1}=\mu
                                               ~\mbox{and}~
                                               \E^\pi|X-X'|^p\le r^p
\big\}.
\vspace{-3mm}
$$
Fix $ 0 \leq \lambda , \lambda' \leq 1 $. 
For a family $ ( \pi_r )_r$ satisfying $\pi_r \in \mathfrak{D}^{\X \times \X}_p(\mu,r)$, we have $
\pi_r \xrightarrow[r \rightarrow 0]{\W_p}  \mu \circ ( X, X )^{-1}
$. 
Then, it follows from Assumption \ref{ass:on g} that $\partial_x \delta_m g $ is continuous with respect to the measure variable, so 
$
\E^{\pi_r} \big[ | \partial_x R ( \bar{\mu}_\lambda, \bar{X}_{\lambda'} ) 
|^{p'} 
\big]
\xrightarrow[r \rightarrow 0]{} 0 
.$
Finally, by Assumption \ref{ass:on g} on $g$, one can dominate $\E [ | 
\partial_x R ( \bar{\mu}_\lambda, \bar{X}_{\lambda'} ) 
|^{p'} 
]$ uniformly in $\lambda$ and $\lambda'$ since $0 \leq \lambda, \lambda' \leq 1$, and we deduce the required result by the dominated convergence theorem.
\ep
\vspace{-4mm}

\subsection{The martingale adapted Wasserstein sensitivity}
We first prove the following lemma. 
\begin{Lemma}\label{lemma:limsup wass ad}
Suppose that Assumption \ref{ass:on g} holds. 
Then,
\vspace{-2mm}
$$
\limsup_{r \rightarrow 0} \frac{
 \upGMad ( r ) -  \upGMad ( 0 )
}{r}
\leq 
\inf_{ h \in \mathbb{L}^{p'}(\mu_1) } U^{\rm M}_{\rm ad}(h) \,\, \text{ where $U^{\rm M}_{\rm ad}$ is defined by \eqref{eq:deriv mart adapted}}
.
\vspace{-2mm}
$$
\end{Lemma}
\proof
First, notice that $ \upGMad ( 0 ) = g( \mu )$. 
Furthermore, for $h \in C^1_b ( S, S )$, by min-max inequality,
\vspace{-2mm}
$$
\frac{
 \upGMad ( r ) -  \upGMad ( 0 )
}{r}
\leq 
\frac{1
}{r}
\sup_{ \mu' \in \Bad ( \mu, r )} \{ \phi_h ( \mu' ) - \phi_h ( \mu )\} ,
\,\,\, \mbox{where} \,\,\,
\phi_h (\mu) := g ( \mu ) + \E^\mu [ h^{\otimes} ]
,
\vspace{-2mm}
$$
and $h^\otimes$ is defined by \eqref{eqdef:otimes}.
However, $ h\in C^1_b(S, S)$ so, $ \phi_h$ satisfies Assumption \ref{ass:on g}. 
The linear approximation of $\phi_h$ at $\mu$ is $\hat{\phi}_h(\mu')
:=
\phi_h ( \mu ) + \int_{\X} w \mathrm{d} (\mu' - \mu)
$ with $ w(x) :=
\delta_mg(\mu,x)+h^\otimes(x) $ since $  \delta_m \E^\mu [h^\otimes ] = h^\otimes $. 
Then, following the proof of Lemma \ref{lemma:reducto al linear}, we get
\vspace{-3mm}
\begin{equation}\label{eq:limsup upbar G}
\limsup_{r \rightarrow 0} \frac{
 \upGMad ( r ) -  \upGMad ( 0 )
}{r}
\leq 
\limsup_{r \rightarrow 0} \frac{1}{r}
 \sup_{  \mu' \in \Bad ( \mu, r ) }
\{ \hat{\phi}_h(\mu') - \hat{\phi}_h(\mu)
\}
.\vspace{-3mm}
\end{equation}
Let $\nu_r \in \Bad ( \mu, r ) $ and $\pi^r$ a bi-causal coupling between $\mu $ and $\nu_r$ satisfying
\vspace{-3mm}
\begin{equation} \label{ineq:distance pi delta}
\!\sup_{ \mu' \in \Bad (\mu, r)} 
\int_{ \X }
\! w \mathrm{d}( \mu' \!-\! \mu)
\!-\! r^2
\!
 \leq 
 \!
 \int_{ \X }
 \!
w \mathrm{ d}( \nu_r  \!-\! \mu),
\, 
\E^{\pi^r} \big[ \vert X \!-\! X' \vert^p \big]^{1/p}
\!\! \leq 
\W_p^{{\rm ad}} ( \nu_r, \mu ) 
\!+\!
r^2.
\vspace{-3mm}
\end{equation}
By differentiability of $\delta_m g $ and $h$, we know that $w$ is differentiable. 
So, by Fubini's Theorem, 
\vspace{-2mm}
$$ \int_{ \X }
w\,\mathrm{ d}( \nu_r  \!-\! \mu) 
=
\int_0^1 
\E^{\pi^r} 
[
\partial_x w ( \bar{X}_\lambda ) \cdot ( X' - X )
]
\mathrm{d}\lambda
\,\, \text{with} \, \bar{X}_\lambda = ( 1 - \lambda ) X + \lambda X' 
.
\vspace{-2mm}
$$
Hence,
\vspace{-2mm}
\begin{equation}\label{ineq:GM}
\!
\frac{1}{r}
\!
 \sup_{  \mu' \in \Bad ( \mu, r ) }
 \{ \hat{\phi}_h(\mu') \!-\! \hat{\phi}_h(\mu) \}
\! \leq \!
\frac{1}{r} \!
\int_0^1  \!\!
\E^{\pi^r} 
[
\partial_x w ( \bar{X}_\lambda ) \!\cdot\! (X' \!-\! X) 
]
\mathrm{d}\lambda
\!+\!
r
=
I_1 \!+\! I_2 \!+\! r,
\vspace{-4mm}
\end{equation}
where 
\vspace{-2mm}
$$
I_1 
:=
\frac{1}{r}
\E^{\pi^r} 
[
\partial_x w ( X ) \!\cdot\! ( X' - X)
]
\,\, \text{and} \,\, 
I_2 :=
\frac{1}{r}
\int_0^1 
\E^{\pi^r} 
\big[
(
\partial_x w (  \bar{X}_\lambda  ) 
-
\partial_x w (  X  )
)
\!\cdot\! ( X' - X )
\big]
\mathrm{d}\lambda
.
\vspace{-2mm}
$$
Since $ \pi^r $ is bi-causal, $ \E^{\pi^r} [ \partial_{x_1} w \cdot X_1'] = \E^{\pi^r} \big[ \E_1^{\mu} [ \partial_{x_1} w ] \cdot X_1' \big] $, hence, applying successively Hölder's inequality and the estimate \eqref{ineq:distance pi delta}:
\begin{align}
I_1 
&=
\frac{1}{r}
\E^{\pi^r} 
\big[
\E^\mu_1 [ 
\partial_{x_1} w ( X  ) ] \cdot ( X'_1 - X_1 )
+
\partial_{x_2} w ( X  )  \cdot ( X'_2 - X_2 )
\big]
\nonumber\\
&\leq 
\frac{\E^{\pi^r} [ \vert X - X' \vert^p ]^{1/p} }{r}
\Vert \partial_x^{\rm ad} w \Vert_{\mathbb{L}^{p'} ( \mu ) } 
\leq 
( 1 + r ) 
\Vert \partial_x^{\rm ad}w \Vert_{\mathbb{L}^{p'} ( \mu )}.
\label{ineq:I1}
\end{align}
Again, by Hölder's inequality and the tower property 
\vspace{-2mm}
$$
I_2
\leq ( 1 +r )
\int_0^1 
\E^{\pi^r}
[
\vert 
\partial_x w \big( X + \lambda ( X' - X ) \big ) 
-
\partial_x w ( X  )
\vert^{p'}
]^{1/p'} 
\mathrm{d} \lambda.
\vspace{-2mm}
$$
Now, since for all $ r > 0$, 
$
\E^{\pi^r} [ \vert X - X' \vert^p ]^{1/p}
\leq 
\W_p^{{\rm ad}} ( \nu_r, \mu ) 
+
r^2 
\leq 
r + r^2 
$
it is clear that $ \pi^r \xrightarrow[ r \rightarrow 0 ]{ \W_p} \mu \circ ( X, X )^{-1} $. 
By Assumption \ref{ass:on g} on $g$ together with the continuity of $\partial_{x_1} h$, it follows that the map 
$
 ( x, x' ) \rightarrow 
 \partial_x w ( x + \lambda ( x' - x ) ) 
-
\partial_x w ( x  )
$
is continuous and $
\vert 
 \partial_x w ( \bar{x}_\lambda  ) 
-
\partial_x w ( x  )
\vert^{p'}
\leq C ( 1 + \vert x \vert^p + \vert x' \vert^p )
$,
for some $C > 0$. 
Then 
\vspace{-3mm}
$$
\E^{\pi^r}
[
\vert 
\partial_x w ( X + \lambda ( X' - X ) ) 
-
\partial_x w ( X  )
\vert^{p'}
]
\xrightarrow[r \rightarrow 0 ]{} 0
.
\vspace{-3mm}
$$
By the dominated convergence Theorem, we deduce that $
I_2 \xrightarrow[r \rightarrow 0 ]{} 0 
$. 
Combined with Inequalities \eqref{ineq:GM} and \eqref{ineq:I1}, this provides
\vspace{-2mm}
$$
\limsup_{r \rightarrow 0} \frac{1}{r}
 \sup_{  \mu' \in \Bad ( \mu, r ) }
 \{ \hat{\phi}_h(\mu') - \hat{\phi}_h(\mu)
\}
\leq 
\Vert \partial_x^{\rm ad} w \Vert_{\mathbb{L}^{p'} ( \mu )}
=
\Vert \partial_x^{\rm ad} \delta_m g + J h( X_1 ) \Vert_{\mathbb{L}^{p'} ( \mu )},
\vspace{-3mm}
$$
as $ \partial_x^{\rm ad} w =  \partial_x^{\rm ad} \delta_m g +\partial_x^{\rm ad} h^{\otimes}  $, $
(\partial_x^{\rm ad} h^{\otimes})_1
=
\E^{\mu}_{1} [ (\partial_{x_1} h)(X_1)\cdot(X_2 -X_1) - h(X_1)]
\,\,\text{and} \,\, 
(\partial_x^{\rm ad} h^{\otimes})_2
=
h(X_1)
$, and, 
by the martingale property, $\E^{\mu}_{1} [ (\partial_{x_1} h)(X_1)\cdot(X_2 -X_1) ]= 0 $, hence, $ \partial_x^{\rm ad} h^{\otimes} = J h( X_1 )$ where $J$ is defined by \eqref{eqdef:caus_grad,J}. 
By Estimate \eqref{eq:limsup upbar G} along with the arbitrariness of $h$ and the density of $C^1_b( S, S ) $ in $\mathbb{L}^{p'}(\mu_1)$,
\vspace{-6mm}
$$
\limsup_{r \rightarrow 0} \frac{
 \upGMad ( r ) -  \upGMad ( 0 )
}{r}
\leq 
\inf_{h \in \mathbb{L}^{p'}( \mu_1 ) }
U_{{\rm ad}}^{\rm M} ( h ) 
.
\vspace{-3mm}
$$
\ep

\smallskip
\noindent {\bf Proof of Proposition \ref{prop:order 1 expansion martingale adapted case}} We proceed in several steps. 
Following Lemma \ref{lemma:reducto al linear}, we will first prove that $\apGMad$ is differentiable at the origin. 
Similarly to \citeauthor{bartlsensitivityadapted} \cite{bartlsensitivityadapted}, we will prove the two reverse inequalities separately. 
While the upper bound does not pose any difficulty, the reverse inequality requires to construct a family of measures that are martingales, and close to $ \mu$ with respect to the adapted Wasserstein distance. 
Once this differentiability is established, together with Lemma \ref{lemma:limsup wass ad}, Proposition \ref{prop:order 1 expansion martingale adapted case} will follow directly.

\smallskip
\noindent {\bf Step $1$.} 
We first prove the differentiability of $\apGMad$ at $0$. 
Following the same line of argument as in Lemma \ref{lemma:limsup wass ad}, we easily obtain 
\vspace{-2mm}
$$
\limsup_{r \rightarrow 0} 
\frac{\apGMad (r) - \apGMad (0) 
}{r} 
\leq 
\inf_{h \in \mathbb{L}^{p'} (\mu_1) } U_{\rm ad}^{\rm M}(h)
.
\vspace{-2mm}
$$

\begin{Remark}\label{rem:non nullity of the opti problem adapted}
{\rm If $ \inf_{h \in \mathbb{L}^{p'} (\mu_1) } U_{\rm ad}^{\rm M}(h) = 0$, then since $ \apGMad $ is increasing, it is differentiable at $0$, with derivative equal to $0$. 
We henceforth assume that $\inf_{h \in \mathbb{L}^{p'} (\mu_1) } U_{\rm ad}^{\rm M}(h) > 0$.}
\end{Remark}

\smallskip

\noindent\textit{Analysis of the optimization problem \eqref{eq:deriv mart adapted}.} Observe that the functional $ U_{{\rm ad}}^{\rm M} $ defined on $\mathbb{L}^{p'}( \mu_1 ) $, a reflexive Banach space, is convex and coercive in the sense of definition \ref{ch01def:coercivity}. 
It then admits a minimizer $h_{\rm ad, M}$ satisfying the first-order condition $ \E^{\mu} [ f ( X_1 ) \cdot \mathbf{N} ( \phi_1(X_1) ) ]=\E^{\mu} [ f( X_1 ) \cdot \mathbf{N} ( \phi_2 ) ]$ for all $ f \in \mathbb{L}^{p'}( \mu_1 )$, where $\mathbf{N}$ is defined by Equation \eqref{eqdef:n and n_ad},  
\vspace{-2mm}
$$
 \phi_1(X_1) = 
\E ^\mu_1 [ \partial_{x_1} \delta_m g ]
-
h_{\rm ad, M}(X_1) 
\,\, \text{and} \,\, 
 \phi_2
 =
\partial_{x_2} \delta_m g 
+
h_{\rm ad, M}(X_1) 
, \text{ yielding Equation \eqref{eq:first order condition martingale}.}
\vspace{-2mm}$$

\smallskip
\noindent \textit{The lower bound.} Set 
$
T^{\rm ad} := ( T^{\rm ad}_1, T^{\rm ad}_2 ) := \frac{1}{c} \mathbf{N}_{\rm ad} ( (\phi_1, \phi_2) )
$ and $
\mu^{rT} := \mu \circ ( X + r T )^{-1}
$
where $c$\footnote{see Remark \ref{rem:non nullity of the opti problem adapted}} is chosen so that $\Vert T^{\rm ad} \Vert_{\mathbb{L}^p (\mu) } = 1$.
 
For $h \in \mathbb{L}^{\infty}( S, S) $, we compute:
\vspace{-2mm}
\begin{align*}
\E^{\mu^{rT}}[ h( X'_1) \!\cdot\! ( X'_2 - X'_1 ) ]
=&
\E^{\mu}[
h( X_1 +  r T_1(X_1) ) \!\cdot\! ( X_2 - X_1 + r (T_2 - T_1) ) ]
\\
=&
\E^{\mu}[ h(X_1 +  r T_1( X_1)) \!\cdot\! (X_2 - X_1) 
             ]
\\
&\hspace{5mm}+ 
r
\E^{\mu} [ h (X_1 +  r T_1(X_1)) \!\cdot\! (T_2 - T_1) ]
\;=\;
0,
\end{align*}
\vspace{-7mm}

\noindent 
where the first term vanishes since $\mu$ is a martingale measure and $T_1 \in \sigma ( X_1 ) $, while the second is also zero since $\E^{\mu}_1 [ T_2 ] = T_1$ by Equation \eqref{eq:first order condition martingale}. 
This shows that $\mu^{rT}$ is a martingale measure. 
However, the coupling $ \pi^r := \Lc ( X, X + r T ( X ) ) $ is causal and not necessarily bi-causal. 
By Lemma $3$ of \citeauthor{blanchet2024empirical} \cite{blanchet2024empirical}, there exists a family $ ( \hat{X}^{\varepsilon} )_\varepsilon$ such that for all $\varepsilon >0 $, $  \hat{X}^{\varepsilon} \in \sigma( X ) $, $\nu_{\varepsilon} := \mathcal{L}(\hat{X}^{\varepsilon})$ is a martingale measure and, 
\vspace{-2mm}
\begin{equation} \label{inter:estim approx mart}
\vert X +r T ( X ) - \hat{X}^\varepsilon \vert \leq \varepsilon \,\, , \,\, \nu_\varepsilon \in \rm M \,\, \text{and} \,\, \mathcal{L} ( X, \hat{X}^\varepsilon ) \in \Pi^{\text{bc}} ( \mu, \nu_\varepsilon ).
\vspace{-1mm}
\end{equation}
Letting $\varepsilon = r \kappa$, and $R_\varepsilon = \E^\mu [ \delta_m g (\mu, \hat{X}^{\varepsilon} ) - \delta_m g ( \mu,  X + r T( X) ) ]$,
for $ \kappa > 0$, we have 
$$
\apGMad ( r( 1 + \kappa )) - \apGMad(0) 
\geq
\E^\mu [ \delta_m g( \mu, X + r T ( X ) ) - \delta_m g ( \mu, X ) ]
+ 
R_\varepsilon
.$$
By Assumption \ref{ass:on g} on $g$,  \eqref{inter:estim approx mart}, and Taylor's formula, we have
$
\vert R_\varepsilon \vert \leq C \varepsilon.
$
Thus, as $r \downarrow 0$,
\vspace{-1mm}
$$
( 1 + \kappa ) \liminf_{r \rightarrow 0} 
\frac{\apGMad (r) - \apGMad (0) 
}{r} 
\geq 
\E^\mu [ \partial_x^{\rm ad} \delta_m g \cdot T ] 
= \Vert \partial_x^{\rm ad} \delta_m g + J h_{\rm ad, M} ( X_1 ) \Vert_{\mathbb{L}^{p'}(\mu)}, 
\vspace{-2mm}
$$ 
where the last equality is a consequence of the first-order condition \eqref{eq:first order condition martingale}. 
Letting $\kappa \rightarrow 0$, we proved the differentiability at 0 of $\apGMad$ at $0$ and, by Lemma \ref{lemma:reducto al linear}, the differentiability at $0$ of $ \lowGMad$.

\noindent {\bf Step $2$.} 
We prove differentiability of $\upGMad$ at $0$. 
First,
$ \limsup_{r \rightarrow 0} \frac{\upGMad(r) - \upGMad(0)}{r} \leq \left. \lowGMad \right.' (0)
$,
by Lemma \ref{lemma:limsup wass ad}.
Furthermore, by weak duality, $\upGMad \geq \lowGMad$ and $\upGMad ( 0 ) =  \lowGMad (0) = g(\mu)$ hence,
$\liminf_{r \rightarrow 0} \frac{\upGMad(r) - \upGMad(0)}{r} \geq \left. \lowGMad \right.' (0)$, 
proving the differentiability at 0 of $\upGMad$ .
\ep
\vspace{-3mm}

\subsection{The standard Wasserstein martingale sensitivity}

In this section, we prove Proposition \ref{prop:order 1 expansion class martingale} by proceeding in three steps as in the previous section. 

\begin{Lemma} \label{lemma:default martingale}
Let $\theta_1,\theta_2 : \X \rightarrow S  $ be such that for some $r_0 >0$:

$\bullet$ $\theta_1$ is compactly supported, continuous and is $C^1$ in $x_1$. 
Furthermore, $(\text{Id} +rJ_1\theta_1)(\Omega)=\Omega$ for all $r < r_0$,

$\bullet$ $\theta_2$ is measurable and continuous in $x_1$. 

\noindent Then, for $\mu$ satisfying Assumption \ref{ass: mu converse inequality}, we have 
\par\vspace{-5mm}
\begin{align*}
&\lim_{r \rightarrow 0} 
\frac1{r}
\Big\| X_1\!+\!r \theta_1 ( X ) 
         - \E^\mu\big[ X_2 \!+\! r \theta_2 ( X ) \big| X_1 \!+\! r \theta_1( X)\big]
\Big\|_{\mathbb{L}^p(\mu)}
\\
&\hspace{20mm}= 
\Big\|
\int_{S} \Big(  \frac{ (q\theta_2)(X_1,x_2)}{q_1(X_1)}
                      -\frac{(x_2-X_1){\rm div}_{x_1}(q\theta_1)(X_1,x_2)}{q_1(X_1)} 
              \Big) \mathrm{d}x_2
\Big\|_{\mathbb{L}^p(\mu_1)}.
\end{align*}
\end{Lemma}

The proof of this Lemma is deferred to the end of this section. 
\begin{Lemma}\label{lemma:density in H}
Let $H^{p'}_{\mu_1}$ be defined by \eqref{eq:deriv martingale class}, $ W^{p'} ( \Omega_1, w )$ be defined by \eqref{eqdef:sobolev} and $\mu$ satisfies \ref{ass: mu converse inequality}.
Then, the closure of $C^1_b$ in $ W^{p'} ( \Omega_1, w )$ w.r.t. the norm $ \Vert \cdot \Vert_{p', w} $ is $H^{p'}_{\mu_1}$.
\end{Lemma}

The proof of this lemma is also deferred to the end of this section. 
The last result is the main ingredient for the existence of the optimal hedge $h_{\rm M}$.

\vspace{3mm}
\noindent {\bf Proof of Proposition \ref{prop:optim problem wass}} 
We first prove the topological properties of $H^{p'}_{\mu_1}$, then move on to the existence of an optimizer.

\noindent\textbf{Step $1$.} We first prove that $H^{p'}_{\mu_1}$ is a reflexive Banach space.
Indeed, both $q_1$ and $v_{p'}$, which are defined by Assumption \ref{ass: mu converse inequality}, are continuous and strictly positive on $\Omega_1$. 
Hence, $1/q_1$ and $1/v_{p'}$ are in $L^{1}_{\rm loc} (\Omega_1)$ in the sense that for all compact $K$ such that $K \subset \Omega_1$ and $ K \cap \partial \Omega_1 = \emptyset $, $\int_K 1/q_1 < \infty$ and $ \int_K 1/v_{p'} <\infty$.  
Consequently, by the remark following equation $(1.2)$ of \citeauthor{1998} \cite{1998}, $ W^{p'} ( \Omega_1,w ) $ is a Banach space. 
Now let 
$
 \psi  :  h \in H^{p'}_{\mu_1} \rightarrow ( \partial_{x_1} h , h )  \in  \mathbb{L}^{p'} (v_{p'}) \times \mathbb{L}^{p'} (\mu_1) 
 $
where $\mathbb{L}^{p'} (v_{p'})$ is the weighted $\mathbb{L}^p$ space with respect to the weighted Lebesgue measure $ v_{p'} (x) \lambda_S(\mathrm{d}x)$. 
Endowing $\mathbb{L}^{p'} (v_{p'}) \times \mathbb{L}^{p'} (\mu_1) $ with the product norm $\vert \cdot \vert $, by definition of $\Vert h \Vert_{H^{p'}_{\mu_1}}$, there exist $C_1 > 0$ and $C_2 > 0$ such that for $h \in H^{p'}_{\mu_1} $
\vspace{-2mm}
$$
C_1 \vert  \psi(h)\vert \leq \Vert h \Vert_{H^{p'}_{\mu_1}} \leq C_2 \vert  \psi(h)\vert.
\vspace{-3mm}
$$
Hence, since $ \mathbb{L}^{p'} (v_{p'}) \times \mathbb{L}^{p'} (\mu_1) $ and $H^{p'}_{\mu_1}$ are both Banach spaces, $\psi$ has closed range (a coercive linear map between two Banach spaces having closed range). 
Hence, $\psi$ is a closed injective embedding.
Since the range of $ \psi$ is a closed subspace of a reflexive Banach space, it is itself reflexive.
Moreover, the operator
$
 \psi : H^{p'}_{\mu_1} \longrightarrow \operatorname{Range}( \psi) $
is a continuous linear isomorphism onto its range, whose inverse is also continuous. 
It follows that $H^{p'}_{\mu_1}$ is a reflexive Banach space.

\noindent{\bf Step $2$.} We prove the existence of a minimizer. 
By the triangle inequality, it is clear that $U^{\rm M}$ is coercive. 
The convexity is also a consequence of the triangle inequality.
Hence $U^{\rm M}$ is a convex, coercive and continuous function and $H^{p'}_{\mu_1}$ is a reflexive Banach space, so the optimization problem \eqref{eq:deriv martingale class} admits a minimizer $h_{\rm M} \in H^{p'}_{\mu_1}$.
\ep

\vspace{3mm}
\noindent We are now ready for the derivation of the martingale model sensitivity. 

\vspace{3mm}
\noindent {\bf Proof of Proposition \ref{prop:order 1 expansion class martingale}}
{\bf Step $1$.} We prove the upper bound.
Assume that $g$ satisfies Assumption \ref{ass:on g},
following the same steps as in the proof of Proposition \ref{prop:order 1 expansion martingale adapted case} (i), we easily get that 
\vspace{-2mm}
$$
\limsup_{r \rightarrow 0} 
 \frac{ 
\apGMcl ( r ) 
-
\apGMcl ( 0 ) }{r} 
\leq 
\inf_{ h \in C^{1}_b ( S , S ) } 
\Vert \partial_{x} (\delta_m g   + h^{\otimes}) \Vert_{\mathbb{L}^{p'} ( \mu )} 
,
\vspace{-2mm}$$
with the usual $h^\otimes$ defined by \eqref{eqdef:otimes}.
Remarking that for $h \in C^{\infty} ( S , S ) $, we have $\Vert \partial_{x} h^{\otimes} \Vert_{\mathbb{L}^{p'}(\mu)} 
=
\E^\mu \big[ \vert (\partial_{x_1} h)^{\otimes} - h \vert^{p'} + \vert h \vert^{p'} \big]^{1/p'}\leq 2 \Vert h \Vert_{H^{p'}_{\mu_1}}$. 
Similarly, we have $
\Vert h \Vert_{H^{p'}_{\mu_1}} \leq 2 \Vert \partial_{x} h^{\otimes} \Vert_{\mathbb{L}^{p'}(\mu)} $. 
Hence the norm $\Vert .\Vert_{H^{p'}_{\mu_1}}$ is the coarsest norm which guarantees the continuity of the map  $h \mapsto \Vert \partial_{x} h^{\otimes} \Vert_{\mathbb{L}^{p'} ( \mu )} $. 
Combined with Lemma \ref{lemma:density in H}, this provides that 
\vspace{-2mm}
$$
\inf_{ h \in C^{1}_b  (S , S ) } 
\Vert \partial_{x} (\delta_m g   + h^{\otimes})  \Vert_{\mathbb{L}^{p'}(\mu)} 
=
\inf_{ h \in H^{p'}_{\mu_1} } 
\Vert \partial_{x} (\delta_m g   + h^{\otimes} ) \Vert_{\mathbb{L}^{p'}(\mu)} 
.
\vspace{-3mm}
$$

\noindent {\bf Step $2$.} We analyse the optimization problem \eqref{eq:deriv martingale class}. 
By Proposition \ref{prop:optim problem wass}, the optimization problem \eqref{eq:deriv martingale class} has a minimizer $h_{\rm M}$ and, with the notations of Assumptions \ref{ass:optimum wass}, we obtain for all $ \phi := (\phi_1,\cdots,\phi_d) \in C_0^{\infty}(S,S)$:
\vspace{-2mm}
$$
0 =\E^\mu [ \partial_{x}\phi^{\otimes} \cdot T ] 
=
\sum_{j= 1}^d \int_{\Omega_1} \nabla \phi_j ( x_1 ) \cdot \alpha_j ( x_1 ) q_1 ( x_1 ) \mathrm{d}x_1 
+ \int_\Omega \phi ( x ) \cdot \big( T_2( x )  - T_1( x ) \big)q \left( x \right) \mathrm{d}x 
.
\vspace{-3mm}
$$
By the arbitrariness of $\phi$ and the regularity Condition \ref{cond:domain and optimum} of Assumptions \ref{ass:optimum wass}, this provides
\vspace{-2mm}
\begin{equation}\label{eq:edp martingale}
-\textnormal{div} ( \alpha_j ) - \frac{ \nabla q_1}{q_1} \cdot \alpha_j + \E^\mu\big[ T_{2, j} - T_{1, j} \vert X_1 = x_1 \big] = 0 
\,\, \text{for all} \,\, j=1,\ldots,d,
\vspace{-2mm}
\end{equation}
where $T _{i} = ( T_{i, 1}, \cdots, T_{i, d} )$. 
Now, let $( T^n_1 )_n $ as defined in Assumption \ref{ass:optimum wass} and $( T^n_2 )_n $ a sequence of smooth functions such that $ \Vert T^n_2 - T_2 \Vert_{\mathbb{L}^p(\mu)} \rightarrow 0 $, set $T^n := ( T^n_1, T^n_2 )$.
By Condition \ref{cond:domain and optimum} of Assumptions \ref{ass:optimum wass}, the support of $T^n_1 $ is strictly included in $\Omega$. 
Furthermore, by Proposition $4.1$ \citeauthor{allaire:hal-02496063} \cite{allaire:hal-02496063}, for $r$ small enough $ \text{Id}_S + r ( T^n_1, 0 ) $ is a global diffeomorphism which is equal to the identity outside $\Omega$ hence $ \big( \text{Id} + r ( T^n_1, 0 ) \big) ( \Omega ) = \Omega $. 
Let 
$
\Delta_{n, r} := X_1 + r T^n_1 - \E^\mu \big[ X_2 + r T_2^n \vert X_1 + r T^n_1 \big] 
$,
then by Lemma \ref{lemma:default martingale}, 
 $ \frac{\Vert \Delta_{n, r} \Vert_{\mathbb{L}^{p} ( \mu ) }}{r} \rightarrow 
R_n 
 $
 where
 \vspace{-1.5mm}
 \begin{equation}\label{eqdef:Rest_Rn}
 R_n
 \!
 :=
 \!
\Big(
\int_{\Omega_1} 
\!
\left\vert
\int_{S} \big( 
T^n_2 \frac{q }{q_1}(x)
\!-\! ( x_2 \!-\! x_1 )\frac{1}{q_1} \textnormal{div}_{x_1} ( q T^n_1 ) ( x)
\big) \mathds{1}_{ \Omega} ( x ) \mathrm{d}x_2 
\right\vert^p
\!\!q_1 ( x_1 )
\mathrm{d} x_1 
\Big)^{1/p}
\!\!\!\!\!\!\!.
\vspace{-1.5mm}
\end{equation}
Define the following random variable, whose law is a martingale measure
\vspace{-1.5mm}
\begin{equation*}
X^n := X + r T^n ( X ) + J_2 \Delta_{n, r}, \quad \text{where $J_2$ is defined by \eqref{eqdef:caus_grad,J}.}
\vspace{-1.5mm}
\end{equation*} 
Notice that $ \W_p ( \mu, \mu^{n}_r ) \leq \gamma_{n, r} $ with $ \gamma_{n, r} := r \Vert T^n \Vert_{\mathbb{L}^{p} ( \mu )} + \Vert \Delta_{n, r}\Vert_{\mathbb{L}^{p} ( \mu )}$ yields
\vspace{-1.5mm}
\begin{align*}
\frac{\apGMcl ( \gamma_{n , r} ) - g ( \mu ) }{ r }
&\geq 
\frac{1}{r}
\E^\mu\big[ \delta_m g ( \mu, X^n   )
-
\delta_m g ( \mu, X ) 
\big]
=
R_{1, n, r} + R_{2, n, r}
\vspace{-1.5mm}
\end{align*}
with 
$
R_{1, n, r} 
:=
\frac{1}{r}
\E^\mu\big[ \delta_m g(\mu, X\!+\! r T^n) 
\!-\!
\delta_m g(\mu, X) 
\big]
$ and 
$
R_{2, n,  r} :=
\frac{1}{r}
\E^\mu \big[ \delta_m g (\mu, X^n) 
\!-\!
 \delta_m g\big( \mu, X \!+\! r T^n (X) \big) 
\big]$.
By Assumption \ref{ass:on g} on $g$ , and dominated convergence, we have 
\vspace{-1.5mm}
$$
R_{1,n, r}  \xrightarrow[r \rightarrow 0]{} \E^\mu \big[ \partial_x \delta_m g ( \mu, X ) \cdot T^n \big]
.
\vspace{-1.5mm}$$
Notice that, by Hölder's inequality together with Assumption \ref{ass:on g} on $\partial_x \delta_m g$:
\vspace{-1.5mm}
\begin{align*}
\vert R_{2, n, r} \vert &=
 \left\vert 
 \frac{1}{r} \int_0^1
  \E^\mu \big[ \partial_{x_2} \delta_m g ( \mu, X + r T^n + \lambda J_2 \Delta_{n ,r} ) \cdot \Delta_{n, r} \mathrm{d} \lambda \big] \right\vert 
\\
&\leq 
 \frac{1}{r} \int_0^1 \E^\mu \big[ 
 \vert \partial_{x_2} \delta_m g ( \mu, X + r T^n ( X ) + \lambda \Delta_{n ,r} )
 \vert^{p'} \big]^{1/p'} 
 \Vert \Delta_{n,r} \Vert_{\mathbb{L}^{p} ( \mu )}  \mathrm{d} \lambda
 \\
 &\leq C_1 \frac{\Vert \Delta_{n, r} \Vert_{\mathbb{L}^{p} ( \mu )} }{r} ( 1 + \int_0^1 ( \Vert X \Vert_{\mathbb{L}^{p} ( \mu )} + r \Vert T^n \Vert_{\mathbb{L}^{p} ( \mu )} + \lambda \Vert \Delta_{n, r} \Vert_{\mathbb{L}^{p} ( \mu )} )^{p/p'} ) \mathrm{d} \lambda
  \\
 &\leq C_2 \frac{\Vert \Delta_{n, r} \Vert_{\mathbb{L}^{p} ( \mu )} }{r}.
\vspace{-3mm}
\end{align*}
\vspace{-7mm}

\noindent 
By this estimate, we get 
\vspace{-4mm}
\begin{equation}\label{ineq:estimate gamma}
\frac{\apGMcl ( \gamma_{n, r} ) - g ( \mu ) }{ r }
\geq 
R_{1,n,r} 
-
C \frac{\Vert \Delta_{n, r} \Vert_{\mathbb{L}^{p} ( \mu )} }{r}.
\vspace{-2mm}
\end{equation}
Since $ \frac{\gamma_{n ,r}}{r} \xrightarrow[r \rightarrow 0]{} \Vert T^n \Vert_{\mathbb{L}^{p} ( \mu )} + R_n $ where $R_n$ is defined by \eqref{eqdef:Rest_Rn}. Then for all $ \kappa > 0 $, there exists $ r_\kappa$ such that for $ r < r_\kappa$, $ \gamma_{n ,r} \leq r (1 + \kappa ) (\Vert T^n \Vert_{\mathbb{L}^{p} ( \mu )} + R_n)$. 
Now, since $\apGMcl$ is increasing, $\apGMcl ( \gamma_{n ,r} ) \leq  \apGMcl ( r (1 + \kappa ) (\Vert T^n \Vert_{\mathbb{L}^{p} ( \mu )} + R_n) ) $ hence, by Inequality \eqref{ineq:estimate gamma},
\vspace{-2mm}
\begin{equation*}
(1 + \kappa ) ( \Vert T^n \Vert_{\mathbb{L}^{p} ( \mu )} + R_n)
\liminf_{r \rightarrow 0} 
\frac{\apGMcl ( r ) - \apGMcl ( 0 )  }{ r }
\geq 
 \E^\mu \big[ \partial_x \delta_m g ( \mu, X ) \cdot T^n \big]
 -
 C R_n .
 \vspace{-2mm}
 \end{equation*}
 Now, letting $ \kappa $ go to $0$, we obtain 
 \vspace{-2mm}
 \begin{equation}\label{ineq:mesure approx}
( \Vert T^n \Vert_{\mathbb{L}^{p} ( \mu )} + R_n)
\liminf_{r \rightarrow 0} 
\frac{\apGMcl ( r ) - \apGMcl ( 0 )  }{ r }
\geq 
 \E^\mu \big[ \partial_x \delta_m g ( \mu, X ) \cdot T^n \big]
 -
 C R_n .
 \vspace{-2mm}
 \end{equation}
As $T^n_1$ is smooth and compactly supported, it follows from Assumption \ref{ass: mu converse inequality} \ref{cond:density}-\ref{cond:support} on $q$ that 
\vspace{-1.5mm}
$$
\textnormal{div} (
\alpha_j^n ) 
=
\int_{S} 
( x_{2, j} - x_{1, j} ) \textnormal{div}_{x_1} ( T^n_1 q ) \frac{1}{q_1}
-
T^n_{1, j} \frac{q}{q_1}
 -
( x_{2, j} - x_{1, j} )( T^n_1 \cdot \nabla_{x_1} q_1)  \frac{q}{q_1^2} \mathrm{d}x_2,
~ j=1,\ldots,d,
\vspace{-1.5mm}
$$
$\alpha_j^n$ being defined by Assumption \ref{ass:optimum wass}. 
Then, letting $(\hat{e}_j)_{1 \leq j \leq d}$ be the canonical basis of $S$,
\vspace{-4mm}
$$
R_n 
=
\Big(
\int_{\Omega_1} 
\Big|
\sum_{j =1}^d ( -\textnormal{div} ( \alpha_j^n ) - \frac{ \partial_{x_1} q_1}{q_1} \cdot \alpha_j^n 
                        + \E^\mu\big[ T^n_{2, j} - T^n_{1, j} \vert X_1 = x_1 \big] )\hat{e}_j
\Big|^p q_1 ( x_1 ) \mathrm{d}x_1
\Big)^{1/p}
.
\vspace{-2mm}
$$
By the convergence conditions of Assumptions \ref{ass:optimum wass} and the Equation \eqref{eq:edp martingale} satisfied by $ ( \alpha_j )_j $ we see that $ R_n \rightarrow 0$. 
Furthermore, it is clear that since $T^n \xrightarrow[ n \rightarrow  +\infty ]{\mathbb{L}^p ( \mu )} T $, letting $n$ go to infinity in \eqref{ineq:mesure approx}, we get 
\vspace{-3mm}
$$
\liminf_{r \rightarrow 0} 
\frac{\apGMcl ( r ) - \apGMcl ( 0 )  }{ r }
\geq 
 \E^\mu \big[ \partial_x \delta_m g ( \mu, X ) \cdot T \big]
 =
  \Vert \partial_x \delta_m g + h_{\rm M}^{\otimes} \Vert_{\mathbb{L}^{p'}(\mu)}. 
 \vspace{-2mm}$$
\ep

\noindent {\bf Proof of Lemma \ref{lemma:default martingale}.}
Assume \ref{ass: mu converse inequality} holds and let $\theta := ( \theta_1, \theta_2 ) $ be as in the statement of Lemma \ref{lemma:default martingale}. 
By Proposition $4.1$ in \citeauthor{allaire:hal-02496063} \cite{allaire:hal-02496063}, the map $ \text{Id} + r( \theta_1, 0 )$ is a diffeomorphism with bounded differential whose inverse also has bounded differential. 
Since  $( \text{Id} + r \theta ) ( \Omega ) = \Omega $ for all $r < r_0$, we deduce that the random variable $ ( X_1 + r \theta_1 (X) , X_2 ) $ admits a density $q_r$ with the same support $\Omega$ and with density: 
 \vspace{-1mm}
\begin{equation}\label{eq:expan qr}
q_r ( x ) 
:= 
q ( u_r (x) , x_2 ) 
| \text{det} ( \partial_{x_1} u_r (x) ) | 
\mathds{1}_{\Omega} (x),
 \vspace{-1mm}
\end{equation}
where $u_r (x)$ is implicitly defined by 
 \vspace{-1mm}
\begin{equation}\label{eq:expan ur}
u_r ( x )
+
r
\theta_1 ( u_r ( x ), x_2 ) = x_1 
~\mbox{for all}~x \in \Omega.
 \vspace{-1mm}
\end{equation}
As $\theta_1$ is compactly supported, $u_r$ is equal to the projection onto the first coordinate outside of the support of $\theta_1$. 
Now, fix $ x := ( x_1, x_2 ) \in \Omega $. 
Since $\theta_1$ is smooth, by the definition of $u_r$, it is clear that 
 \vspace{-1mm}
$$
u_r ( x ) = x_1 -   r\theta_1 ( x) +  \circ( r )  \,\,\, \text{and} \,\,\, \partial_{x_1}u_r ( x ) = \text{Id} - r \partial_{x_1} \theta_1 ( x ) + \circ ( r ), 
 \vspace{-1mm}
$$
where both $\circ(r)$ are uniform in $x$.
By direct calculation and using the fact that $ \theta_1$ is smooth with compact support, we see that 
 \vspace{-1mm}
$$
q_r ( x ) = q( x ) - r \textnormal{div}_{x_1} ( q \theta_1 ) + \circ( r ) 
~\mbox{and}~
\sup_{ x \in \Omega } \vert \text{det} ( \partial_{x_1}u_r ) \left( x \right) - 1 \vert \xrightarrow[ r \rightarrow 0]{} 0.
 \vspace{-1mm}
$$
Recall from Assumption \ref{ass: mu converse inequality}-\ref{cond:reg density} that $q$ is Lipschitz in its first variable, uniformly in the second variable. 
Then, it follows that 
 \vspace{-1mm}
$$ 
\Vert q_r - q \Vert_{\mathbb{L}^\infty(\Omega)} \xrightarrow[r \rightarrow 0]{} 0
.
 \vspace{-1mm}
 $$
Now, since $q > 0 $ and is continuous, $ \text{inf}_{x \in \text{supp} ( \theta_1 ) } q ( x ) > 0 $  which implies that there exists $r_c$ such that for $r \leq r_c$, and $ x $ in the support of $\theta_1$, 
$
q_r ( x ) \geq \frac{1}{2} q ( x )  
.$
Also, outside the support of $\theta_1$, $q_r = q$. 
This proves that for all $ x \in \Omega $ and for $r \leq r_c$, we have 
 $
  q_r ( x ) \geq \frac{1}{2} q ( x )  
.$
By a similar argument, we easily have the following inequality for $ r $ small enough, 
$$
\frac{1}{r} \vert q_r - q \vert \leq 2 ( \vert \theta_1 \vert \vert \partial_1 q  \vert + q \vert \partial_1 \theta_1 \vert ).
$$
Now, for $x_1 \in \Omega_1$, since $\mu$ is a martingale measure, we have $ x_1 = \frac{\int_S x_2 q ( x_1, x_2 ) \mathrm{d}x_2 }{ \int_S q ( x_1, x_2 ) \mathrm{d}x_2 }$ \textit{a.e.} and  
$$
\E^{\mu_r}[X_2|X_1=x_1]-x_1
=
\int_S 
\frac{x_2q_r (x) 
}{
\int_S 
q_r (x) \mathrm{d}x_2
}
\mathrm{d}x_2
-
x_1 
=
\int_S 
x_2 ( 
\frac{q_r (x) 
}{
\int_S 
q_r ( x ) \mathrm{d}x_2
}
-
\frac{q (x) 
}{
\int_S 
q (x) \mathrm{d}x_2
}
) 
\mathrm{d}x_2.
$$
By the first-order expansions \eqref{eq:expan qr}, \eqref{eq:expan ur}, together with the dominated convergence Theorem, one easily sees that 
\begin{equation}\label{eq:conver term un}
\begin{split}
&\lim_{r \rightarrow 0} \!
\frac{1}{r}
\Big( \frac{q_r (x)}
                {\int q_r (  x_1 , z_2  ) \mathrm{d}z_2}
        \!-\! \frac{q (x)}
                 {\int q ( x_1 , z_2  ) \mathrm{d}z_2}
\Big) 
\!=\!
\frac{q(x)\!\int \!\textnormal{div}_{x_1} \!( q \theta_1 \! )\mathds{1}_{\Omega} ( x_1, z_2 ) \mathrm{d}z_2 }
        {q_1( x_1 )^2}
\!-\!
\frac{\textnormal{div}_{x_1}(q\theta_1) (x) }
        {q_1( x_1 )}.
    \end{split}
\end{equation}
Hence 
\begin{align*}
\frac{\E^{\mu_r}[X_2|X_1=x_1]-x_1}{r}
&\rightarrow 
\int_S 
x_2 ( \frac{q(x)\!\int \!\textnormal{div}_{x_1} ( q \theta_1 )\mathds{1}_{\Omega} ( x_1, z_2 ) \mathrm{d}z_2 }
        {q_1( x_1 )^2}
-\frac{\textnormal{div}_{x_1}(q\theta_1) (x) }
        {q_1( x_1 )}
) 
\mathrm{d}x_2.
\\
&=
\big( \int_S 
x_2 q(x) \mathrm{d}x_2 \big)\frac{ \!\int \!\textnormal{div}_{x_1} ( q \theta_1 )\mathds{1}_{\Omega} ( x_1, z_2 ) \mathrm{d}z_2 }
        {q_1( x_1 )^2}
-
\int_S 
x_2 \frac{\textnormal{div}_{x_1}(q\theta_1) (x) }
        {q_1( x_1 )}
\mathrm{d}x_2.
\\
&=
x_1 \frac{ \!\int \!\textnormal{div}_{x_1} ( q \theta_1 )\mathds{1}_{\Omega} ( x_1, z_2 ) \mathrm{d}z_2 }
        {q_1( x_1 )}
-
\int_S 
x_2 \frac{\textnormal{div}_{x_1}(q\theta_1) (x) }
        {q_1( x_1 )}
\mathrm{d}x_2.
\\
&=
\int_S (x_1 - x_2) 
 \frac{ \textnormal{div}_{x_1} ( q \theta_1 )\mathds{1}_{\Omega} ( x_1, x_2 ) }{q_1( x_1 )}
 \mathrm{d}x_2 
\end{align*}
Since $\theta_2 $ is continuous with respect to its first variable and $u_r$ differs from identity only in a compact set, we have 
 \vspace{-3.5mm}
\begin{equation}\label{eq:conver term deux}
 \theta_2 ( u_r ( x ), x_2 ) 
\frac{q_r (  x  ) 
}{
\int_S 
q_r (  x_1 , z_2  ) \mathrm{d}z_2
}
\longrightarrow
\theta_2 ( x ) \frac{q ( x ) }{q_1 ( x_1 ) }
.
 \vspace{-1mm}
\end{equation}
Then expressing $ N_r := \Vert 
\E^\mu \big[ X_2 + r \theta_2 \vert X_1 + r \theta_1 \big] - X_1 - r \theta_1  \Vert_{\mathbb{L}^{p} ( \mu )}  $ in terms of the density $q$, we have 
$$
N_r = \Big( \int_{\Omega} \Big\vert 
\E^{\mu_r}[X_2|X_1=x_1] + r\E^{\mu_r}[\theta_2|X_1=x_1] - x_1 \Big\vert^p q_r (x_1, z) \mathrm{d}x_1 \mathrm{d}z \Big)^{1/p}
$$
we deduce from the Convergences \eqref{eq:conver term un} and \eqref{eq:conver term deux} together with the dominated convergence Theorem that
$$
\frac{N_r}{r} \rightarrow 
\Big(
\int_{\Omega_1} 
\Big|
\int_{S} ( - ( x_2 - x_1 )\frac{1}{q_1} \textnormal{div}_{x_1} ( q \theta_1 ) ( x )
+
\theta_2 ( x ) \frac{q ( x ) }{q_1 ( x_1 ) }
) \mathds{1}_{ \Omega} ( x ) \mathrm{d}x_2 
\Big|^p
q_1 ( x_1 )
\mathrm{d} x_1 
\Big)^{1/p}
.
 \vspace{-3mm}
$$
\ep

\vspace{2mm}

\noindent 
{\bf Proof of Lemma \ref{lemma:density in H}}
By definition, $H^{p'}_{\mu_1}$ is the closure of $C^\infty ( S, S ) \cap \mathbf{W}^{p}(\Omega, (q_1, v_{p'})) $ with respect to the norm $ \Vert \cdot \Vert_{p', w} $. Since $ C^\infty_b ( S, S )  \subset C^1_b $ , we only need to prove that any function in $C^1_b$ can be approximated by a sequence of $C^\infty ( S, S )$ with respect to the norm $ \Vert \cdot \Vert_{p', w } $. Let $ f \in C^1_b $ , by convolution, there exists a sequence of smooth functions $( \phi_n ) $ such that $ \phi_n \rightarrow f $ , $\partial_x \phi_n \rightarrow \partial_x f $ uniformly in every compact. 
Furthermore, since $f$ is bounded, one can choose $( \phi_n )_n $ such that $ \text{max} ( \Vert \phi_n \Vert_\infty,  \Vert \partial_x \phi_n \Vert_\infty  ) \leq \text{max} ( \Vert f \Vert_\infty,  \Vert \partial_x f \Vert_\infty  ) $, so by uniform integrability, we have $\Vert f - \phi_n \Vert_{p', w } \rightarrow 0$.
\ep

\subsection{Fredholm integral equation defined by the operator $\Kc$}

In this section, we study the computation of the martingale model risk optimal hedge $h_{\rm M}$ in the one-dimensional case $d=1$ with $p = 2$. 
Our objective is to characterize $h_{\rm M}$ as the solution of the integro-differential Equation \eqref{eqdef:fred hedge}. 

\begin{Lemma}\label{lemma:injectivity kernel}
Let  $\Kc$ be the Hilbert-Schmidt operator defined by \eqref{eqdef:kernel}. 
Under Assumptions \ref{ass: mu converse inequality} and \ref{ass:fredholm}, the spectrum of $\Kc $ is $\lbrace 0 \rbrace $, and hence $I - \Kc $ is invertible.
\end{Lemma}

\proof By the classical Hilbert-Schmidt operator theory, we know that $ \Kc $ is compact and that $0$ is in the spectrum. 
Let $ \lambda \neq 0 $ with corresponding eigenvector $f$. 
By the Fredholm alternative, we only need to check that $ f $ is $0$. 
We have 
$
\lambda f( x_1 ) 
=
2\int_\ell^{x_1 }
K ( x_1, y_1 ) f ( y_1 ) q_1 ( y_1 ) \mathrm{d} y_1,
$
hence, since $k $ is non decreasing and $q_1$ is continuous, 
$$
\vert \lambda \vert \vert f( x_1 ) \vert 
\leq 
C
\int_\ell^{x_1 }
\big( k ( b ) - k ( y_1 ) \big)^+ \sqrt{ q_1 ( y_1 ) }
\vert f ( y_1 )\vert \mathrm{d} y_1,
$$
where $\ell$ and $b$ are defined in Assumption \ref{ass:fredholm}. Hence, by the Gronwall Lemma, $f$ is $0$ whenever $\lambda \neq 0 $.
\ep

\vspace{5mm}
\noindent {\bf Proof of Proposition \ref{prop:fredholm equation}}
The existence of the optimal hedge $h_{\rm M}$ is established in Proposition \ref{prop:optim problem wass}. 
Let $\mu$ satisfy Assumption \ref{ass:fredholm} and denote $\psi:=\delta_m g(\mu, \cdot)+h_{\rm M}^\otimes$. Applying the standard calculus of variation arguments for the minimization problem \eqref{eq:deriv martingale class}, the map $\eps\longmapsto U^{\rm M} (h_{\rm M}+\eps f)$ has an interior minimum at the point $\eps=0$ for all perturbations $f\in C^1_b$ with compact support. 
The first-order condition then provides
 \vspace{-1mm}
\begin{equation}\label{eqdef:first order cold Wass}
0
=
\E^\mu\Big[\partial_x(f^\otimes)(X)
                   \cdot
                   \partial_x\psi(X)
           \Big]
~\mbox{for all}~
f\in C^1_b.
 \vspace{-1mm}
\end{equation}
Now, notice that $ \partial_{x_1} f^{\otimes} = \left. f' \right.^\otimes - f $ and $ \partial_{x_2} f^{\otimes} = f $, hence 
 \vspace{-1mm}
$$
0 = \E^\mu [ f ( X_1 ) ( \partial_{x_2} - \partial_{x_1} ) \psi(X)+ f' ( X_1 ) ( X_2 - X_1 ) \partial_{x_1}\psi(X)  ]
.
 \vspace{-1mm}
 $$
Replacing $\psi$ by its expression yields and simplifying since $\mu$ is a martingale measure yields, 
\begin{align*}
0 &= \E^\mu 
	\big[ 
			f ( X_1 ) (
				\partial_{x_2} - \partial_{x_1} ) \delta_m g (\mu, X) + 2h_{\rm M}(X_1) 
	\\
	&\hspace{15mm}
	+ f' ( X_1 ) ( X_2 - X_1 ) ( \partial_{x_1} \delta_m g (\mu, X)
							   	+ h_{\rm M}'(X_1)(X_2 - X_1) )
	\big]
\\ 
&= \int
	\Big(
			f ( x_1 ) \big( 
			 (\partial_{x_2} - \partial_{x_1} ) \delta_m g (\mu, x) + 2h_{\rm M}(x_1) 
				\big)
		\\
	&\hspace{15mm}+ f' ( x_1 ) ( x_2 - x_1 ) \big( \partial_{x_1} \delta_m g (\mu, x)
							   	+ h_{\rm M}'(x_1)(x_2 - x_1) \big)
	\Big) q(x) \mathrm{d}x.
\end{align*}
By integration by parts, this provides 
\begin{align*}
&\int f ( x_1 ) \big( (\partial_{x_2} - \partial_{x_1} ) \delta_m g (\mu, x) + 2h_{\rm M}(x_1) 
				\big)
			 	q(x) \mathrm{d}x
			 	\\
		&=
-
\int f'(x_1) \Big( \int_{-\infty}^{x_1} \int  \big(
				\partial_{x_2} - \partial_{x_1} ) \delta_m g (\mu, \xi_1, x_2) + 2h_{\rm M}(\xi_1) \big)
			 	q(\xi_1 , x_2 ) \mathrm{d}x_2 \mathrm{d} \xi_1 \Big) \mathrm{d}x_1
.
\end{align*}
Hence the equality 
\begin{align*}
&\int f'(x_1) \Big( \int_{-\infty}^{x_1} \int  \big(
				(\partial_{x_2} - \partial_{x_1} ) \delta_m g (\mu, \xi_1, x_2) + 2h_{\rm M}(\xi_1) \big)
			 	q(\xi_1 , x_2 ) \mathrm{d}x_2 \mathrm{d} \xi_1 \Big) \mathrm{d}x_1
\\
&=
\int
f' ( x_1 ) ( x_2 - x_1 ) \big( \partial_{x_1} \delta_m g (\mu, x)
							   	+ h_{\rm M}'(x_1)(x_2 - x_1) \big)
	\Big) q(x) \mathrm{d}x,
\end{align*}
or, denoting $ \E^{\mu}_{1, \xi_1}  := \E^{\mu} [ \cdot \vert X_1 = \xi_1 ]$, 
\begin{align*}
&\int f'(x_1) \Big( \int_{\ell}^{x_1} 
				\big(
						\E^{\mu}_{1, \xi_1} \big[
								( \partial_{x_2} - \partial_{x_1} ) \delta_m g (\mu, \xi_1, X_2) \big] 
						+ 2h_{\rm M}(\xi_1) \big)q_1 ( \xi_1 ) \mathrm{d} \xi_1 
						\Big) \mathrm{d}x_1
\\
&=
\int
f' ( x_1 ) \Big( \E^{\mu}_{1, x_1} \big[ ( X_2 - X_1 ) \partial_{x_1} \delta_m g (\mu, X) \big] q_1(x_1)
	+ h_{\rm M}'(x_1) v_2 (x_1) 
	\Big)  \mathrm{d}x_1
.\end{align*}
By the arbitrariness of $f \in C_b^1$, we deduce that there exists a constant $c_1$ such that
\begin{align*}
-c_1 + 
&\int_{-\infty}^{x_1} 
				\big(
						\E^{\mu}_{1, \xi_1} \big[
								( \partial_{x_2} - \partial_{x_1} ) \delta_m g (\mu, \xi_1, X_2) \big] 
						+ 2h_{\rm M}(\xi_1) \big)q_1 ( \xi_1 ) \mathrm{d} \xi_1 
\\
&=
 \E^{\mu}_{1, x_1}\big[ ( X_2 - X_1 ) \partial_{x_1} \delta_m g (\mu, X)  \big]
	+ h_{\rm M}'(x_1) v_2 (x_1 ) 
\,\,\, \text{\textit{a.e.} on } \,\, I, 
\end{align*}
which is exactly the integro-differential equation \eqref{eqdef:fred hedge}.

\noindent We next continue under the additional Assumption \ref{ass:fredholm}. Dividing the last equation by $v_2$ which is assumed to be strictly positive and integrating, we obtain the Fredholm equation:  
\begin{align*}
h_{\rm M}(x_1)
 &=
 c_0 + c_1k(x_1)+ u(x_1)+
 2\E^{\mu} \Big[
  \big( k( x_1 ) - k ( X_1 ) \big)^+
h_{\rm M}( X_1 ) \Big],
\end{align*}
where $u$ is defined by \eqref{eqdef:def u}.
Since $ I - \Kc $ is an isomorphism, we may decompose $h_{\rm M} = \sum_{i = 0}^1 c_i \phi_i  + \Psi $ where  $\phi_0, \phi_1, \Psi$ are solutions of $\phi_0 - \Kc [ \phi_0 ] = 1 $ where \(1\) denotes the constant function equal to \(1\), and $\phi_1 - \Kc [ \phi_1 ] = k $, and $ \Psi- \Kc [ \Psi ] = u  $.
In order to obtain the constants $c_0,c_1$, we observe that the perturbation of the solution $h_{\rm M}$ by the constant $f=1$ and $f = \text{Id}$ induce after direct manipulation both Equations \eqref{f=1} and \eqref{f=Id}. These perturbations are indeed admissible as 
the first-order condition \eqref{eqdef:first order cold Wass} holds for all $f$ in $ H^{2}_{\mu_1}$, defined in \eqref{eq:deriv martingale class}.

\noindent Conversely, it is straightforward to verify that any solution $h_{\rm M}$ of the Fredholm equation solves our optimization problem due to its convexity. 
\ep

\subsection{Sensitivities under additional first marginal constraint}

{\bf Proof of Proposition \ref{prop:order 1 expansion marginalandmartingale adapted case}}
We follow the lines of the proof of Proposition \ref{prop:order 1 expansion martingale adapted case} and we only consider the proof of the sensitivity of $G_{{\rm ad}}^{\rm M, {{\rm m}_1}}$, the remaining sensitivities follow the same line of argument. 
Let $g$ satisfy Assumption \ref{ass:on g}. 

\vspace{3mm}
\noindent {\bf Step $1$.} 
We simplify the problem. 
As the set of $\mu' \!\in\! \Pc_p ( \X ) $ in $ \BadM ( \mu, r )$ satisfying the constraint $\mu'\circ X_1^{-1} = \mu_1$ is included in $  \Bcl ( \mu, r ) $, the result of Lemma \ref{lemma:reducto al linear} still holds: 
$$
\underline{G}_{{\rm ad}}^{\rm M, {{\rm m}_1}} ( r ) 
=
\apGMmiad ( r ) 
+ \circ ( r ),
~\mbox{with}~
\apGMmiad ( r ) 
=
g ( \mu ) 
+
\sup_{\mu' \in \BadMmi ( \mu, r ) } 
\E^{\mu'} [\delta_m g ]. 
$$

\noindent{\bf Step $2$.} 
We prove the upper bound. 
As the constraint $\mu'\circ X_1^{-1} = \mu_1$ is equivalent to 
$\E^{\mu'} [ f ( X_1 ) ] = \E^{\mu} [ f ( X_1 ) ] $ for all $f \in C^{1}_b$, we obtain by following the steps of the proof of Proposition \ref{prop:order 1 expansion martingale adapted case} that 
 \vspace{-2mm}
$$
\limsup_{r \rightarrow 0} \frac{\apGMmiad ( r ) - \apGMmiad ( 0 )}{r}
\leq 
\big\| 
\partial_x^{\rm ad} \delta_m g + J h ( X_1 ) + J_1 f' ( X_1 ) \big\|_{\mathbb{L}^{p'}(\mu)}
,
 \vspace{-2mm}$$
by a density argument, this implies that
 \vspace{-2mm}
 $$
\limsup_{r \rightarrow 0} \frac{\apGMmiad ( r ) - \apGMmiad ( 0 )}{r}
\leq 
\inf_{f, h \in \mathbb{L}^{p'}(\mu_1)}
\big\|
\partial_x^{\rm ad} \delta_m g + J h ( X_1 ) + J_1 f ( X_1 ) \big\|_{\mathbb{L}^{p'}(\mu)}
. \vspace{-1mm}
$$
\noindent{\bf Step $3$.} 
We now prove the lower bound. 
Again, the problem \eqref{eqref:deriv marginal} is convex, coercive, continuous and admits a minimizer $(h_{\rm M, m_1}, f_{\rm M, m_1})$. Letting, $\mathbf{N}$ be defined by \eqref{eqdef:n and n_ad}, $(h, f)$ satisfies the first-order condition
 \vspace{-1mm}
\begin{equation}\label{eqref:first order condition marginal martingal}
\begin{split}
\mathbf{N} \big( 
&\E^\mu_1 [ \partial_{x_1} \delta_m g ] - h_{\rm M, m_1} ( X_1 ) + f_{\rm M, m_1}( X_1 ) \big)
=
\E^\mu_1 \Big[ 
\mathbf{N} \big(
\partial_{x_2} \delta_m g 
+
h( X_1 ) \big)
\Big]
\\
&\E ^\mu_1\big[ \partial_{x_1} \delta_m g \big] - h_{\rm M, m_1} + f_{\rm M, m_1}
=
0
.
\end{split}
\end{equation}
Set $\phi_1 ( X ) :=  \E ^\mu_1\big[ \partial_{x_1} \delta_m g \big] - h_{\rm M, m_1} + f_{\rm M, m_1}$ and $\phi_2 ( X ) := 
\partial_{x_2} \delta_m g  + h$. 
By the first-order condition \eqref{eqref:first order condition marginal martingal}, $\phi_1  = 0 $. 
Then the coupling $ \mu^{rT} := \mu\circ\big( X + r T ( X ) \big)^{-1} $ where $ T := \frac{1}{c} \Nad ( \phi ) $, where $c$ is a normalization constant satisfies $ \mu^{rT} \circ X_1^{-1} = \mu_1$ by Equation \eqref{eqref:first order condition marginal martingal}. 
Furthermore, by the first-order condition \eqref{eqref:first order condition marginal martingal}, $ \mu^{rT}$ is martingale. 
Finally, the coupling $ \pi_r := \mu \circ \big(  X ,  X + r T ( X ) \big)^{-1}$ is clearly bi-causal since $ \phi_1  = 0$. 
The required result is now obtained by following the line of argument of the proof of Proposition \ref{prop:order 1 expansion martingale adapted case} (i), we get the desired result. 
\ep

\subsection{Model risk sensitivity for the optimal stopping problem}

\noindent {\bf Proof of Proposition \ref{prop:deriv optimal stopping mart adapted}} Let $\ell$ satisfy Assumptions \ref{ass:optimal stopping}. 
The proof of all results \eqref{eqref:deriv american option marginal}, \eqref{eqref:deriv american option martingale and marginal}, \eqref{eqref:deriv american option martingale} consists in following the same line of argument as in \citeauthor{backhoff_adapted_2020} \cite{backhoff_adapted_2020} (proof of Theorem $2.8$). 
The only difference is that the family of almost optimal couplings has to be modified in order to derive the lower bound.
The family of couplings to consider is given in step $3$ of the proof of Proposition \ref{prop:order 1 expansion martingale adapted case} (i) and step $3$ of the proof of Proposition \ref{prop:order 1 expansion marginalandmartingale adapted case}.
\ep

\printbibliography

@incollection{allaire:hal-02496063,
	title        = {Shape and topology optimization},
	author       = {Allaire, Gr{\'e}goire and Dapogny, Charles and Jouve, Fran{\c{c}}ois},
	year         = 2021,
	booktitle    = {Handbook of numerical analysis},
	publisher    = {Elsevier},
	volume       = 22,
	pages        = {1--132}
}

@misc{Bach,
  title        = {Learning Theory from First Principles},
  author       = {Bach, Francis},
  year         = {2025},
  note         = {Version of May 26, 2025},
  url          = {https://www.di.ens.fr/~fbach/ltfp_book.pdf}
}

@article{backhoff_adapted_2020,
	title        = {Adapted Wasserstein distances and stability in mathematical finance},
	author       = {Backhoff-Veraguas, Julio and Bartl, Daniel and Beiglb{\"o}ck, Mathias and Eder, Manu},
	year         = 2020,
	journal      = {Finance and Stochastics},
	publisher    = {Springer},
	volume       = 24,
	number       = 3,
	pages        = {601--632}
}

@article{backhoff_all_2020,
	title        = {All adapted topologies are equal},
	author       = {Backhoff-Veraguas, Julio and Bartl, Daniel and Beiglb{\"o}ck, Mathias and Eder, Manu},
	year         = 2020,
	journal      = {Probability Theory and Related Fields},
	publisher    = {Springer},
	volume       = 178,
	pages        = {1125--1172}
}

@article{backhoff_estimating_2022,
	title        = {Estimating processes in adapted Wasserstein distance},
	author       = {Backhoff, Julio and Bartl, Daniel and Beiglb{\"o}ck, Mathias and Wiesel, Johannes},
	year         = 2022,
	journal      = {The Annals of Applied Probability},
	publisher    = {Institute of Mathematical Statistics},
	volume       = 32,
	number       = 1,
	pages        = {529--550}
}

@article{bartl2021sensitivity,
	title        = {Sensitivity analysis of Wasserstein distributionally robust optimization problems},
	author       = {Bartl, Daniel and Drapeau, Samuel and Ob{\l}{\'o}j, Jan and Wiesel, Johannes},
	year         = 2021,
	journal      = {Proceedings of the Royal Society A},
	publisher    = {The Royal Society},
	volume       = 477,
	number       = 2256,
	pages        = 20210176
}

@article{bartlsensitivityadapted,
	title        = {Sensitivity of Multiperiod Optimization Problems with Respect to the Adapted Wasserstein Distance},
	author       = {Bartl, Daniel and Wiesel, Johannes},
	year         = 2023,
	journal      = {SIAM Journal on Financial Mathematics},
	volume       = 14,
	number       = 2,
	pages        = {704--720}
}

@article{bayraktar_nonparametric_2023,
	title        = {Nonparametric adaptive robust control under model uncertainty},
	author       = {Bayraktar, Erhan and Chen, Tao},
	year         = 2023,
	journal      = {SIAM Journal on Control and Optimization},
	publisher    = {SIAM},
	volume       = 61,
	number       = 5,
	pages        = {2737--2760}
}

@article{blanchet_distributionally_2018,
	title        = {Distributionally robust mean-variance portfolio selection with Wasserstein distances},
	author       = {Blanchet, Jose and Chen, Lin and Zhou, Xun Yu},
	year         = 2022,
	journal      = {Management Science},
	publisher    = {INFORMS},
	volume       = 68,
	number       = 9,
	pages        = {6382--6410}
}

@article{blanchet_distributionally_2024,
author = {Jose Blanchet and Jiajin Li and Sirui Lin and Xuhui Zhang},
title = {{Distributionally Robust Optimization and Robust Statistics}},
volume = {40},
journal = {Statistical Science},
number = {3},
publisher = {Institute of Mathematical Statistics},
pages = {351 -- 377},
year = {2025},
doi = {10.1214/24-STS955},
URL = {https://doi.org/10.1214/24-STS955}
}

@article{blanchet_quantifying_2016,
	title        = {Quantifying distributional model risk via optimal transport},
	author       = {Blanchet, Jose and Murthy, Karthyek},
	year         = 2019,
	journal      = {Mathematics of Operations Research},
	publisher    = {INFORMS},
	volume       = 44,
	number       = 2,
	pages        = {565--600}
}

@article{blanchet_robust_2019,
	title        = {Robust Wasserstein profile inference and applications to machine learning},
	author       = {Blanchet, Jose and Kang, Yang and Murthy, Karthyek},
	year         = 2019,
	journal      = {Journal of Applied Probability},
	publisher    = {Cambridge University Press},
	volume       = 56,
	number       = 3,
	pages        = {830--857}
}

@inproceedings{blanchet_statistical_2023,
	title        = {Statistical limit theorems in distributionally robust optimization},
	author       = {Blanchet, Jose and Shapiro, Alexander},
	year         = 2023,
	booktitle    = {2023 Winter Simulation Conference (WSC)},
	pages        = {31--45},
	organization = {IEEE}
}

@article{blanchet2024empirical,
  title={Empirical martingale projections via the adapted Wasserstein distance},
  author={Jose Blanchet and Johannes Wiesel and Erica Zhang and Zhenyuan Zhang},
  journal={The Annals of Applied Probability},
  year={2024},
}

@book{CarmonaDelarue,
  title     = {Probabilistic Theory of Mean-Field Games with Applications I--II},
  author    = {Carmona, Ren{\'e} and Delarue, Fran{\c{c}}ois},
  year      = {2018},
  publisher = {Springer}
}

@article{duoandikoetxea_forty_2013,
	title        = {Forty years of Muckenhoupt weights},
	author       = {Duoandikoetxea, Javier},
	year         = 2013,
	journal      = {Function Spaces and Inequalities},
	publisher    = {Charles University and Academy of Sciences Prague},
	pages        = {23--75}
}

@article{farokhi_distributionally-robust_2023,
	title        = {Distributionally robust optimization with noisy data for discrete uncertainties using total variation distance},
	author       = {Farokhi, Farhad},
	year         = 2023,
	journal      = {IEEE Control Systems Letters},
	publisher    = {IEEE},
	volume       = 7,
	pages        = {1494--1499}
}

@article{fuhrmann_wasserstein_2023,
	title        = {Wasserstein perturbations of Markovian transition semigroups},
	author       = {Fuhrmann, Sven and Kupper, Michael and Nendel, Max},
	year         = 2023,
	journal    = {Annales de l'Institut Henri Poincare (B) Probabilites et statistiques},
	volume       = 59,
	number       = 2,
	pages        = {904--932},
	organization = {Institut Henri Poincar{\'e}}
}

@book{greeks,
  author       = {Hull, John C.},
  title        = {Options, Futures, and Other Derivatives},
  publisher    = {Pearson},
  edition      = {9},
  year         = {2015}
}

@article{han2022distributionally,
  title={Distributionally robust risk evaluation with a causality constraint and structural information},
  author={Han, Bingyan},
  journal={Mathematical Finance},
  year={2022},
  publisher={Wiley Online Library}
}

@article{jiang2024sensitivity,
	title        = {Sensitivity of causal distributionally robust optimization},
	author       = {Jiang, Yifan and Ob{\l}{\'o}j, Jan},
	year         = 2024,
	journal      = {arXiv preprint arXiv:2408.17109}
}

@article{jiang2024duality,
  title={Duality of causal distributionally robust optimization: the discrete-time case},
  author={Jiang, Yifan},
  journal={arXiv preprint arXiv:2401.16556},
  year={2024}
}

@article{lam_robust_2016,
	title        = {Robust sensitivity analysis for stochastic systems},
	author       = {Lam, Henry},
	year         = 2016,
	journal      = {Mathematics of Operations Research},
	publisher    = {INFORMS},
	volume       = 41,
	number       = 4,
	pages        = {1248--1275}
}

@article{lanzetti_first-order_2022,
	title        = {First-Order Conditions for Optimization in the Wasserstein Space},
	author       = {Lanzetti, Nicolas and Bolognani, Saverio and D{\"o}rfler, Florian},
	year         = {2025},
	journal      = {SIAM Journal on Mathematics of Data Science},
	volume       = {7},
	number       = {1},
	pages        = {274--300},
	doi          = {10.1137/23M156687X}
}

@phdthesis{margheriti_sur_2020,
	title        = {Sur la stabilit{\'e} du probl{\`e}me de transport optimal martingale},
	author       = {Margheriti, William},
	year         = 2020,
	school       = {Universit{\'e} Paris-Est}
}

@article{mohajerin_esfahani_data-driven_2018,
	title        = {Data-driven distributionally robust optimization using the Wasserstein metric: Performance guarantees and tractable reformulations},
	author       = {Mohajerin Esfahani, Peyman and Kuhn, Daniel},
	year         = 2018,
	journal      = {Mathematical Programming},
	publisher    = {Springer},
	volume       = 171,
	number       = 1,
	pages        = {115--166}
}

@article{nendel_parametric_2022,
	title        = {A parametric approach to the estimation of convex risk functionals based on Wasserstein distance},
	author       = {Nendel, Max and Sgarabottolo, Alessandro},
	year         = {2025},
	journal      = {Applied Mathematics \& Optimization},
	volume       = {93},
	number       = {1},
	doi          = {10.1007/s00245-025-10352-5}
}

@article{neufeld_robust_2024,
  title        = {Robust SGLD algorithm for solving non-convex distributionally robust optimisation problems},
  author       = {Neufeld, Ariel and En, Matthew Ng Cheng and Zhang, Ying},
  year         = {2024},
  journal      = {arXiv preprint arXiv:2403.09532}
}

@book{pflug_multistage_2014,
	title        = {Multistage stochastic optimization},
	author       = {Pflug, Georg Ch and Pichler, Alois},
	year         = 2014,
	publisher    = {Springer},
	volume       = 1104
}

@article{tolle_uniqueness_2012,
	title        = {Uniqueness of weighted Sobolev spaces with weakly differentiable weights},
	author       = {T{\"o}lle, Jonas M},
	year         = 2012,
	journal      = {Journal of Functional Analysis},
	publisher    = {Elsevier},
	volume       = 263,
	number       = 10,
	pages        = {3195--3223}
}

@article{von1928theorie,
	title        = {Zur theorie der gesellshaftsphiele},
	author       = {von Neuman, J},
	year         = 1928,
	journal      = {Math. Ann},
	volume       = 100
}

@article{1998,
	title        = {Weighted sobolev spaces},
	author       = {Zhikov, Vasilii Vasil’evich},
	year         = 1998,
	journal      = {Sbornik: Mathematics},
	publisher    = {IOP Publishing},
	volume       = 189,
	number       = 8,
	pages        = {1139--1139}
}

\end{document}